\documentclass[amsmath,nofootinbib,notitlepage,prd,superscriptaddress,twocolumn]{revtex4-1}

\usepackage[usenames,dvipsnames]{xcolor}
\usepackage{aas_macros,graphicx,hyperref,orcidlink,txfonts,array,multirow}

\hypersetup{colorlinks=true,citecolor=RoyalBlue,linkcolor=RoyalBlue,urlcolor=RoyalBlue}

\newcommand{\ab}[1]{\left|#1\right|}
\newcommand{\av}[1]{\left\langle#1\right\rangle}
\newcommand{\br}[1]{\left[#1\right]}
\newcommand{\cu}[1]{\left\{#1\right\}}
\newcommand{\pa}[1]{\left(#1\right)}
\newcommand{\ed}{\mathop{}\!\mathrm{d}}
\newcommand{\pd}{\mathop{}\!\partial}
\DeclareMathOperator\arcsinh{arcsinh}
\DeclareMathOperator\cn{cn}
\DeclareMathOperator\sn{sn}
\DeclareMathOperator\re{Re}
\DeclareMathOperator\sign{sign}
\DeclareMathOperator\bigO{\mathcal{O}}

\begin{document}

\title{Adaptive Analytical Ray Tracing of Black Hole Photon Rings\texorpdfstring{\vspace{-8pt}}{}}

\author{Alejandro C\'ardenas-Avenda\~no\,\orcidlink{0000-0001-9528-1826}} 
\affiliation{Princeton Gravity Initiative, Princeton University, Princeton, New Jersey 08544, USA}
\affiliation{Programa de Matem\'atica, Fundaci\'on Universitaria Konrad Lorenz, 110231 Bogot\'a, Colombia}

\author{Alexandru Lupsasca\,\orcidlink{0000-0002-1559-6965}}
\affiliation{Princeton Gravity Initiative, Princeton University, Princeton, New Jersey 08544, USA}
\affiliation{Department of Physics \& Astronomy, Vanderbilt University, Nashville, Tennessee 37212, USA}

\author{Hengrui Zhu\,\orcidlink{0000-0001-9027-4184}}
\affiliation{Princeton Gravity Initiative, Princeton University, Princeton, New Jersey 08544, USA}

\begin{abstract}
\vspace{-10pt}
Recent interferometric observations by the Event Horizon Telescope have resolved the horizon-scale emission from sources in the vicinity of nearby supermassive black holes.
Future space-based interferometers promise to measure the ``photon ring''---a narrow, ring-shaped, lensed feature predicted by general relativity, but not yet observed---and thereby open a new window into strong gravity.
Here we present \texttt{AART}: an Adaptive Analytical Ray-Tracing code that exploits the integrability of light propagation in the Kerr spacetime to rapidly compute high-resolution simulated black hole images, together with the corresponding radio visibility accessible on very long space-ground baselines.
The code samples images on a nonuniform adaptive grid that is specially tailored to the lensing behavior of the Kerr geometry and is therefore particularly well-suited to studying photon rings.
This numerical approach guarantees that interferometric signatures are correctly computed on long baselines, and the modularity of the code allows for detailed studies of equatorial sources with complex emission profiles and time variability.
To demonstrate its capabilities, we use \texttt{AART} to simulate a black hole movie of a stochastic, non-stationary, non-axisymmetric equatorial source; by time-averaging the visibility amplitude of each snapshot, we are able to extract the projected diameter of the photon ring and recover the shape predicted by general relativity.
\end{abstract}

\maketitle

\section{Introduction}
\vspace{-8pt}

According to general relativity, black holes display unique lensing behavior: for instance, any two spatial points outside the event horizon are connected by infinitely many light rays, each of which executes a different number of orbits around the black hole under its extreme gravitational pull \cite{Darwin1959,Luminet1979,Ohanian1987,Bozza2010,GrallaLupsasca2020a}.
As a result, images of a black hole surrounded by a non-spherical, optically thin emission region decompose into a sequence of superimposed layers indexed by photon half-orbit number $n$, with each layer consisting of a full lensed image of the main emission  \cite{Gralla2019,Johnson2020,GrallaLupsasca2020a,Hadar2021,Chael2021,Paugnat2022,Vincent2022}.
The direct ($n=0$) layer typically displays a central dark area---the black hole---encircled by the weakly lensed, primary image of the accretion flow onto the hole, whose details depend sensitively on astrophysical conditions.
On the other hand, the higher-$n$ layers arise from photons on highly bent trajectories that are strongly lensed to form a series of narrow ``photon rings.''\footnote{For an animation of this effect, see \url{https://youtu.be/4-DvyMPs-gA}.}
These rings are usually stacked on top of the broader $n=0$ emission and their shape rapidly converges (exponentially fast in $n$) to that of the ``Kerr critical curve''\footnote{Spherical emission creates a ``shadow'' inside of this curve \cite{Falcke2000,Narayan2019,Vincent2022}, which is often called the ``shadow edge'' even when its interior is not dark \cite{Volker2022}.} \cite{Bardeen1973}: a theoretical curve in the observer sky corresponding to the apparent image of asymptotically bound photon orbits.
In contrast to the astrophysics-dependent $n=0$ image, this ``$n\to\infty$ photon ring'' is completely determined by the Kerr black hole---depending only on its mass, spin, and inclination---and delineates its cross-sectional area in the sky.

Recently, 1.3\,mm interferometric observations by the Event Horizon Telescope have resolved the horizon-scale emission from sources in the immediate vicinity of two nearby supermassive black holes: M87* \cite{EHT2019a}, the central compact object at the core of our neighboring galaxy Messier 87, and Sgr~A* \cite{EHT2022a}, our own black hole at the center of the Milky Way. 
Their reconstructed images display a central brightness depression within a thick ring consistent with theoretical expectations for the direct image of the surrounding accretion flow \cite{EHT2019e,Arras2022,Carilli2022,Lockhart2021,EHT2022e}.

However, these observations have thus far not provided any evidence for the presence of a lensed photon ring \cite{Lockhart2022}.
They are instead dominated by $n=0$ photons \cite{Gralla2019,Johnson2020}, which form the image layer that is more sensitive to the astrophysics of the flow than to purely relativistic effects \cite{Gralla2021,Lara2021,Bauer2021}.
As a result, the bounds placed on possible deviations from general relativity are on the order of several percent \cite{EHT2019f,EHT2022f} and comparable to the constraints derived from gravitational-wave observations of stellar-mass, binary black holes \cite{CardenasAvendano2020}, or x-ray spectroscopy measurements of low-mass binaries and active galactic nuclei \cite{Ayzenberg2021}.
By contrast, future detections of orbiting ($n\geq1$) photons will open a new window into strong gravity and enable higher-precision probes of the Kerr geometry, since it is this orbiting light which forms the part of the image---the photon ring---that belongs to the black hole itself, rather than to its plasma.

There are three major obstacles to measuring a photon ring.
First, since the photon rings are exponentially narrow (in $n$) features, resolving them requires interferometric observations on exponentially long baselines \cite{Johnson2020}.
At the current observing frequency of 230\,GHz, even Earth-spanning baselines are too short to resolve the first ring, but it should become accessible to next-generation space-based interferometers.
In particular, SALTUS (the Single Aperture Large Telescope for Universe Studies) is a bold proposal---currently a contender for NASA's upcoming Probe mission---to launch within the next decade a spacecraft far enough to access the first two rings of M87*.

Optical depth poses a second hurdle: even though photons could in principle circumnavigate the black hole indefinitely (albeit unstably), in practice, those that traverse its emission region multiple times are eventually reabsorbed by the matter they intersect---an effect that cuts off image layers past some $n>0$.
Nevertheless, since absorptivity decreases with photon energy, the first few rings still ought to be present in images taken at sufficiently high frequencies.
Simple models suggest that at 230\,GHz, the $n=1$ ring is always visible while the $n=2$ ring may be only marginally observable, whereas both rings should be clearly visible at 345\,GHz \cite{Vincent2022}.
State-of-the-art simulations of general-relativistic magnetohydrodynamic (GRMHD) flows \cite{Wong2022a} also confirm these expectations \cite{Wong2022b}.
For this reason, SALTUS is slated to simultaneously observe at both frequencies.
M87* makes for a particularly exciting prospective target because a measurement of its $n=2$ ring diameter could deliver a stringent test of the Kerr hypothesis, which predicts a definite shape for its higher-$n$ rings: photons orbiting just outside the horizon of a black hole can probe its extreme gravity and carry away information about its spacetime geometry, encoded in the observable shape of the rings that these photons produce in their observer's sky \cite{Gralla2020,GrallaLupsasca2020c,GLM2020}.

Time variability introduces a third significant complication.
While time-averaged GRMHD-simulated movies have shown that the photon rings are persistent, sharp features that come to dominate observations with very-long-baseline interferometry (VLBI) after averaging over sufficiently long timescales \cite{Johnson2020}, it remains to be understood how clearly visible  the rings will be to a realistic, near-future, space-VLBI mission like SALTUS, which will be limited in the number of snapshots it can collect.
In each snapshot, such an interferometer---with a single space leg---can only sample the radio visibility on one space-ground baseline, thereby only measuring the projected diameter $d_\varphi$ of the photon ring at one angle $\varphi$ in the image.\footnote{The angle in the image corresponds to the space element's baseline angle in the Fourier plane \cite{Gralla2020,GrallaLupsasca2020c,GLM2020}, while the baseline length determines the index $n$ of the subring whose interferometric signature dominates the signal \cite{Johnson2020,Paugnat2022}.}
To compensate for its sparse baseline coverage, the instrument can observe at regular intervals along its orbit around the Earth, eventually filling in every angle $\varphi$ in the Fourier domain, with each $d_\varphi$ thus sampled twice per orbit.
The baseline lengths over which the ring signature dominates the signal fix the orbital radius (about lunar distance for the $n=2$ ring of M87*) and hence the orbital period ($\sim\!1$ month), which in turn sets the cadence of these snapshots: $\sim\!40M_{\rm M87^*}$, or roughly every two weeks.
As it is evidently impractical to maintain coherence over such timescales, the snapshots must be incoherently time-averaged;
moreover, only $\sim\!24$ snapshots of $d_\varphi$ can be sampled per year.

In a single snapshot, the ``clean'' interferometric signature of the ring---a periodic ringing in the visibility amplitude---is typically ``polluted'' by noise from both the instrument and from astrophysical fluctuations (plasma flares, emission ropes, or other ring mimickers), which can obscure the signal.
The key question is then:
\textit{Can the interferometric signature of a photon ring---and hence its projected diameter---be recovered from an incoherent time-average over $N\approx20$ snapshots of its visibility amplitude on very long space-ground baselines?}

An affirmative answer to this question would open the door to a consistency test of the Kerr hypothesis---a cornerstone of general relativity (GR) in the strong-field regime---via space-VLBI measurements of the photon ring shape.
The paper~\cite{GLM2020} (henceforth: GLM) took the first steps toward establishing the viability of such a test for M87*.
GLM examined a range of models of stationary, axisymmetric, equatorial disks that reproduce the time-averaged observational appearance of GRMHD-simulated flows, and found that the observable shape of their $n=2$ ring always follows a specific functional form, independent of the source model.
They concluded that this ring shape is a robust prediction of strong-field GR.

In other words, observations of the $n=2$ photon ring can \textit{in principle} disentangle gravitational and astrophysical effects that are otherwise commingled in the direct image.
Moreover, GLM simulated interferometric data of the kind that could be collected by a mission like SALTUS, and were able to extract this ring shape from the visibility amplitude on space-ground baselines.
Their experimental forecast achieved a sub-percent level of precision for the resulting test of the Kerr hypothesis, suggesting that M87* holds exceptional promise as a target for such a test \textit{in practice}.
This analysis was recently reviewed in depth and extended to an even larger selection of equatorial disk models, supporting these conclusions \cite{Paugnat2022}.

While these early results are encouraging, demonstrating the feasibility of a ring shape measurement requires further theoretical work.
Crucially, even though the GLM analysis did include realistic instrument noise, it only considered time-averaged images of equatorial disks.
The latter limitation was recently tackled with a study of geometric thick-disk models \cite{Vincent2022}, but to date a detailed investigation of source fluctuations and time variability has yet to be carried out.

The present work is the first attempt to remedy this lacuna.
The main obstruction is technical: as high-order photon rings are exponentially narrow compared to the overall structure of a black hole image, resolving them in the image plane incurs a large computational cost.
More precisely, their presence in the image introduces a large separation of scales between the pixel grid size (which must be large enough to capture the entire field of view) and the pixel spacing (the grid must achieve a sufficiently fine resolution to see the narrow rings).
While a brute-force approach---pumping millions of pixels in the grid---can overcome this scale separation for a handful of images, it becomes intractable when dealing with a black hole \textit{movie} consisting of several hundred snapshots, in which case adaptive ray tracing is necessary \cite{Wong2021,Gelles2021a} (and less wasteful).

\begin{figure*}
    \centering
    \includegraphics[width=\textwidth]{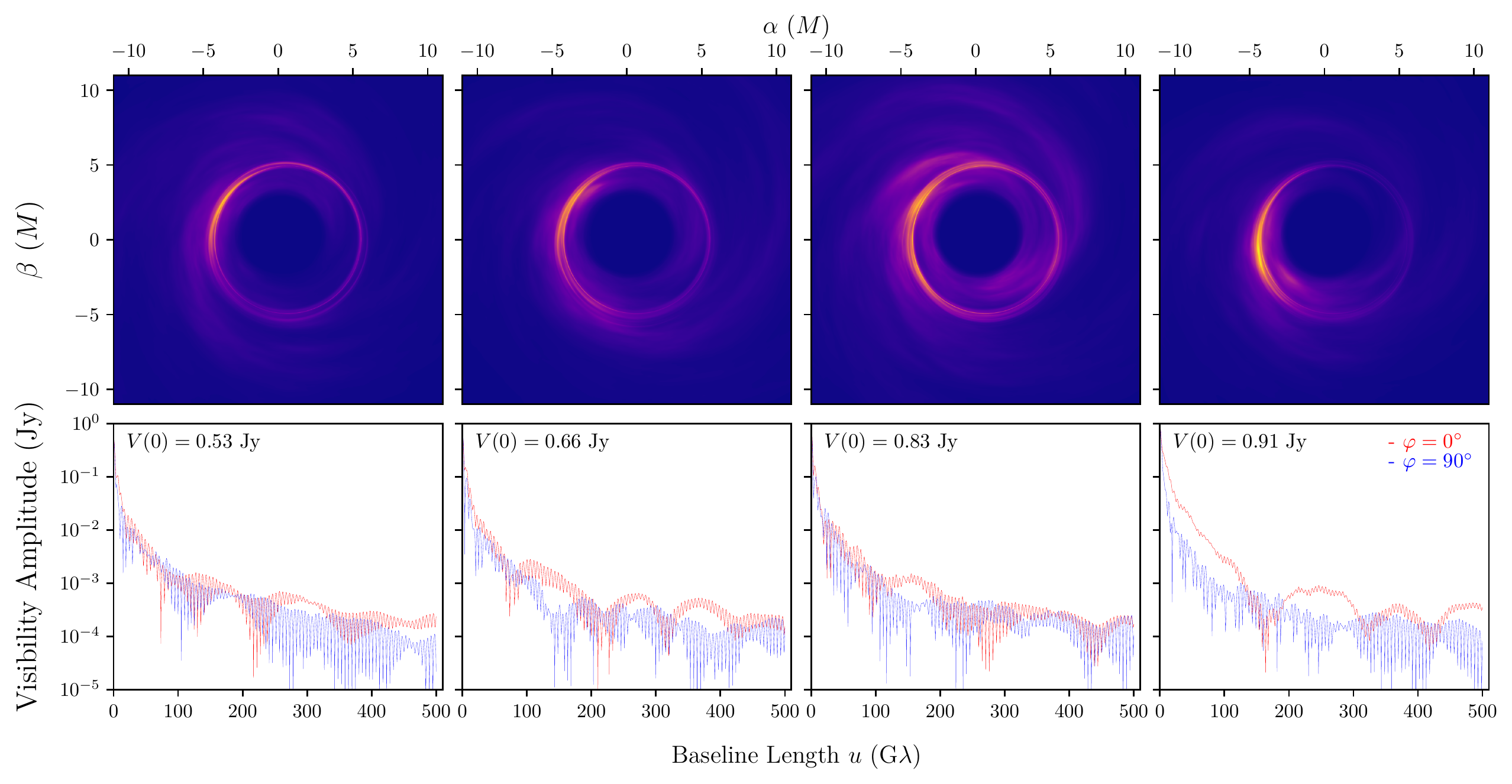}
    \caption{Top: Snapshots of a non-stationary, non-axisymmetric, equatorial source around a Kerr black hole ray-traced with \texttt{AART}.
    Most image features in these snapshots arise from fluctuations in the source, with the exception of the strikingly bright and narrow photon ring: a persistent, sharp feature that comes to dominate the image formed by time-averaging many of these snapshots (Fig.~\ref{fig:TimeAverage}).
    These images were produced using the parameters listed in Table~\ref{tbl:Parameters} at regular time intervals of $250M$ .
    Bottom: The corresponding visibility amplitudes for spin-aligned (blue) and spin-perpendicular (red) cuts across each of the above images.
    The black hole spin is $a/M=94\%$ and the observer inclination is $\theta_{\rm o}=17^\circ$.}
    \label{fig:Snapshots}
\end{figure*}

\begin{figure*}
    \centering
    \includegraphics[width=\textwidth]{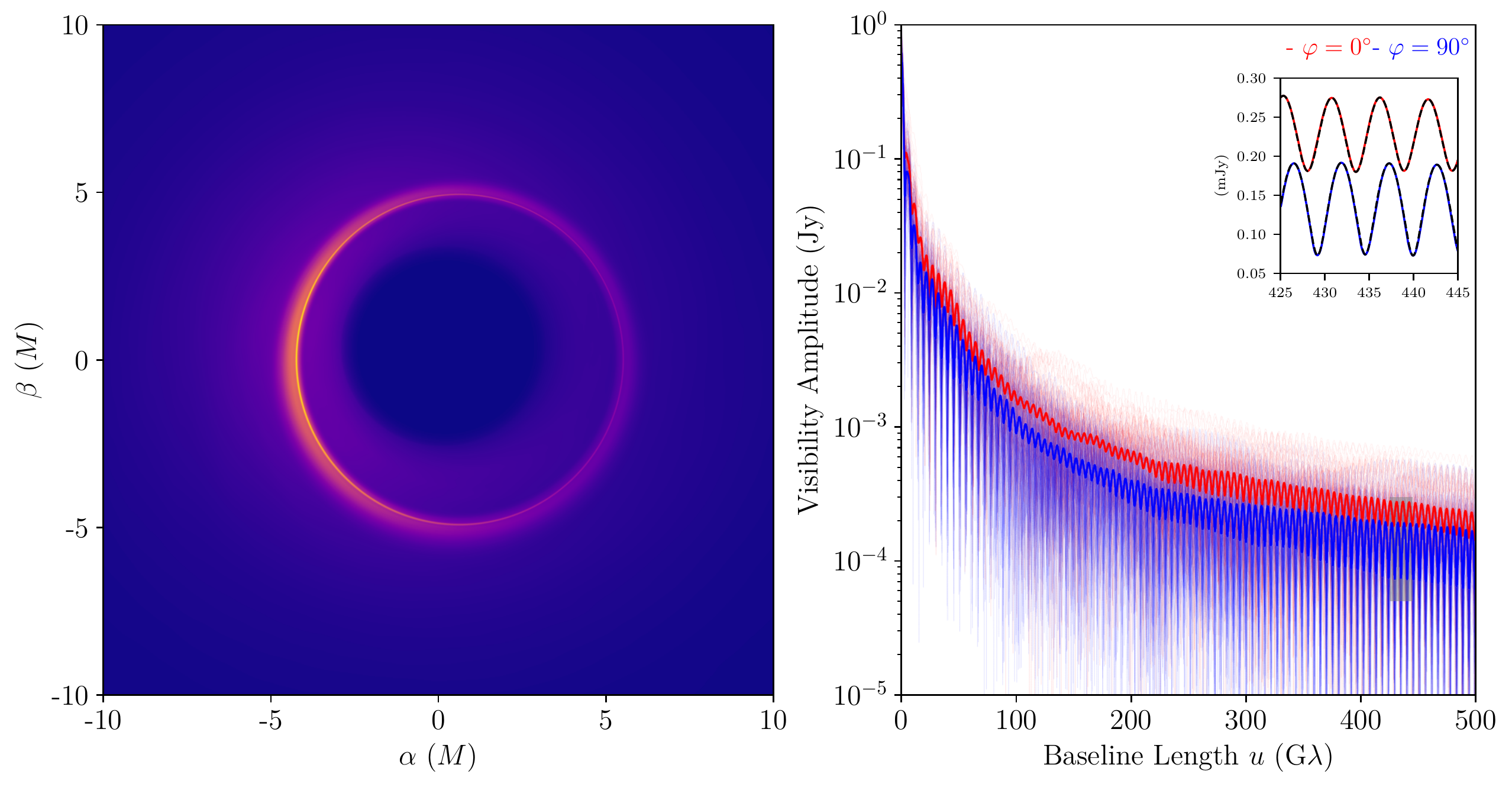}
    \caption{Left: Time average of 100 ray-traced snapshots of \texttt{inoisy} equatorial profiles uniformly sampled over a time interval of $1000M$.
    The \texttt{inoisy} parameters take the ``best-guess'' values for M87* presented in Table~\ref{tbl:Parameters}. 
    Hence, this image is directly comparable to the time-averaged image of M87* presented in Fig.~1 of Ref.~\cite{Johnson2020}, for which the underlying GRMHD simulation's parameters were chosen to be consistent with the 2017 Event Horizon Telescope observations of M87*.
    Right: Visibility amplitudes of all 100 snapshots along cuts parallel ($\varphi=0^\circ$, red) and perpendicular ($\varphi=90^\circ$, blue) to the black hole spin axis.
    The solid lines correspond to the incoherently time-averaged amplitudes.
    Embedded is a panel zooming into the average visibility amplitude in the baseline range $u\in[425,445]\,$G$\lambda$, with the best-fit ring signature overlaid (black).}
    \label{fig:TimeAverage}
\end{figure*}

Here, we present a numerical tool\footnote{The code is publicly available at~\url{https://github.com/iAART/aart.}} designed to efficiently compute high-resolution ``slow-light'' movies of generic (non-stationary and non-axisymmetric) equatorial sources around a Kerr black hole, together with their associated radio visibility on very long baselines; we present example outputs in Fig.~\ref{fig:Snapshots}.

The code was developed with the intent to maximize speed while still guaranteeing the accuracy of its output, particularly the radio visibility on very long baselines, which encodes the high-frequency components of a snapshot's Fourier transform and is therefore extremely sensitive to its most minute image features.
The code's structure is highly modular and most of its individual components reproduce pre-existing capabilities; its main novelty is arguably to combine all these routines into one single and convenient-to-use (we hope!) package.

One new and important technique from which \texttt{AART} derives much of its power deserves special mention: it turns the very feature of photon rings that makes them so difficult to fully resolve---namely, their thinness---to its advantage.
It does so by decomposing the full image into multiple layers labeled by half-orbit number $n$, and ray tracing in each layer the image of the $n^\text{th}$ photon ring with exponentially high (in $n$) resolution.

More precisely, the lensing behavior of a black hole forces the $n^\text{th}$ photon ring to lie within an exponentially small (in $n$) region of the image plane, dubbed the $n^\text{th}$ \textit{lensing band}, with each band completely fixed by the Kerr geometry \cite{GLM2020,Paugnat2022}.
A key innovation of \texttt{AART} is to first compute (once and for all) the lensing bands associated with a given black hole spin and inclination, and then to ray trace, for the $n^\text{th}$ layer, only pixels lying in the $n^\text{th}$ lensing band.
Since each lensing band contains an exponentially demagnified image of the main emission, by also increasing the resolution in each band exponentially,\footnote{Successive subrings are demagnified by an analytically known, angle-and-spin-dependent factor $e^{-\gamma(\varphi)}$, where $\gamma(\varphi(\tilde{r}))$ is the Lyapunov exponent that governs the instability of nearly bound photons at orbital radius $\tilde{r}$ \cite{Johnson2020,GrallaLupsasca2020a}.} the code is guaranteed to resolve the source with roughly the same effective resolution in every layer.
In other words, \texttt{AART} adapts its ray-tracing resolution (grid spacing) to each layer, but also adjusts the ray-tracing region (grid size), so as to resolve the increasingly fine features present in higher layers using only a fixed number of pixels per layer.

Other ray tracers use adaptive mesh refinement to increase pixel density in regions where they detect long geodesic path lengths \cite{Wong2021} or large gradients \cite{Gelles2021a,White2022}.
This results in a new, non-uniform grid for every new snapshot, which is recursively refined until a desired criterion is met, or else the number of recursions exceeds a hard-set limit.
Therefore, small image features can sometimes be missed when this limit is hit.

By contrast, the adaptiveness provided by the lensing bands is determined by the Kerr geometry alone and results in the same, uniform grid for every image of a given black hole spin and inclination.
These grids therefore need be computed only once, and provided that their resolution increases at the proper rate set by the demagnification factor, they cannot miss any feature that is already resolved in the direct $n=0$ image.

Finally, high-$n$ layers comprise photons that execute many orbits in the Kerr photon shell \cite{Teo2021} where their radial potential almost develops a double root and geodesic integrals diverge logarithmically, leading to a growing risk of numerical error.
To minimize error, \texttt{AART} performs analytical ray tracing using an exact solution of the Kerr null geodesic equation recently given in terms of Legendre elliptic integrals and Jacobi elliptic functions \cite{GrallaLupsasca2020b}, similar in spirit to previous implementations based on Carlson symmetric forms \cite{Dexter2009,Yang2014}.

To summarize: \texttt{AART} uses lensing bands to prevent a large separation of scales (between the grid size and its spacing) and thereby maintain its speed as it increases the pixel density in higher-$n$ photon rings, which is necessary to ray trace the fine image features that (due to the nonlocal character of the Fourier transform) influence the visibility on long baselines; moreover, the rings are ray traced analytically to avoid errors.

These features of \texttt{AART} are specially tailored to the photon ring and its interferometric signature; with this tool in hand, it becomes feasible to investigate the effects of time variability on measurements of the $n=2$ ring shape, and we can begin to answer the key experimental question posed above.

To study the effects of source fluctuations on the observed visibility amplitude, we call upon both \texttt{AART} and \texttt{inoisy} \cite{Lee2021}: a code that can rapidly generate realizations of a 2D Gaussian random field with Mat\'ern covariance.
Such a field can provide a simple statistical model for a generic (non-stationary and non-axisymmetric) stochastic source in the equatorial plane of a Kerr black hole, with a prescribed two-point function.
Since a ``realistic'' choice of autocorrelation structure is not yet known (and will require additional research into the physics of the plasma), our goal will be to vary the statistics of the model within a wide range of ``reasonable'' possibilities (informed by GRMHD simulations) so as to parameterize our uncertainty in the expected variability of signals from sources such as M87*.

Completing such a parameter survey is a large undertaking beyond the scope of this first paper.
Here, we will be content with a proof-of-concept demonstration that \texttt{AART} is up to the task for such a study.
To showcase its capabilities, we ray trace 100 snapshots of an \texttt{inoisy} source with statistics set by the ``best-guess'' parameters for M87* (listed in Table~\ref{tbl:Parameters}).
Sample snapshots and their associated visibility amplitude are shown in Fig.~\ref{fig:Snapshots}, while Fig.~\ref{fig:TimeAverage} includes all of the snapshots, with the left panel displaying their time-averaged image and the right panel all the individual visibility amplitudes together with their incoherent time average (solid lines).
As expected, we find that the astrophysical fluctuations wash out from the time average, leaving an image that is visibly dominated by a prominent photon ring with clear $n=1$ and $n=2$ subrings.

Correspondingly, the incoherently time-averaged visibility amplitude is dominated on long baselines by the perfectly clean interferometric signature of the $n=2$ ring.
In particular, we find that the signal in the range $u\in[425,445]\,$G$\lambda$---which a satellite at lunar distance from the Earth could access with $345$\,GHz observations---exactly follows the periodic ringing pattern predicted for a thin ring (black overlay in panel inset).

\begin{figure*}
    \centering
    \includegraphics[width=\textwidth]{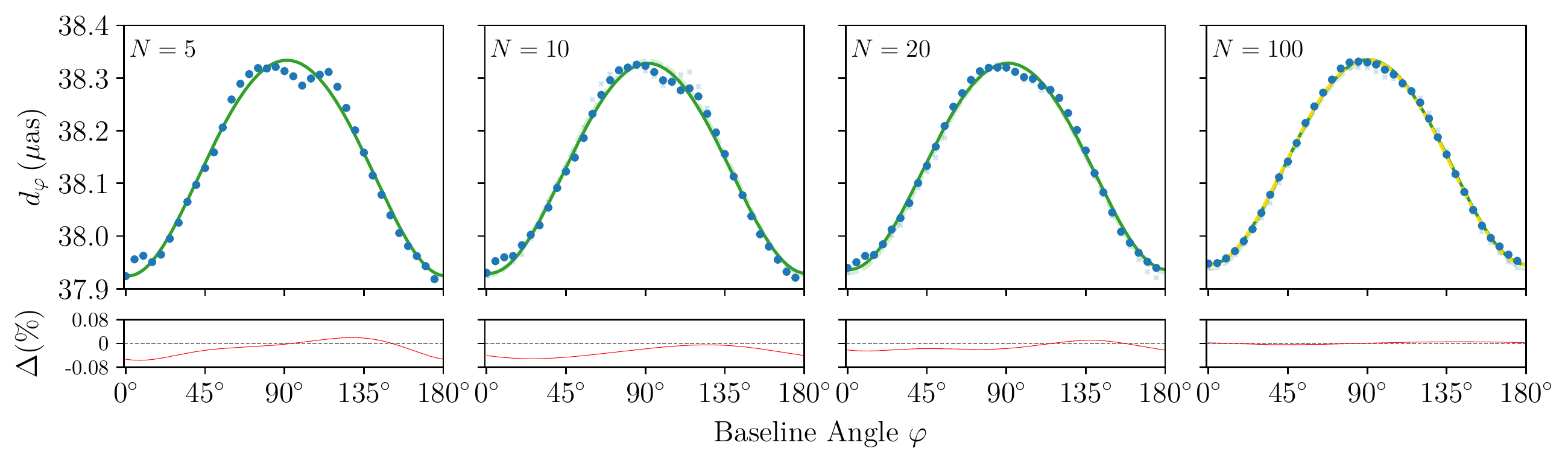}
    \caption{Top: The projected diameter $d_\varphi$ of the $n=2$ photon ring inferred from an incoherent time average over $N$ snapshots of the source.
    The green curves are best fits of the GR-predicted shape [Eq.~\eqref{eq:Circlipse}] to the synthetic data.
    To show how the fits improve with increasing $N$, each panel displays the data of its predecessor in the background.
    Bottom: The difference between the best-fitting curve at a given $N$ and the ``true'' shape of the ring (solid gold line in last panel) to which the signal converges as $N\rightarrow\infty$.
    The parameters of the model are listed in Table~\ref{tbl:Parameters}.}
    \label{fig:ProjectedDiameter}
\end{figure*}

As we show in Fig.~\ref{fig:ProjectedDiameter}, the projected diameter $d_\varphi$ of the $n=2$ ring can then be extracted from the periodicity of this ringing in the visibility amplitude $|V(u,\varphi)|$.
The top panels display the ring diameter $d_\varphi$ inferred from an average over $N$ snapshots, with $N=5,10,20$, and finally 100, by which time the smooth shape of the ring has emerged.
The bottom panels display the relative deviation from this fiducial shape that is induced by astrophysical fluctations, whose noise clearly averages out of the image; it is encouraging to see this noise is also beat down in measurements of the ring shape using only a few snapshots.

Of course, whether these conclusions are likely to hold for M87* has yet to be established, and a systematic investigation of astrophysical fluctuations remains to be done.
In a soon-to-be-released paper, we will initiate such a study by repeating this analysis for multiple models of M87* in which we vary the parameters both of the black hole (its spin and inclination) and of the source (the \texttt{inoisy} model).
We also hope to report on the $n=1$ ring's signature and how it may encode the spin.

To guide the reader, we now give a summary of the rest of the paper, which---like \texttt{AART}---is written in a modular way.

\subsection*{Summary}

In Sec.~\ref{sec:Lensing}, we review the problem of light propagation in the Kerr spacetime.
We write down the null geodesic equation and its exact analytical solution as it is implemented in \texttt{AART}, and describe the key concept of lensing bands (Fig.~\ref{fig:LensingBands}).
We then illustrate the lensing behavior of the black hole by plotting its ``transfer functions'': mappings of directions in the observer sky to the spacetime points where the corresponding light rays intersect the equatorial plane.
The transfer functions for polar coordinates $(r_{\rm s},\phi_{\rm s})$ in the plane are shown in Figs.~\ref{fig:n0TransferFunctions}, \ref{fig:n1TransferFunctions}, and \ref{fig:n2TransferFunctions} for the $n=0$, $n=1$, and $n=2$ images, respectively.
Likewise, the time lapse $\Delta t$ between source and observer is also shown in Figs.~\ref{fig:n0Time} and \ref{fig:n1Time} for the $n=0$ and $n=1$ images, respectively.
The $n^\text{th}$ image always fills out the $n^\text{th}$ lensing band, as expected.

We also give an approximate formula for light bending that was derived by Beloborodov \cite{Beloborodov2002} for nonrotating black holes in the weak-deflection regime.
We express his result in a very simple form [Eq.~\eqref{eq:BeloborodovApproximation}] that proves to be remarkably accurate for the computation of $n=0$ images of axisymmetric sources, \textit{for most black hole spins and inclinations}.
This observation, which is illustrated in Fig.~\ref{fig:BeloborodovApproximation}, highlights the fact that $n=0$ photons barely carry an imprint of the black hole spin, as they do not spend enough time near it to be strongly affected by its gravitational field.
Beloborodov's approximation can thus be regarded as a generalization to all inclinations of the ``just add one'' prescription for the transfer function of a spin-aligned observer, for whom the impact parameter $\rho$ is simply related to emission radius $r_{\rm s}$ by adding one: $\rho\approx r_{\rm s}+M$ \cite{GrallaLupsasca2020a,Gates2020}.
\texttt{AART} can also parallel transport linear polarization.
We check that Beloborodov's approximation is adequate for ray tracing $n=0$ polarimetric images (Fig.~\ref{fig:BeloborodovComparison}), which can be done using simple algebraic equations that we write down explicitly.

In Sec.~\ref{sec:Model}, we describe our model of equatorial emission.
First, we review stationary and axisymmetric equatorial disk profiles that can reproduce the time-averaged observational appearance of M87* in GRMHD simulations \cite{GLM2020,Chael2021}.
Then, we add in astrophysical fluctuations with prescribed statistics.
We model these fluctuations using Gaussian random fields and describe their statistical properties in great detail.
In Sec.~\ref{sec:Applications}, we use \texttt{inoisy} to simulate a variable source.
With \texttt{AART}, we ray trace its instantaneous snapshots (Fig.~\ref{fig:inoisySnapshotM87}) and  compute its light curve (Fig.~\ref{fig:LightCurves}).
Great care must be taken in the choice of resolution and field of view used in each layer, and Sec.~\ref{sec:Requirements} discusses these issues in depth (Figs.~\ref{fig:Profiles}, \ref{fig:FieldOfView}, and \ref{fig:ResolutionCheck}).

We can then produce movies of a stochastic, non-stationary, non-axisymmetric source. In Sec.~\ref{sec:Forecast}, we compute movies of the associated visibility amplitude and use this synthetic data to reconstruct the projected diameter $d_\varphi$ of the $n=2$ ring, before concluding with a brief discussion of future prospects in Sec.\ref{Conclusions}.
We relegate some details to Apps.~\ref{app:GeodesicIntegrals}, \ref{app:FourVelocity}, and \ref{app:MaternCovariance}.
Throughout the paper, we work in $(-,+,+,+)$ metric signature with geometric units in which $G_{\rm N}=c=1$.
Our conventions for Legendre elliptic integrals are listed in App.~A of Ref.~\cite{Kapec2020}.

\section{Theoretical framework}
\label{sec:Lensing}

The special integrability properties of the Kerr spacetime reduce its geodesic equation to a problem of quadratures \cite{Carter1968}, resulting in elliptic integrals that are expressible in Legendre normal form \cite{Rauch1994,Vazquez2004,Li2005,GrallaLupsasca2020b}.
Modern computers can evaluate the Legendre elliptic integrals very fast and to arbitrary precision, reducing the computational cost of ray tracing in Kerr.
In this section, we review the exact solution of the Kerr null geodesic equation in Legendre form \cite{GrallaLupsasca2020a,GrallaLupsasca2020b} as well as the approximate solution derived in Schwarzschild by Beloborodov \cite{Beloborodov2002}, and we use them to plot the transfer functions mapping Bardeen's coordinates in the observer sky \cite{Bardeen1973} to the equatorial plane.

\subsection{Null geodesics of the Kerr exterior}

Astrophysical, rotating black holes of mass $M$ and angular momentum $J=aM$ are subject to the Kerr bound $|a|\leq M$ and are described by the Kerr geometry.

The Kerr metric is written in Boyer-Lindquist coordinates $(t,r,\theta,\phi)$ in terms of functions $\Sigma(r,\theta)=r^2+a^2\cos^2{\theta}$ and $\Delta(r)=r^2-2Mr+a^2$.
The roots of $\Delta(r)$ define the radii of the outer and inner event horizons:
\begin{align}
    \label{eq:Horizons}
	r_\pm=M\pm\sqrt{M^2-a^2}.
\end{align}
With respect to Mino time $\tau$, a photon with four-momentum $p^\mu$ follows a null geodesic $x^\mu(\tau)$ obtained by solving
\begin{align}
	\frac{dx^\mu}{d\tau}=-\frac{\Sigma}{p_t}p^\mu.
\end{align}
The resulting trajectory is independent of the photon energy $-p_t$ and can be parameterized by two conserved quantities: the energy-rescaled angular momentum and Carter constant
\begin{align}
    \label{eq:ConservedQuantities}
	\lambda=-\frac{p_\phi}{p_t},\quad
	\eta=\frac{p_\theta^2}{p_t^2}-a^2\cos^2{\theta}+\lambda^2\cot^2{\theta}. 
\end{align}
We are interested in solving for the trajectories $x^\mu(\tau)$ that connect two spacetime events $(t_{\rm s},r_{\rm s},\theta_{\rm s},\phi_{\rm s})$ and $(t_{\rm o},r_{\rm o},\theta_{\rm o},\phi_{\rm o})$, where the labels `s' and `o' stand for `source' and `observer', respectively.
We will specialize to distant observers ($r_{\rm o}\gg M$) at nonzero inclination $\theta_{\rm o}\in(0,\pi/2)$ above the equatorial plane $\theta_{\rm s}=\pi/2$ where we place the source.
(The measure-zero cases $\theta_{\rm o}\in\{0,\pi/2\}$ technically require separate treatments~\cite{GrallaLupsasca2020a}, but in practice one can simply change the inclination slightly and use the same analytical expressions.)
The symmetries of the Kerr geometry---its stationarity and axisymmetry---allow us to set $t_{\rm s}=0$ and $\phi_{\rm o}=0$ without loss of generality.

A photon that reaches an observer with four-momentum $p_{\rm o}^\mu$ and conserved quantities $(\lambda,\eta)$ appears in the sky at Cartesian position $(\alpha,\beta)$ given in terms of $\pm_{\rm o}=\sign{p_{\rm o}^\theta}$ by Bardeen \cite{Bardeen1973},
\begin{align}
	\label{eq:BardeenCoordinates}
	\alpha=-\frac{\lambda}{\sin{\theta_{\rm o}}},\quad
	\beta&=\pm_{\rm o}\sqrt{\eta+a^2\cos^2{\theta_{\rm o}}-\lambda^2\cot^2{\theta_{\rm o}}},
\end{align}
for $r_{\rm o}\to\infty$.
If the photon carries a linear polarization, then its observed electric vector polarization angle (EVPA) is \cite{Li2009,Lupsasca2020,Himwich2020,Narayan2021}
\begin{align}
	\label{eq:EVPA}
	\chi=\arctan\pa{\frac{\nu\kappa_1-\beta\kappa_2}{\beta\kappa_1+\nu\kappa_2}},\quad
	\nu=-(\alpha+a\sin\theta_{\rm o}),
\end{align}
where $\kappa$ denotes the complex-valued Penrose--Walker constant
\begin{align}
    \label{eq:PenroseWalker}
	\kappa&=\kappa_1+i\kappa_2
	=(\mathcal{P}_\mathcal{A}-i\mathcal{P}_\mathcal{B})(r-ia\cos{\theta}),\\
	\mathcal{P}_\mathcal{A}&=\pa{p^t f^r-p^r f^t}+a\sin^2{\theta}\pa{p^r f^\phi-p^\phi f^r},\\
	\mathcal{P}_\mathcal{B}&=\br{\pa{r^2+a^2}\pa{p^\phi f^\theta-p^\theta f^\phi}-a\pa{p^t f^\theta-p^\theta f^t}}\sin{\theta},
\end{align}
a quantity that is also conserved along null geodesics---thanks to the Petrov type D nature of the Kerr metric \cite{Chandrasekhar1983}---and can be used to \textit{algebraically} solve the parallel transport problem for a linear polarization vector $f$ by evaluating $\kappa$ at the source.

The separability of Kerr geodesic motion allows the $r$ and $\theta$ trajectories to be decoupled \cite{Carter1968},
These independent motions are then controlled by radial and angular geodesic potentials whose zeros give the turning points of their respective motion:
\begin{align}
	\label{eq:RadialPotential}
	\mathcal{R}(r)&=\pa{r^2+a^2-a\lambda}^2-\Delta(r)\br{\eta+\pa{\lambda-a}^2},\\
	\label{eq:AngularPotential}
	\Theta(\theta)&=\eta+a^2\cos^2{\theta}-\lambda^2\cot^2{\theta}.
\end{align}

Both potentials have exactly four (not always real) roots
\begin{align}
	r_1&=-z-\sqrt{-\frac{\mathcal{A}}{2}-z^2+\frac{\mathcal{B}}{4z}},
	&&\theta_1=\arccos\pa{\sqrt{u_+}},\\
	r_2&=-z+\sqrt{-\frac{\mathcal{A}}{2}-z^2+\frac{\mathcal{B}}{4z}},
	&&\theta_2=\arccos\pa{\sqrt{u_-}},\\
	r_3&=z-\sqrt{-\frac{\mathcal{A}}{2}-z^2-\frac{\mathcal{B}}{4z}},
	&&\theta_3=\arccos\pa{-\sqrt{u_-}},\\
	r_4&=z+\sqrt{-\frac{\mathcal{A}}{2}-z^2-\frac{\mathcal{B}}{4z}},
	&&\theta_4=\arccos\pa{-\sqrt{u_+}},
\end{align}
which depend only on the conserved quantities $(\lambda,\eta)$ via\footnote{Here, $\sqrt[3]{x}$ denotes the real cube root of $x$ if $x$ is real, or else, the principal value of the function $x^{1/3}$ (that is, the cubic root with maximal real part).}
\begin{gather}
    z=\sqrt{\frac{\omega_++\omega_--\mathcal{A}/3}{2}},\quad 
    \omega_\pm=\sqrt[3]{-\frac{\mathcal{Q}}{2}\pm\sqrt{\frac{\mathcal{P}^3}{27}+\frac{\mathcal{Q}^2}{4}}},\\
	\mathcal{A}=a^2-\eta-\lambda^2,\quad
	\frac{\mathcal{B}}{2M}=\eta+(\lambda-a)^2,\quad
	\mathcal{C}=-a^2\eta,\\
	\mathcal{P}=-\frac{\mathcal{A}^2}{12}-\mathcal{C},\quad
	\mathcal{Q}=-\frac{\mathcal{A}}{3}\br{\pa{\frac{\mathcal{A}}{6}}^2-\mathcal{C}}-\frac{\mathcal{B}^2}{8},\\
	u_\pm=\triangle_\theta\pm\sqrt{\triangle_\theta^2+\frac{\eta}{a^2}},\quad
	\triangle_\theta=\frac{1}{2}\pa{1-\frac{\eta+\lambda^2}{a^2}}.
\end{gather}
We will now restrict our attention to positive spins $0\le a\le M$.

\subsection{Kerr critical curve}

The radial potential \eqref{eq:RadialPotential} develops a double root at $\tilde{r}\ge r_+$ (that is, $\mathcal{R}(\tilde{r})=\mathcal{R}'(\tilde{r})=0$) if and only if \cite{Bardeen1973,GrallaLupsasca2020a}
\begin{align}
    \label{eq:CriticalParameters}
	\tilde{\lambda}&=a+\frac{\tilde{r}}{a}\br{\tilde{r} -\frac{2\Delta(\tilde{r})}{\tilde{r}-M}},\quad
	\tilde{\eta}=\frac{\tilde{r}^3}{a^2}\br{\frac{4M\Delta(\tilde{r})}{(\tilde{r}-M)^2}-\tilde{r}},
\end{align}
where $\tilde{r}\in[\tilde{r}_-,\tilde{r}_+]$ with
\begin{align}
	\tilde{r}_\pm&=2M\br{1+\cos\pa{\frac{2}{3}\arccos\pa{\pm\frac{a}{M}}}}.
\end{align}
A geodesic with critical conserved quantities $\lambda=\tilde{\lambda}$ and $\eta=\tilde{\eta}$ asymptotes to an unstably bound orbit at radius $\tilde{r}$ in the Kerr photon shell.
For a distant observer, the Kerr critical curve $\mathcal{C}$ is the image in the sky of these asymptotically bound orbits:
\begin{align}
    \label{eq:CriticalCurve}
    \mathcal{C}=\cu{\pa{\tilde{\alpha},\tilde{\beta}}:(\lambda,\eta)=\pa{\tilde{\lambda},\tilde{\eta}}}.
\end{align}
Though this purely theoretical curve is not in itself observable, it does play a key role in the study of lensing by a Kerr black hole~\cite{Bardeen1972}.
It is traced using Eqs.~\eqref{eq:BardeenCoordinates} evaluated on Eqs.~\eqref{eq:CriticalParameters} for all the values $\tilde{r}\in[\tilde{r}_-,\tilde{r}_+]$ such that $\tilde{\beta}^2\ge0$.
It is always closed, convex, and reflection-symmetric about the $\alpha$ axis \cite{GrallaLupsasca2020c}.

\subsection{Analytical backwards ray tracing}
\label{subsec:RayTracing}

The character (real or complex) and ordering of the radial roots $\{r_1,r_2,r_3,r_4\}$ (when they are real) lead to a classification of radial motion into four types \cite{Hackmann2010,GrallaLupsasca2020b,Compere2022}.
We are, however, only interested in the rays that connect a distant observer to an equatorial source.
This excludes two of the motion types, leaving only two others corresponding to rays that start from infinity at one end point before they either cross the horizon or return to infinity at the other.\footnote{The angular motion also has two possible behaviors according to whether $\eta\gtrless0$.
We ignore vortical rays with $\eta<0$ as they cannot reach the equator \cite{Kapec2020}.
Such rays always lie within the apparent image of the horizon.}
The boundary between these two behaviors in the phase space of null geodesics constitutes the Kerr photon shell of asymptotically bound orbits, a special locus in phase space with emergent conformal symmetry \cite{Hadar2022}.
We now describe how this separation is manifested in the sky.

Given a black hole spin and observer inclination $(a,\theta_{\rm o})$, we can pick a direction $(\alpha,\beta)$ in the observer sky and shoot a light ray backwards into the geometry in that direction.
Such a ray will cross the equatorial plane a total number of times $N(\alpha,\beta)$, which can be determined by computing the total Mino time $\tau$ elapsed along the entire development of the trajectory.
Rays in the interior of the critical curve all fall into the black hole; that is, they encounter no radial turning point and terminate their motion across the Kerr exterior on the event horizon at $r=r_+$.
Rays in the exterior of the critical curve are all deflected back to infinity; that is, their radial motion encounters a turning point at $r=r_4$, whereupon they bounce back toward $r\to\infty$.
Thus, the critical curve \eqref{eq:CriticalCurve} delineates the boundary between photon capture (its interior) and photon escape (its exterior), and may be regarded as the cross-sectional area of the hole.
In-between rays that lie exactly on $\mathcal{C}$ are trapped in the photon shell where they can in principle orbit forever; in practice, this never occurs because such orbits are unstable (or equivalently, because $\mathcal{C}$ is infinitely thin).

Let $r_{ij}=r_i-r_j$ and define $A=\sqrt{r_{32}r_{42}}$, $B=\sqrt{r_{31}r_{41}}$, and
\begin{align}
	\label{eq:IrType2}
	\mathcal{I}_0^{(2)}&=\frac{2}{\sqrt{r_{31}r_{42}}}F\pa{\left.\arcsin{\sqrt{\frac{r-r_4}{r-r_3}\frac{r_{31}}{r_{41}}}}\right|\frac{r_{32}r_{41}}{r_{31}r_{42}}},\\
	\label{eq:IrType3}
	\mathcal{I}_0^{(3)}&=\frac{1}{\sqrt{AB}}F\pa{\left.\arccos\pa{\frac{1-\frac{r-r_2}{r-r_1}\frac{B}{A}}{1+\frac{r-r_2}{r-r_1}\frac{B}{A}}}\right|\frac{(A+B)^2-r_{21}^2}{4AB}},
\end{align}
where $F(\varphi|k)$ is an incomplete elliptic integral of the first kind.
These functions are the antiderivatives \eqref{eq:Ir2} and \eqref{eq:Ir3}, and all the radial geodesic integrals $I_k$ are definite integrals that can be obtained from their respective antiderivative $\mathcal{I}_k$ as follows.

If a ray lies inside of $\mathcal{C}$, then its motion is, following the labeling introduced in Ref.~\cite{GrallaLupsasca2020b}, of type (2) when all roots are real (in which case $r_+>r_->r_4>r_3>r_2>r_1$), or else of type (3) when $r_3=\bar{r}_4$ are complex-conjugate roots with $r_+>r_->r_2>r_1$.\footnote{There exist type (4) rays with both $r_1=\bar{r}_2$ and $r_3=\bar{r}_4$, but they are vortical.}
In either case, the definite integrals down to radius $r_{\rm s}$ on the ray are
\begin{align}
	\label{eq:RadialIntegralInsideC}
	I_k(r_{\rm s})=\mathcal{I}_k^{(2,3)}(r_{\rm o})-\mathcal{I}_k^{(2,3)}(r_{\rm s}).
\end{align}

If a ray lies outside of $\mathcal{C}$, then its motion is always of type (2), and the definite integrals down to radius $r_{\rm s}$ on the ray are
\begin{align}
	\label{eq:RadialIntegralOutsideC}
	I_k(r_{\rm s})=\mathcal{I}_k^{(2)}(r_{\rm o})\mp\mathcal{I}_k^{(2)}(r_{\rm s}),
\end{align}
with sign $-/+$ before/after reaching the turning point at $r=r_4$.

The Mino time $\tau(r_{\rm s})$ elapsed along a ray inside $\mathcal{C}$ is thus
\begin{align}
	\tau^-(r_{\rm s})=\int_{r_{\rm s}}^{r_{\rm o}}\frac{\ed r}{\sqrt{\mathcal{R}(r)}}
	\equiv I_0(r_{\rm s})
	=\mathcal{I}_0^{(2,3)}(r_{\rm o})-\mathcal{I}_0^{(2,3)}(r_{\rm s}).
\end{align}
Hence, the total Mino time elapsed along the full light ray is
\begin{align}
    \label{eq:MinoTimeToHorizon}
	\tau_{\rm max}^-=\int_{r_+}^{r_{\rm o}}\frac{\ed r}{\sqrt{\mathcal{R}(r)}}
	=\tau^-(r_+).
\end{align}
Likewise, the Mino time $\tau(r_{\rm s})$ elapsed along a ray outside $\mathcal{C}$ is
\begin{align}
	\tau^+(r_{\rm s})=\br{\int_{r_{\rm s}}^{r_{\rm o}}+2w\int_{r_4}^{r_{\rm s}}}\frac{\ed r}{\sqrt{\mathcal{R}(r)}}
	=\mathcal{I}_0^{(2)}(r_{\rm o})\mp\mathcal{I}_0^{(2)}(r_{\rm s}),
\end{align}
where the sign is $-$ before the turning point at $r=r_4$ ($w=0$) and $+$ after the bounce ($w=1$).
Hence, the total Mino time elapsed along the full light ray is
\begin{align}
    \label{eq:MinoTimeToInfinity}
	\tau_{\rm max}^+=2\int_{r_+}^{r_{\rm o}}\frac{\ed r}{\sqrt{\mathcal{R}(r)}}
	=2\mathcal{I}_0^{(2)}(r_{\rm o}),
\end{align}
and its radial turn occurs at the ``half-way'' Mino time
\begin{align}
	\label{eq:HalfwayMinoTime}
	\tau_4=\int_{r_+}^{r_{\rm o}}\frac{\ed r}{\sqrt{\mathcal{R}(r)}}
	=\mathcal{I}_0^{(2)}(r_{\rm o})
	=\frac{\tau_{\rm max}^+}{2},
\end{align}
We will use $\tau_{\rm max}$ to denote the appropriate choice of $\tau_{\rm max}^\pm$.

At last, the total number of equatorial crossings $N(\alpha,\beta)$ is
\begin{align}
	\label{eq:MaxCrossing}
	N=\left\lfloor\frac{\tau_{\rm max}\sqrt{-u_-a^2}+\sign(\beta)F_{\rm o}}{2K}\right\rfloor-H(\beta)+1,
\end{align}
where $H(x)$ denotes the Heaviside function while $K$ and $F_{\rm o}$ are defined in Eqs.~\eqref{eq:K}--\eqref{eq:Fo}; see also App. A of Ref.~\cite{Paugnat2022}.

A light ray crosses the equatorial plane for the $(n+1)^\text{th}$ time at Mino time $\tau=\tau_{\rm s}^{(n)}$, where
\begin{align}
	\label{eq:CrossingMinoTimes}
	\tau_{\rm s}^{(n)}=G_\theta^{(n)}
	\in\br{0,\tau_{\rm max}},\quad
	n\in\cu{0,\ldots,N-1},
\end{align}
with $G_\theta^{(n)}$ given in Eq.~\eqref{eq:Gtheta}.
This equatorial crossing occurs at
\begin{align}
	\label{eq:rs}
    r_{\rm s}^{(n)}&=r_{\rm s}^{(2,3)}\pa{\tau_{\rm s}^{(n)}},\\
	\phi_{\rm s}^{(n)}&=\phi_{\rm o}-I_\phi^{(n)}-\lambda G_\phi^{(n)},\\
	\label{eq:ts}
    t_{\rm s}^{(n)}&=t_{\rm o}-I_t^{(n)}-a^2G_t^{(n)},
\end{align}
with $r_{\rm s}^{(2,3)}(\tau)$ given in Eqs.~\eqref{eq:SourceRadius2} and \eqref{eq:SourceRadius3} for geodesics of types (2) and (3), respectively, while the angular integrals $G_{t,\phi}^{(n)}$ are given in Eqs.~\eqref{eq:Gphi}--\eqref{eq:Gt}.
Meanwhile, the radial integrals
\begin{align}
	I_{t,\phi}^{(n)}=I_{t,\phi}\pa{r_{\rm s}^{(n)}}
\end{align}
decompose via Eqs.~\eqref{eq:Iphi}--\eqref{eq:It} into definite integrals that are to be evaluated via Eq.~\eqref{eq:RadialIntegralInsideC} for rays inside $\mathcal{C}$ or via Eq.~\eqref{eq:RadialIntegralOutsideC} for rays outside $\mathcal{C}$, using the antiderivatives $\mathcal{I}_k^{(2,3)}$ given by Eqs.~\eqref{eq:Ir2}--\eqref{eq:Ipm2} for type (2) geodesics or Eqs.~\eqref{eq:Ir3}--\eqref{eq:Ipm3} for type (3) geodesics.
In practice, we take $r_{\rm o}\gg M$ to be very large but not infinite, for reasons described in Eq.~\eqref{eq:RenormalizedTime} below.

\begin{figure}
    \includegraphics[width=\columnwidth]{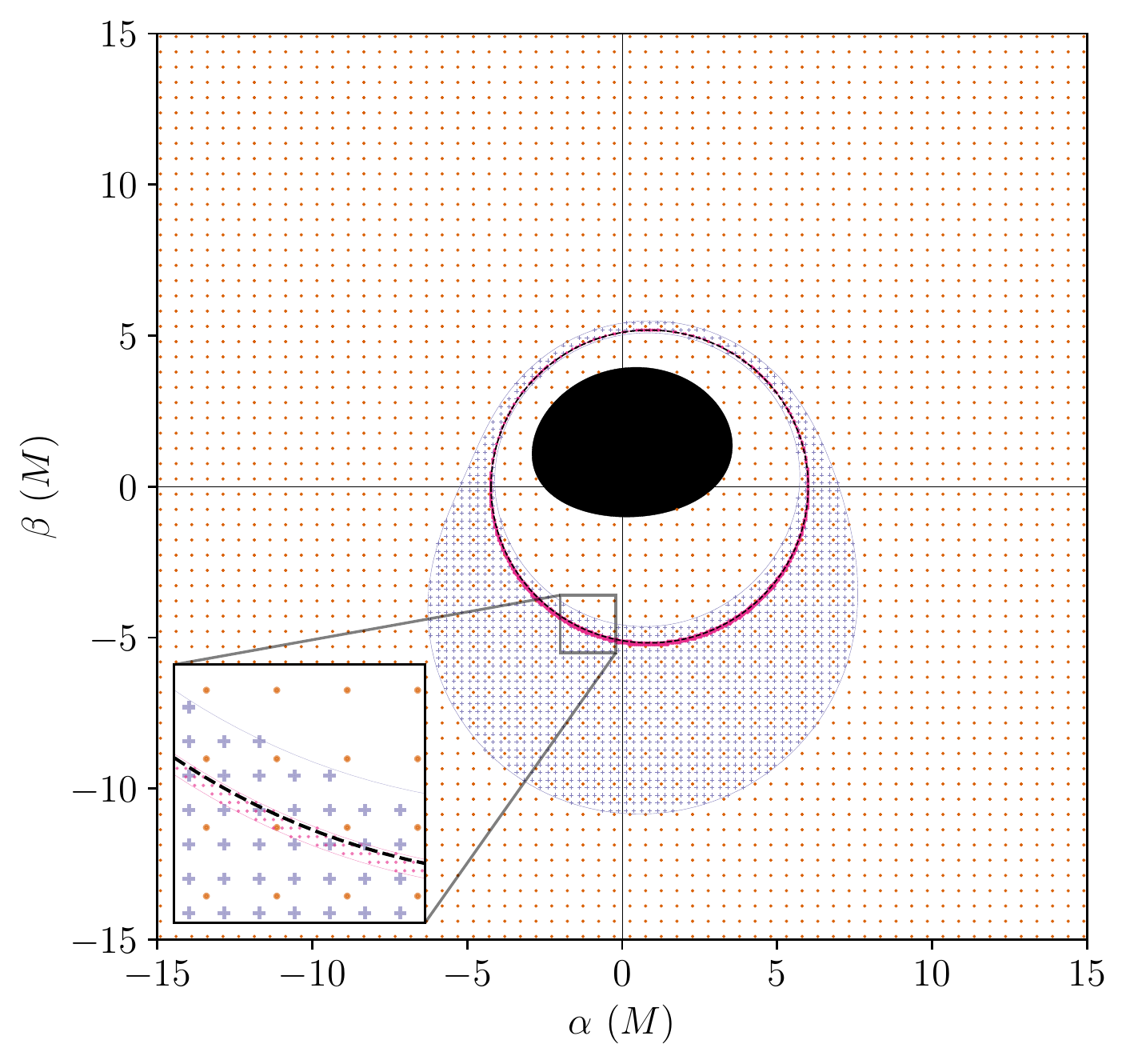}
    \caption{Illustration of lensing bands with different grid resolutions, for black hole spin $a/M=50\%$ and observer inclination $\theta_{\rm o}=60^\circ$.
    Orange, purple and magenta marks depict light rays that intersect the equatorial plane once ($n=0$ band), twice ($n=1$ band) and thrice ($n=2$ band), respectively.
    The black dashed line is the critical curve \eqref{eq:CriticalCurve}.
    Embedded is a panel showcasing the increasing grid resolution in higher-$n$ bands.
    The region inside the lensed $n=0$ equatorial horizon, shown in black, is always excluded from our calculations.}
    \label{fig:LensingBands}
\end{figure}

\subsection{Kerr lensing bands}
\label{subsec:LensingBands}

The functions $r_{\rm s}^{(n)}$, $\phi_{\rm s}^{(n)}$, and $t_{\rm s}^{(n)}$ defined in Eqs.~\eqref{eq:rs}--\eqref{eq:ts} are the ``transfer functions'' that map the equatorial plane to its $(n+1)^\text{th}$ lensed image in the observer sky, where the index $n\ge0$ may be thought of as a photon half-orbit number \cite{Johnson2020,GrallaLupsasca2020a}.

To describe the lensing behavior of a Kerr black hole, it is helpful to draw contour plots of these function---that is, level sets of fixed $r_{\rm s}^{(n)}$, $\phi_{\rm s}^{(n)}$, and $t_{\rm s}^{(n)}$ within each image layer $n$.
First, however, one must determine the regions of the image plane in which these functions have support.

This is a nontrivial problem because these functions always evaluate to some value: for instance, $r_{\rm s}^{(n)}(\alpha,\beta)$ always returns some source radius, even when the ray shot back from $(\alpha,\beta)$ does not in fact intersect the equatorial plane $n+1$ times.
In such cases, $r_{\rm s}^{(n)}(\alpha,\beta)$ may sometimes be obviously unphysical (it could, for example, take a negative value), but not always.

We define the $n^\text{th}$ lensing band as the image-plane subregion consisting of those rays that cross the equatorial plane at least $n+1$ times after being shot back from the observer, before either terminating on the horizon (if they are shot back from inside the critical curve $\mathcal{C}$) or else returning to infinity (if they are shot back from outside $\mathcal{C}$) after an elapsed Mino time $\tau_{\rm max}$.

By definition, the $n^\text{th}$ lensing band is the physical domain of the transfer functions $r_{\rm s}^{(n)}$, $\phi_{\rm s}^{(n)}$, and $t_{\rm s}^{(n)}$.
This definition also implies that it is the subregion of the image plane in which the function $N(\alpha,\beta)$ defined in Eq.~\eqref{eq:MaxCrossing} is precisely equal to $n+1$.

This last observation leads us to a formula that characterizes the boundary of the $n^\text{th}$ lensing band: it is the set of points $(\alpha,\beta)$ in the image plane for which
\begin{align}
	\frac{\tau_{\rm max}\sqrt{-u_-a^2}+\sign(\beta)F_{\rm o}}{2K}-H(\beta)=n,
\end{align}
such that $N(\alpha,\beta)$ jumps by one across the boundary of each lensing band.
This condition can equivalently be written as
\begin{align}
	\label{eq:LensingBandBoundaries}
	\tau_{\rm max}=G_\theta^{(n)},
\end{align}
with $G_\theta^{(n)}$ given in Eq.~\eqref{eq:Gtheta}.
The $n\ge1$ lensing bands have a different character than the $n=0$ layer---we now focus on the former and defer a discussion of the latter to the next section.

Each $n\ge1$ lensing band is annular in shape and foliated by contours of fixed $r_{\rm s}^{(n)}$, with every source radius mapping onto a unique closed curve within the band.
Moreover, this bijective map is order-preserving: moving radially outwards within a band, one crosses contours of monotonically increasing $r_{\rm s}^{(n)}$.
Thus, the inner edge of the $n^\text{th}$ lensing band is the $(n+1)^\text{th}$ image of the equatorial event horizon (the contour of fixed $r_{\rm s}^{(n)}=r_+$), while its outer edge is the $(n+1)^\text{th}$ image of the equatorial circle at infinity (the contour of fixed $r_{\rm s}^{(n)}=\infty$).

Rays that connect a distant observer to the event horizon cannot encounter a radial turning point along their trajectory, so the inner edge must always appear inside the critical curve.
Conversely, rays that connect a distant observer to infinity have to make a turn along their radial trajectory, so the outer edge must always appear outside the critical curve.
Hence, the $n\ge1$ lensing bands are annuli that always straddle the critical curve.
Since successive images of a source are demagnified and appear exponentially closer to the critical curve \cite{Johnson2020,GrallaLupsasca2020a}, the lensing bands form a stack of nested annuli, as seen in Fig.~\ref{fig:LensingBands}.

To summarize, the inner edge of the $n^\text{th}$ lensing band is the solution of Eq.~\eqref{eq:LensingBandBoundaries} with $\tau_{\rm max}=\tau_{\rm max}^-$ given in Eq.~\eqref{eq:MinoTimeToHorizon}, while its outer edge is the solution with $\tau_{\rm max}=\tau_{\rm max}^+$ given in Eq.~\eqref{eq:MinoTimeToInfinity}.
The two sides of Eq.~\eqref{eq:LensingBandBoundaries} are continuous functions $G_\theta^{(n)}(\alpha,\beta)$ and $\tau_{\rm max}^\pm(\alpha,\beta)$ of the image-plane position $(\alpha,\beta)$.

In practice, we use the following procedure to determine these edges.
We first select a set of polar angles $\varphi_i\in[0,2\pi]$ around the image plane, with associated positions $(\tilde{\alpha}_i,\tilde{\beta}_i)$ on the critical curve \eqref{eq:CriticalCurve}---that is, $(\tilde{\alpha}_i,\tilde{\beta}_i)\in\mathcal{C}$ is the point where the critical curve is intersected by the ray emanating from the origin at polar angle $\varphi_i$.
Next, we parameterize these rays as $(\alpha_i(\epsilon),\beta_i(\epsilon))=\epsilon(\tilde{\alpha}_i,\tilde{\beta}_i)$.
For each $\varphi_i$, we compute $G_\theta^{(n)}(\epsilon)$ along the corresponding ray over the range $\epsilon\in(0.5,3)$.
We also compute $\tau_{\rm max}^-(\epsilon)$ on the range $\epsilon\in(0.5,1)$ and $\tau_{\rm max}^+(\epsilon)$ on the range $\epsilon\in(1,3)$.
We then determine the unique parameters $\epsilon^\pm$ within each of these ranges such that $G_\theta^{(n)}(\epsilon^\pm)=\tau_{\rm max}^\pm(\epsilon^\pm)$.
These values correspond to the points $(\alpha_i^\pm,\beta_i^\pm)=\epsilon^\pm(\tilde{\alpha}_i,\tilde{\beta}_i)$ along the ray of constant $\varphi_i$ where it intersects the lensing band's outer and inner edges, respectively.
Connecting the points $(\alpha_i^\pm,\beta_i^\pm)$ thus obtained at every angle $\varphi_i$ traces out the edges of the lensing band, which is finally obtained as the annular region between these two closed curves.
These boundaries---and hence, the lensing band---can be determined to arbitrarily high precision by sampling sufficiently many $\varphi_i$.

To check the computation, one can verify that for each $i$, $r_{\rm s}^{(n)}(\alpha_i^-,\beta_i^-)\approx r_+$ and $r_{\rm s}^{(n)}(\alpha_i^+,\beta_i^+)\approx\infty$ up to numerical error.\footnote{In principle, one could also determine the inner and outer boundaries of the $n^\text{th}$ lensing band as those contours of fixed $r_{\rm s}^{(n)}(\alpha,\beta)=r_-$ (inside $\mathcal{C}$) or fixed $r_{\rm s}^{(n)}(\alpha,\beta)=\infty$ (outside $\mathcal{C}$), respectively, which are closest to $\mathcal{C}$.
However, this method is impractical because $r_{\rm s}^{(n)}(\alpha,\beta)$ can vary significantly across these contours.
Yet, it is this very behavior that makes this check effective.}
Further discussion of lensing bands may be found in Ref.~\cite{Paugnat2022}.

\subsection{Apparent image of the horizon}
\label{subsec:ApparentHorizon}

We now return to the $n=0$ lensing band, which (unlike its higher-$n$ counterparts) is noncompact.
This difference can be understood intuitively: if the black hole mass shrinks to zero, the $n\ge1$ lensing bands must disappear with the hole, whereas the $n=0$ band must extend to fill the entire image plane.

Mathematically, the equations in the previous section still hold.
The inner and outer edges of the $n=0$ band are still the contours of fixed $r_{\rm s}^{(0)}=r_+$ and fixed $r_{\rm s}^{(0)}=\infty$, respectively.
These curves respectively correspond to the primary image of the equatorial event horizon (appearing inside $\mathcal{C}$) and of the  circle of infinite radius (appearing outside $\mathcal{C}$).
The inner edge can be determined via the same root-finding procedure outlined above (with $n=0$), resulting in a closed curve $\mathcal{H}$ contained within the critical curve $\mathcal{C}$.
On the other hand, the outer edge is at infinity in the image plane (as it would be in flat space), so the $n=0$ lensing band is no longer an annulus.
Instead, it consists of the entire (unbounded) exterior of $\mathcal{H}$.

Meanwhile, the interior of $\mathcal{H}$ may technically be viewed as the $n=-1$ band, since it consists of rays that never reach the equator before terminating on the horizon.
In an equatorial disk model, this region would be completely dark, giving rise to an ``inner shadow'' feature \cite{Chael2021}, and may naturally be viewed as the apparent image of the horizon.
In Fig.~\ref{fig:LensingBands}, we display this image as a shaded black region for the case of a black hole with spin $a/M=50\%$ observed from an inclination $\theta_{\rm o}=60^\circ$.

\subsection{Analytic grid adaptiveness}
\label{subsec:GridAdaptiveness}

It has long been known that images of the equatorial plane of a Kerr black hole are lensed into increasingly demagnified, compact regions of the image plane \cite{Luminet1979,Beckwith2005}.
More recently, this lensing behavior has been exploited to efficiently ray trace equatorial disk models in which the $n=0$, 1, and 2 image components are all fully resolved \cite{GLM2020}.

AART is designed with this goal in mind and is built around the lensing behavior of a Kerr black hole.
It ray traces images of an equatorial source layer-by-layer, using a non-uniform grid adapted to this layered structure.
The $n^\text{th}$ image layer, which consists of the $(n+1)^\text{th}$ image of the source, is only ray traced on pixels within the $n^\text{th}$ lensing band, which is precisely the region of the image plane occupied by this image.
As a result, AART avoids computing unneeded pixels in each layer, substantially improving its efficiency.

This layer-by-layer approach naturally allows for different resolutions to be used in each image layer.
This is important because the exponential-in-$n$ demagnification of successive images of a source requires the use of exponentially higher resolutions to resolve the source at the same level of detail in higher-$n$ layers.
AART increases the resolution in successive lensing bands by roughly the same demagnification factor $e^\gamma$ that shrinks them ($\gamma$ is a Lyapunov exponent governing the orbital instability of bound photon orbits~\cite{Johnson2020}) such that every image of the source is resolved by roughly the same number of pixels---see Sec.~\ref{sec:Requirements} for a more detailed discussion of the resolution requirements in each layer.
This approach enables ray tracing up to arbitrarily high $n$ and ensures that all images of the source are equally well resolved, with a roughly fixed computational cost per layer.

To summarize, AART makes use of a non-uniform adaptive grid composed of multiple layers, each of which it ray traces at a resolution adapted to the lensing band that makes up the layer.
Since the lensing bands are completely determined by the Kerr geometry via the analytic formula \eqref{eq:LensingBandBoundaries}, this form of adaptiveness is analytic, with the ray tracing also carried out analytically using the exact transfer functions \eqref{eq:rs}--\eqref{eq:ts}.
This adaptiveness is purely geometrical and independent of image features or gradients (unlike, for example, the method in Refs.~\cite{Wong2021,Gelles2021a}).

In practice, to produce the $n=0$ grid points, AART creates a Cartesian grid covering the desired area of the image plane, but excluding points in the apparent image of the horizon (that is, the interior of $\mathcal{H}$).
In each $n\ge1$ layer, the code produces a Cartesian grid of points in an area centered around the critical curve and then discards those points which do not lie within the $n^\text{th}$ lensing band.
Numerically, the code only keeps those points which lie within the concave hull of the points $(\alpha_i^+,\beta_i^+)$ and outside the concave hull of the points $(\alpha_i^-,\beta_i^-)$.

We caution the reader that the $n\ge1$ bands are not always convex: at high inclinations and moderate to high spins, their edges may become non-convex curves, as may be seen for instance in the right column of Fig.~\ref{fig:n1TransferFunctions} for the $n=1$ band or in Fig.~\ref{fig:n2TransferFunctions} for the $n=2$ band.
In such cases, using the convex hull of the boundary points $(\alpha_i^\pm,\beta_i^\pm)$ to define the lensing band can produce an incorrect grid that  overestimates the size of the band and includes points which do not really belong in it, which is why must use the concave hull instead.

In Fig.~\ref{fig:LensingBands}, we display an example of an AART grid (with a coarse resolution for visualization purposes), which includes points in the $n=0$ (orange dots), $n=1$ (purple crosses) and $n=2$ (magenta dots) lensing bands.

\subsection{Visualization of the equatorial transfer functions}
\label{subsec:Visualization}

\begin{figure*}
    \centering
    \includegraphics[width=\textwidth]{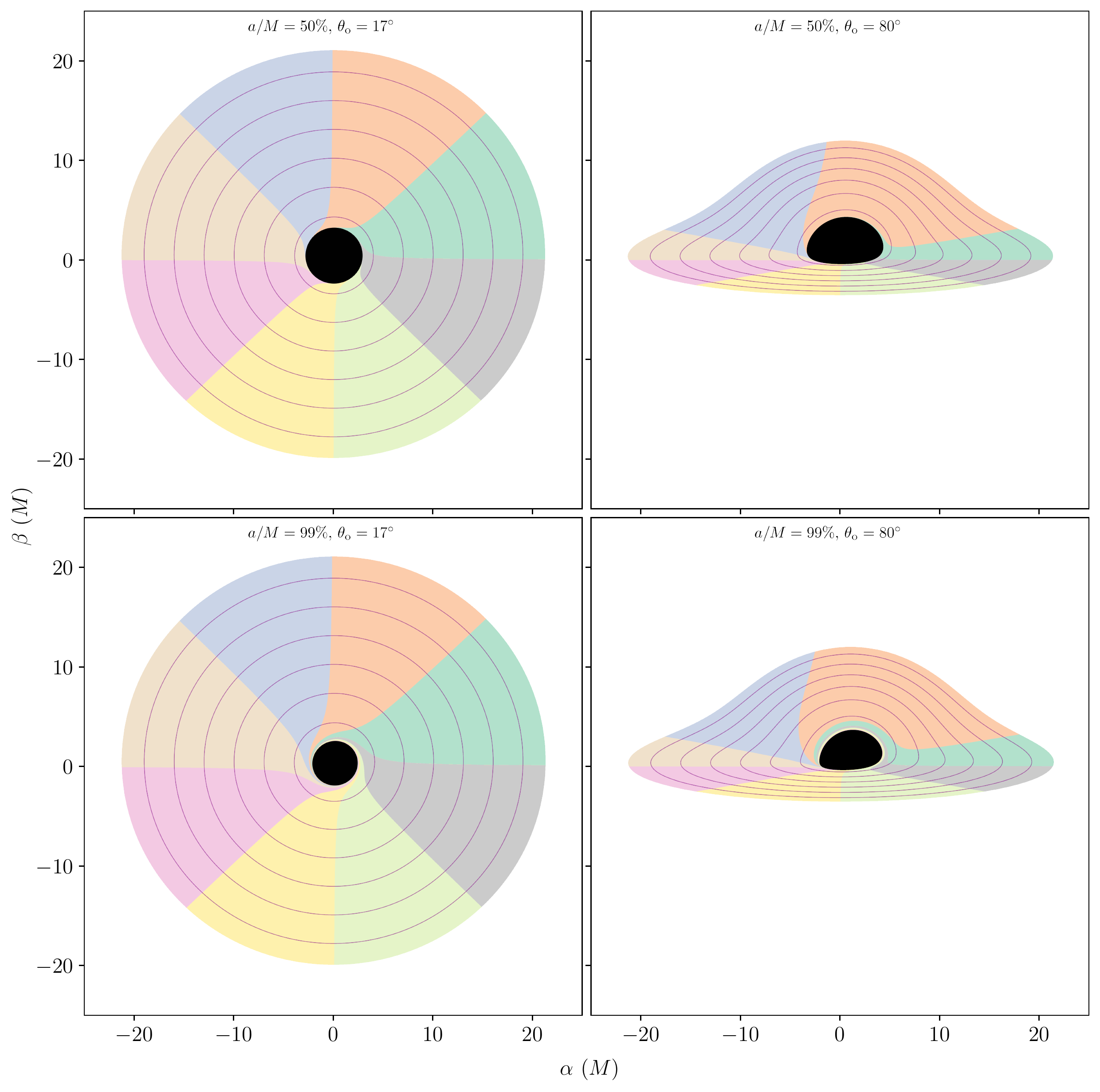}
    \caption{Direct ($n=0$) images of the equatorial plane $\theta_{\rm s}=\pi/2$ for Kerr black holes with spins $a/M\in\cu{50\%,99\%}$ and observer inclinations $\theta_{\rm o}\in\cu{17^\circ,80^\circ}$.
    The isoradial curves (purple) correspond to rings of constant Boyer-Lindquist radius $r_{\rm s}/M\in\cu{3,6,9,12,15,18}$ between $r_{\rm s}=r_+$ (the apparent image of the equatorial horizon, in black) and the cutoff at $r_{\rm s}=20M$.
    The colors change every $45^\circ$ across curves of constant $\phi_{\rm s}$.}
    \label{fig:n0TransferFunctions}
\end{figure*}

\begin{figure*}
    \centering
    \includegraphics[width=\textwidth]{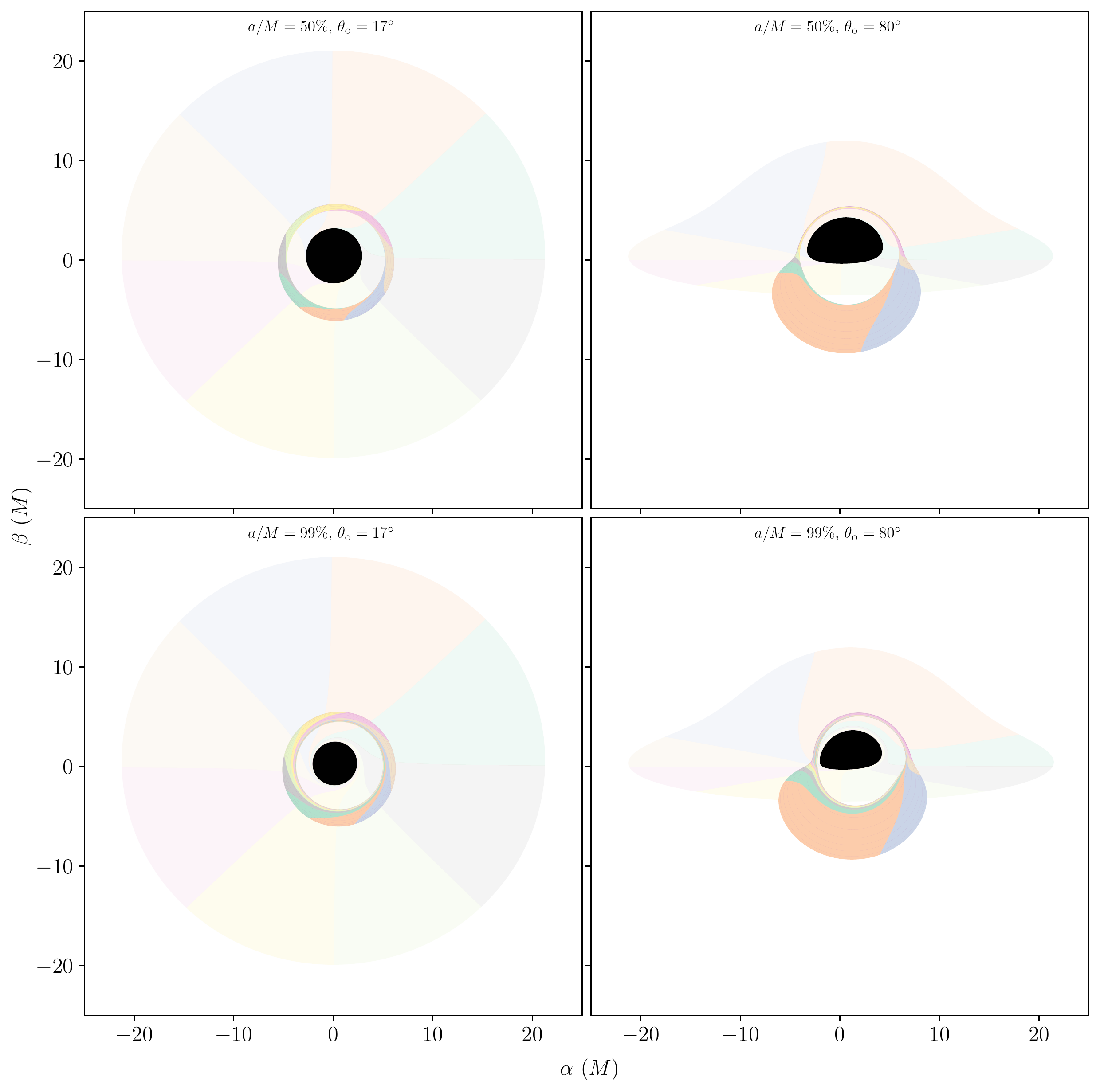}
    \caption{Same as Fig.~\ref{fig:n0TransferFunctions} for the $n=1$ image, which is a demagnified and rotated copy of the direct $n=0$ image (shown as a faded background).}
    \label{fig:n1TransferFunctions}
\end{figure*}

\begin{figure*}
    \centering
    \includegraphics[width=\textwidth]{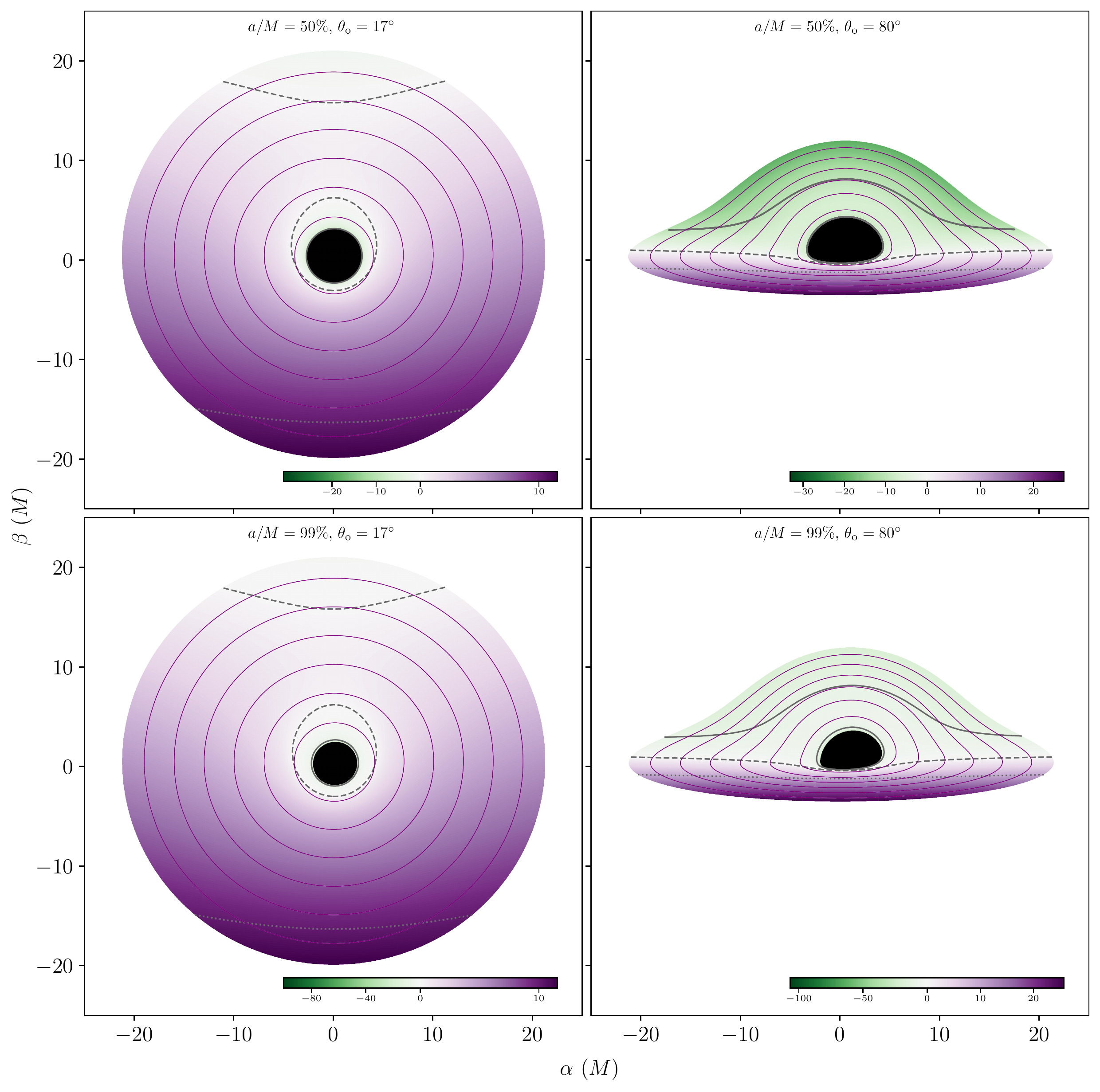}
    \caption{Same as Fig.~\ref{fig:n0TransferFunctions} for the renormalized time $\tilde{t}^{(0)}$ relative to an observer at $r_{\rm o}=10000M$ [Eq.~\eqref{eq:RenormalizedTime}], with the same isoradial curves (purple).
    The solid, dashed, and dotted grey lines respectively denote contours of fixed $\tilde{t}/M\in\cu{-10,0,10}$.
    Each panel has its own color scale for clarity.}
    \label{fig:n0Time}
\end{figure*}

\begin{figure}
    \centering
    \includegraphics[width=\columnwidth]{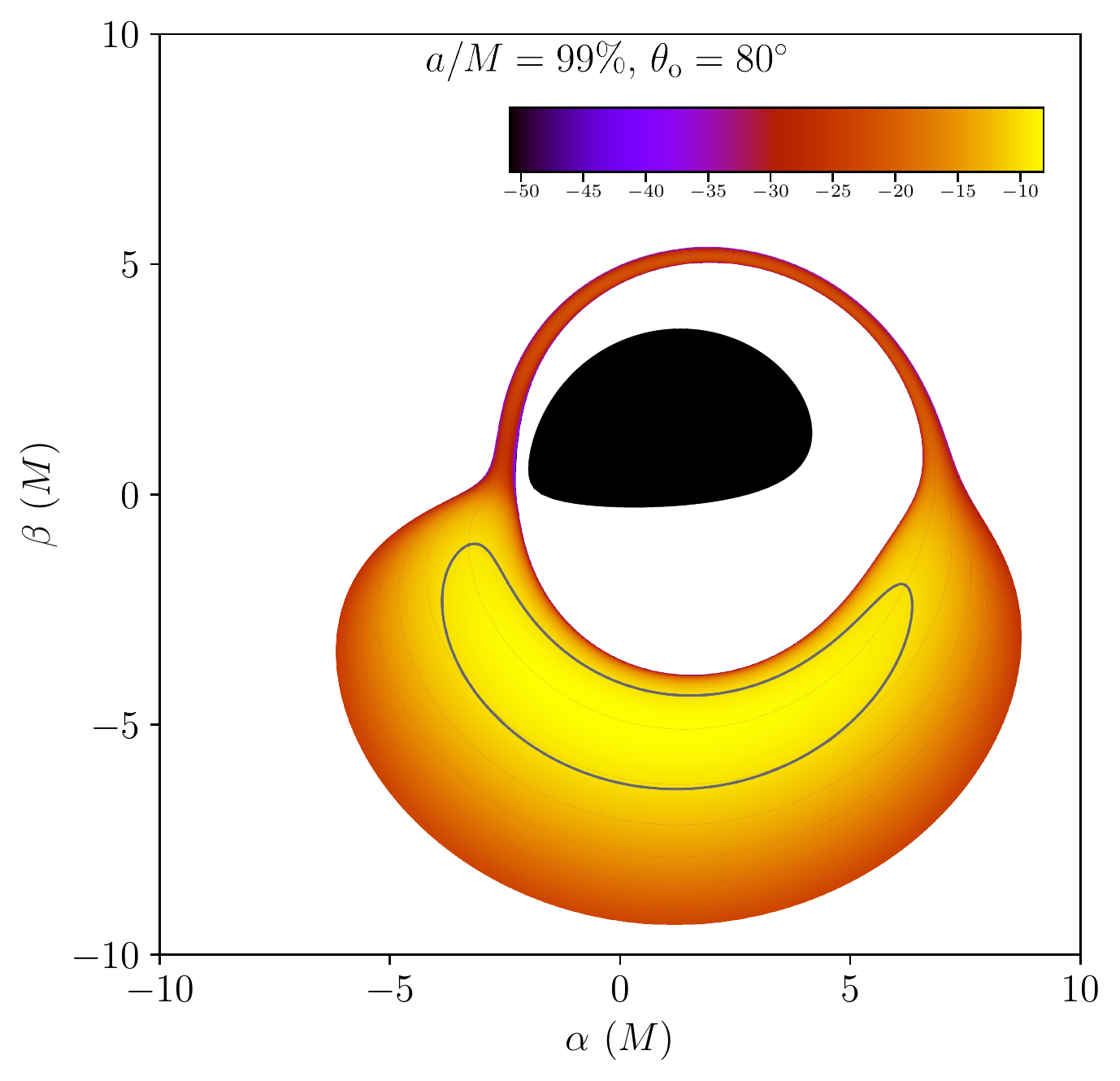} 
    \caption{Same as Fig.~\ref{fig:n0Time} for the $n=1$ image.
    The solid grey line is an isochronous curve of fixed renormalized time $\tilde{t}^{(1)}=-10M$.
    The color scale is chosen to highlight the rapid variation in $\tilde{t}^{(1)}$ at the edges of the lensing band (which correspond to the horizon and infinity).}
    \label{fig:n1Time}
\end{figure}

\begin{figure}
    \centering
    \includegraphics[width=\columnwidth]{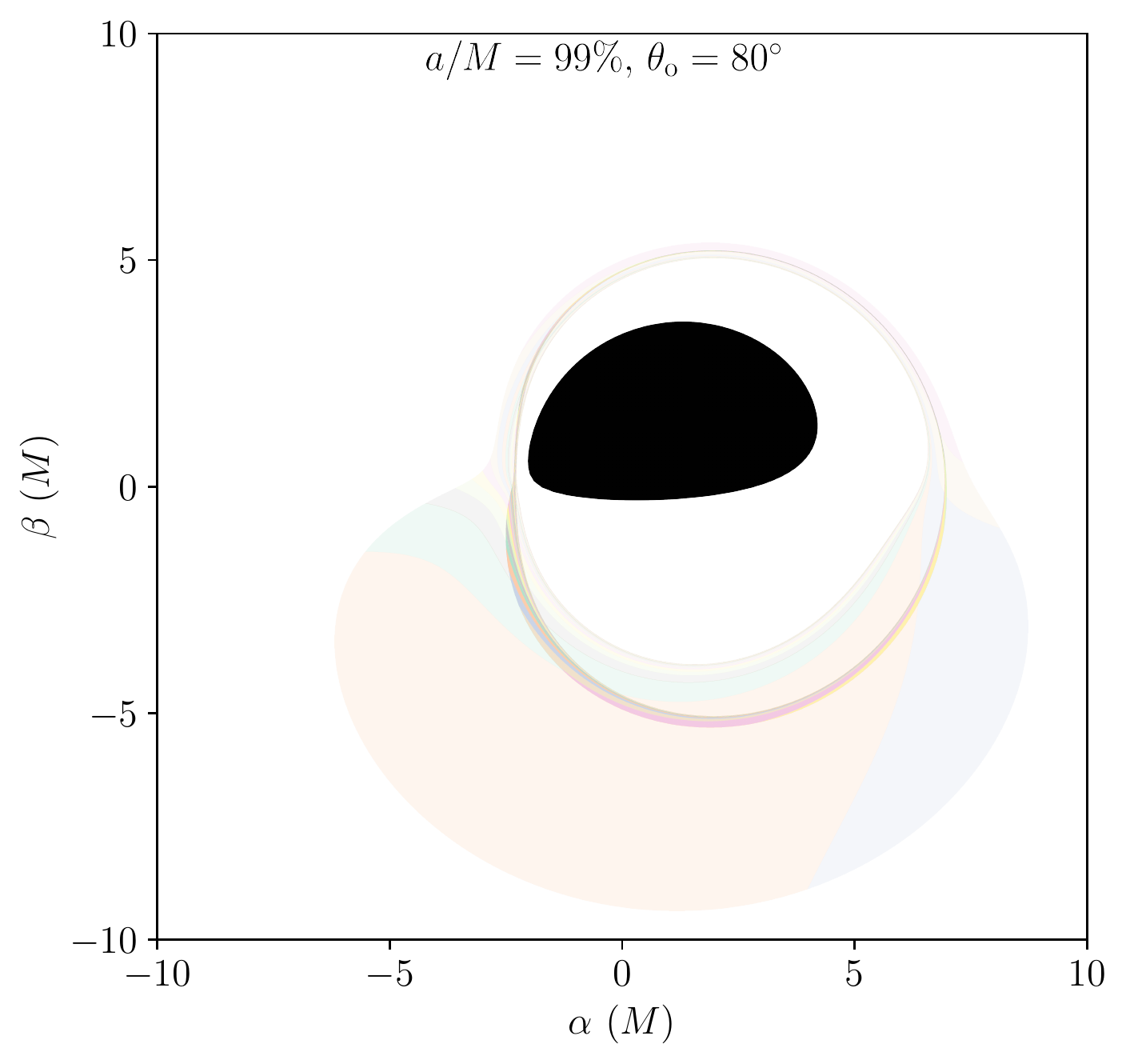}
    \caption{Same as Figs.~\ref{fig:n0TransferFunctions} and \ref{fig:n1TransferFunctions} for the $n=2$ image, with the $n=1$ image shown as a faded background.
    Despite a significant zoom, the extremely narrow $n=2$ lensing band is still barely visible.}
    \label{fig:n2TransferFunctions}
\end{figure}

\begin{figure*}
    \centering
    \includegraphics[width=\textwidth]{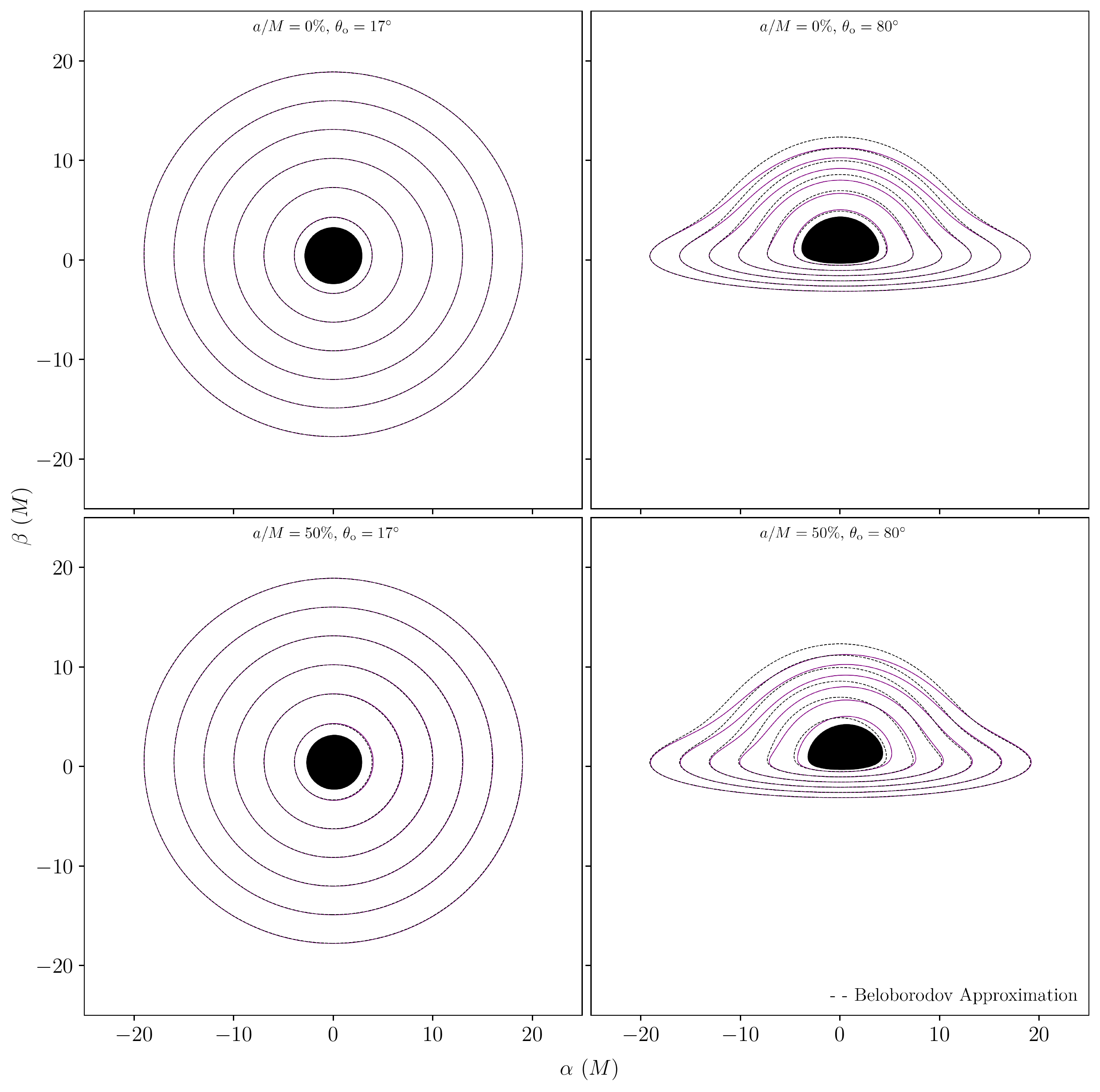}
    \caption{Apparent positions of source rings of constant Boyer-Lindquist radius $r_{\rm s}$ in the equatorial plane ($\theta_{\rm s} = \pi/2$), for different values of the spin and inclination within the $n=0$ band, using the exact expression [Eq.~\eqref{eq:rs}] and the Beloborodov approximation [Eq.~\eqref{eq:BeloborodovApproximation}].
    The Beloborodov approximation is still good even for moderate values of the spin and high inclination values.}
    \label{fig:BeloborodovApproximation}
\end{figure*}

Having defined the transfer functions \eqref{eq:rs}--\eqref{eq:ts} that map the equatorial plane of a Kerr black hole onto the image plane of a distant observer, we will now illustrate their behavior with the help of contour plots.
Though we have yet to describe their range, we have already elucidated their domain of definition: as described in Sec.~\ref{subsec:LensingBands}, the functions $r_{\rm s}^{(n)}$, $\phi_{\rm s}^{(n)}$, and $t_{\rm s}^{(n)}$ are only physical within the corresponding $n^\text{th}$ lensing band.

Following Refs.~\cite{Luminet1979,GrallaLupsasca2020a}, in Figs.~\ref{fig:n0TransferFunctions} and \ref{fig:n1TransferFunctions}, we display contours of fixed $r_{\rm s}^{(0)}$ and $r_{\rm s}^{(1)}$, which foliate the $n=0$ and $n=1$ lensing bands, respectively.
Physically, these ``isoradial'' curves are the direct ($n=0$) and first relativistic ($n=1$) images of the rings of constant Boyer-Lindquist radius $r_{\rm s}$ in the equatorial plane $\theta_{\rm s}=\pi/2$.
Since the $n=0$ lensing band is unbounded, $r_{\rm s}^{(0)}$ has noncompact support, while $r_{\rm s}^{(1)}$ maps the entire Kerr equatorial plane into a finite annulus: the $n=1$ lensing band.
In each band, $r_{\rm s}^{(n)}$ spans the entire range $[r_+,\infty)$ once and only once, so every equatorial ring produces precisely one image in each lensing band.
Moreover, the map $r_{\rm s}^{(n)}$ is order-preserving: the source radius grows monotonically in the radial direction.

These two figures also illustrate the behavior of $\phi_{\rm s}^{(0)}$ and $\phi_{\rm s}^{(1)}$ by painting the Kerr equatorial plane with a color wheel, as in Ref.~\cite{James2015}.
This wheel changes colors across ``isopolar'' curves of fixed $\phi_{\rm s}$, which are drawn at every $45^\circ$ in the source plane.
Their corresponding images---the curves of fixed $\phi_{\rm s}^{(n)}$---form a swirling pattern that illustrates the effects of frame-dragging.
We emphasize that this swirl is a purely geometric effect, and that gravitational redshift is not yet included at this stage (its effects will be described in Sec.~\ref{sec:Model} below).

Unlike $r_{\rm s}^{(n)}$, $\phi_{\rm s}^{(n)}$ has a rather complicated range.
While $\phi_{\rm s}^{(0)}$ spans the entire range $[0,2\pi)$ once and only once, so that every point source produces precisely one direct image (as in flat spacetime), as $n$ grows large, the maps $\phi_{\rm s}^{(n)}$ cover this range an increasing (linearly divergent) number of times \cite{Gralla2018,Gates2020,Gates2021}.

As a result, the equatorial plane is unfolded increasingly many times in higher lensing bands, which therefore contain multiple images of a single point source!
This surprising and intricate lensing behavior is connected to the development of caustics \cite{Bozza2008} and will be explored elsewhere.

While the transfer functions $r_{\rm s}^{(n)}$ and $\phi_{\rm s}^{(n)}$ suffice to describe the lensing of a static source, one still needs $t_{\rm s}^{(n)}$ to describe a time-dependent source and account for its time-delay effects.
Since light takes forever to reach a distant observer that is truly at asymptotic infinity, $t_{\rm s}^{(n)}$ is technically infinite when $r_{\rm o}\to\infty$.
More precisely, the null geodesic equation in Kerr dictates that
\begin{align}
	\label{eq:TimeAsymptotics}
	\frac{dt}{dr}\stackrel{r\to\infty}{\approx}1+\frac{2M}{r}+\bigO\pa{\frac{1}{r^2}}.
\end{align}
Hence, the coordinate time elapsed along a geodesic grows as
\begin{align}
	\label{eq:TimeDivergence}
	t_{\rm s}^{(n)}\stackrel{r_{\rm o}\to\infty}{\approx}r_{\rm o}+2M\log{r_{\rm o}}+\bigO\pa{1},
\end{align}
with the leading linear divergence arising from the first term in Eq.~\eqref{eq:TimeAsymptotics}, and the subleading log-divergence from the second.
Because of these divergences, $t_{\rm s}^{(n)}$ is inherently dependent on the observer radius.
However, subtracting these (``infrared'') divergences yields a ``renormalized'' time
\begin{align}
	\label{eq:RenormalizedTime}
	\tilde{t}^{(n)}\equiv t_{\rm s}^{(n)}-\pa{r_{\rm o}+2M\log{r_{\rm o}}},
\end{align}
which remains finite even as $r_{\rm o}\to\infty$.
Moreover, in this limit, the dependence on $r_{\rm o}$ drops out, and $\tilde{t}^{(n)}$ tends to a constant along any outward-propagating light ray.
In the limit $M\to0$, $\tilde{t}^{(0)}\to t_{\rm s}^{(0)}-r_{\rm o}$ and $\tilde{t}_{\rm s}^{(0)}=0$ is the future light cone of the origin.
In other words, in flat space, a ray emitted radially outwards from $r_{\rm s}=0$ at $t_{\rm s}=0$ reaches null infinity $r_{\rm o}\to\infty$ with $\tilde{t}^{(0)}=0$.

We display density plots of $\tilde{t}^{(0)}$ and $\tilde{t}^{(1)}$ in Figs.~\ref{fig:n0Time} and \ref{fig:n1Time}, respectively.
In practice, to take the limit $r_{\rm o}\to\infty$, we evaluate Eq.~\eqref{eq:RenormalizedTime} at a large value of $r_{\rm o}\gg M$.
The color scale in Fig.~\ref{fig:n0Time} ensures that $\tilde{t}^{(0)}=0$ is white and sits in the middle of the scale.
At higher spins and inclinations, the range of $\tilde{t}^{(0)}$ grows wider and more skewed toward negative values.
Accordingly, we use a different color scale in each panel.
We also display three contours of fixed time: $\tilde{t}^{(0)}=-10M$ (solid), $\tilde{t}^{(0)}=0$ (dashed), and $\tilde{t}^{(0)}=10M$ (dotted).
These ``isochronous'' curves may be either open or closed, and in some cases, they may even be composed of multiple disconnected segments, highlighting the warped nature of the black hole spacetime.

The range of $\tilde{t}^{(n)}$ is always infinite in every lensing band.
More precisely, it is unbounded below but bounded above: the reason is that, if the entire equatorial plane emits light forever, then an observer can see light from arbitrarily far back in the past (emitted from regions that are either very close to the horizon or very far behind it) but it does take some minimum time for any light to reach an observer, even when it is emitted from the nearest point in the plane.
In Fig.~\ref{fig:n0Time}, we do not see arbitrarily large negative values of $\tilde{t}^{(0)}$ because we have a finite resolution (and therefore do not have pixels that resolve rays arbitrarily close to the horizon) and because we cut off the disk at $r_{\rm s}=20M$ (and therefore do not see rays emitted from farther behind the black hole).

The infinite range of $\tilde{t}^{(1)}$ is more readily apparent in Fig.~\ref{fig:n1Time}, where we use a different color scale to highlight the divergent time elapsed along light rays at the edges of the lensing band: the inner edge consists of rays that asymptote to the horizon infinitely far back in the past, while the outer edge consists of rays that bounce back towards null infinity, incurring another time lapse that diverges as Eq.~\eqref{eq:TimeDivergence}. 
The rapidly changing colors at the edges of the lensing band illustrate this behavior.
Again, the finite resolution near the inner edge, together with the cutoff at $r_{\rm s}=20M$ near the outer edge, preclude us from resolving light rays emitted arbitrarily far back in the past.
In Fig.~\ref{fig:n1Time}, we chose a color scale spanning 99.9\% of the range of time lapses sampled by the pixels making up the $n=1$ grid. 

Comparing Figs.~\ref{fig:n0Time} and \ref{fig:n1Time}, we see that successive images of the equatorial plane are delayed by a time of order $\tau\sim15M$.
Likewise, comparing the $n=1$ image of the equatorial plane in Fig.~\ref{fig:n1TransferFunctions} to its $n=2$ image in  Fig.~\ref{fig:n2TransferFunctions} shows that successive images are not only time-delayed but also demagnified and rotated.
This lensing behavior results in a self-similar photon ring substructure that is governed by Kerr critical exponents $\gamma$, $\delta$, and $\tau$, which respectively control the demagnification, rotation, and time delay of successive subring images \cite{Johnson2020,GrallaLupsasca2020a}.

\subsection{Analytic ray tracing with Beloborodov's approximation}
\label{subsec:Beloborodov}

Direct ($n=0$) light rays experience the smallest deflection.
In Schwarzschild, Beloborodov used an ingenious expansion to derive an excellent analytic expression that approximates the trajectories of such rays \cite{Beloborodov2002}. His small-deflection-angle expansion is remarkably effective because its leading, linear term only receives its first subleading correction from a cubic term with a small coefficient.
To describe this result, we define
\begin{align}
	b=\sqrt{\alpha^2+\beta^2},\quad
	\cos{\psi}=-\frac{\beta\tan{\theta_{\rm o}}}{\sqrt{b^2+\beta^2\tan^2{\theta_{\rm o}}}}
    \in[-1,1].
\end{align}
In terms of these variables, a straightforward manipulation of Beloborodov's formulas leads to the approximate expression
\begin{align}
	\label{eq:BeloborodovApproximation}
	\frac{r_{\rm s}^{(0)}}{M}\approx\sqrt{\pa{\frac{1-\cos{\psi}}{1+\cos{\psi}}}^2+\frac{b^2}{1-\cos^2{\psi}}}-\pa{\frac{1-\cos{\psi}}{1+\cos{\psi}}},
\end{align}
which is highly accurate for low-to-moderate inclinations $\theta_{\rm o}$, as shown in Fig.~\ref{fig:BeloborodovApproximation}.
The agreement may seem surprising in the Kerr case, since this formula is derived for Schwarzschild.
In practice, the approximation \eqref{eq:BeloborodovApproximation} is excellent for rays that never get within $\sim\!4M$ of the black hole---for rays that come closer and experience greater deflection, it remains good along the portion of the ray up to the first equatorial crossing, but breaks down afterward, once the deflection angle grows large.

Intuitively, the reason Eq.~\eqref{eq:BeloborodovApproximation} holds even for nonzero spins is that $n=0$ rays are only weakly lensed, and do not spend sufficient time orbiting the black hole to explore its geometry.
Typically, $n=0$ rays only come close enough to the black hole to probe the leading, monopole moment of its gravitational field (the mass), but are largely insensitive to its subleading, dipole moment (the spin), which contributes small corrections to the transfer function in an inverse-radius expansion.\footnote{This observation also underlies the ``just add one'' prescription $r_{\rm s}^{(0)}\approx b-M$, which gives an excellent approximation to the transfer function for a polar observer at $\theta_{\rm o}=0$ and for any value of the spin, as first noticed in Ref.~\cite{GrallaLupsasca2020a} and then derived, along with subleading $\bigO\pa{M/b}$ corrections, in Ref.~\cite{Gates2020}.}

With a little effort, under the Beloborodov approximation for the direct image, one can show that the sign of the quantity
\begin{align}
	\label{eq:BeloborodovSign}
	\tau_{\rm B}=\cos{\psi}-\frac{1}{1-\frac{3b}{\tilde{b}}\cos\br{\frac{1}{3}\arccos\pa{-\frac{\tilde{b}}{b}}}},
\end{align}
where $\tilde{b}=\sqrt{\tilde{\alpha}^2+\tilde{\beta}^2}=3\sqrt{3}M$ is the Schwarzschild critical impact parameter (apparent image radius of the critical curve), determines whether a ray shot backward from an image-plane position $(\alpha,\beta)$ with $b>\tilde{b}$, which must necessarily encounter a radial turning point, does so before ($\tau_{\rm B}>0$) or after ($\tau_{\rm B}<0$) first crossing the equatorial plane at radius \eqref{eq:BeloborodovApproximation}.\footnote{Tsupko provides an analytic expression for the shape of $n\ge2$ images \cite{Tsupko2022}.}

\subsection{Photon four-momentum at equatorial crossings}
\label{subsec:SourceMomentum}

Last, we describe how to compute the source momentum of a photon that is loaded onto a ray as it crosses the equator.

The four-momentum $p=p_\mu\ed x^\mu$ of a Kerr photon is
\begin{align}
	\label{eq:NullMomentum}
	p=E\pa{-\ed t\pm_r\frac{\sqrt{\mathcal{R}(r)}}{\Delta(r)}\ed r\pm_\theta\sqrt{\Theta(\theta)}\ed\theta+\lambda\ed\phi},
\end{align}
where $E=-p_t$ is the photon energy, while $\mathcal{R}(r)$ and $\Theta(\theta)$ are the radial and angular geodesic potentials \eqref{eq:RadialPotential}--\eqref{eq:AngularPotential}, which depend on the specific (energy-rescaled) angular momentum $\lambda$ and Carter constant $\eta$ of the photon defined in Eq.~\eqref{eq:ConservedQuantities}.

A ray shot backwards from image-plane position $(\alpha,\beta)$ has conserved quantities $(\lambda,\eta)$ obtained by inverting Eq.~\eqref{eq:BardeenCoordinates}.
When such a ray crosses the equatorial plane for the $(n+1)^\text{th}$ time at the radius $r_{\rm s}^{(n)}$ given by Eq.~\eqref{eq:rs}, a photon of energy $E$ loaded onto the ray will have a geodesic momentum given by Eq.~\eqref{eq:NullMomentum} evaluated at $r=r_{\rm s}^{(n)}$ and $\theta=\pi/2$ with those values of $(\lambda,\eta)$.
This specifies the momentum up to discrete signs $\pm_{r,\theta}$.
We now describe how to also determine these two signs.

First, $\pm_\theta=\sign{p_{\rm s}^\theta}$ is trivial to compute, since $n=0$ photons must be emitted toward the observer, and so they always have $\pm_\theta=-\sign\pa{\cos{\theta_{\rm o}}}$.
This sign must flip at every crossing, so
\begin{align}
	\pm_\theta=(-1)^{n+1}\sign\pa{\cos{\theta_{\rm o}}}
\end{align}
at the $(n+1)^\text{th}$ equatorial crossing.
As for $\pm_r=\sign{p_{\rm s}^r}$, two possibilities arise.
Rays that appear inside the critical curve can never encounter a radial turning point and are therefore always outgoing; as such, for rays inside of $\mathcal{C}$, we always have
\begin{align}
	 \text{Inside }\mathcal{C}:\quad
	 \pm_r=+.
\end{align}
On the other hand, rays that appear outside the critical curve do encounter a radial turning point at radius $r=r_4$, which they reach at Mino time $\tau=\tau_4$ given in Eq.~\eqref{eq:HalfwayMinoTime}.
Hence, for rays outside of $\mathcal{C}$,
$\pm_r$ depends on whether the Mino time $\tau_{\rm s}^{(n)}=G_\theta^{(n)}$ of the $(n+1)^\text{th}$ equatorial crossing, which is given in Eq.~\eqref{eq:Gtheta}, precedes or follows the radial turn at Mino time $\tau_4$; that is,
\begin{align}
	 \text{Outside }\mathcal{C}:\quad
	 \pm_r=\sign\pa{\tau_4-G_\theta^{(n)}}.
\end{align}

When using the Beloborodov approximation from Sec.~\ref{subsec:Beloborodov} to ray trace $n=0$ images, one merely replaces the exact $r_{\rm s}^{(0)}$ in the preceding discussion by its approximation \eqref{eq:BeloborodovApproximation}, which specifies the geodesic momentum at first equatorial crossing up to signs $\pm_{r,\theta}$.
The angular sign is still $\pm_\theta=-\sign\pa{\cos{\theta_{\rm o}}}$.
For the radial sign $\pm_r$, we must again distinguish between rays inside and outside $\mathcal{C}$, but this time the relevant critical curve is that of Schwarzschild: a circle of constant radius $b=\tilde{b}$, where $b=\sqrt{\alpha^2+\beta^2}$ and $\tilde{b}=3\sqrt{3}M$.
Rays inside $\mathcal{C}$ (with $b<\tilde{b}$) have $\pm_r=+$ as always, while rays outside $\mathcal{C}$ with ($b>\tilde{b}$) have
\begin{align}
	\pm_r=\sign{\tau_{\rm B}}.
\end{align}

\section{Equatorial emission model}
\label{sec:Model}

In the previous section, we described how Kerr black holes lens light.
We will now introduce a simple, phenomenological model of electromagnetic emission from the equatorial plane and describe how to compute its observational appearance as seen by distant observers such as ourselves.

\subsection{Intensity images}

Our goal is to compute an observed intensity $I_{\rm o}$ at every image-plane position $(\alpha,\beta)$.
For each pixel in the image, we compute $I_{\rm o}(\alpha,\beta)$ by tracing the corresponding ray backwards into the geometry; each time it passes through the emission region---in this case, the equatorial plane---we load additional photons onto the ray according to the local source intensity $I_{\rm s}$.
Since $I/\nu^3$ is the invariant number of photons of frequency $\nu$ along a ray, it follows that 
\begin{align}
	\label{eq:ObservedIntensity}
	I_{\rm o}(\alpha,\beta)=\sum_{n=0}^{N(\alpha,\beta)-1}\zeta_ng^3\pa{r_{\rm s}^{(n)},\alpha,\beta}I_{\rm s}\pa{r_{\rm s}^{(n)},\phi_{\rm s}^{(n)}, t_{\rm s}^{(n)}},
\end{align}
which generalizes to non-stationary and non-axisymmetric sources the analogous formula in Ref.~\cite{GLM2020}.
In this expression, $\zeta_n$ is a geometric ``fudge'' factor and $g=\nu_{\rm o}/\nu_{\rm s}$ is the observed redshift, while the transfer functions $r_{\rm s}^{(n)}$, $\phi_{\rm s}^{(n)}$, and $t_{\rm s}^{(n)}$, which are given in Eqs.~\eqref{eq:rs}--\eqref{eq:ts}, denote the spacetime position of the ray's $(n+1)^\text{th}$ equatorial crossing.
Here, $n$ ranges from 0 to $N-1$, with $N$ the
number of crossings given in Eq.~\eqref{eq:MaxCrossing}.

The factor $\zeta_n$ is meant to account for the (neglected) effects of geometrical thickness, which we can mimick with \cite{GLM2020}
\begin{align}
	\zeta_n=
	\begin{cases}
		1 & n=0,\\
		\zeta & n>0,
	\end{cases}
	\quad
	0<\zeta\le1.
\end{align}
The inclusion of this geometric factor significantly improves the agreement of this simplified equatorial model of emission with time-averaged radiative GRMHD simulations \cite{Chael2021}.

Finally, to derive the observed redshift $g$, we must prescribe a four-velocity for the accretion flow of radiating matter.
We consider flows with a four-velocity of the general form
\begin{align}
	\label{eq:GeneralFourVelocity}
    u=u^t\pa{\pd_t-\iota\pd_r+\Omega\pd_\phi},
\end{align}
where the angular and radial-infall velocities are defined as
\begin{align}
    \label{eq:OmegaAndIota}
    \Omega=\frac{u^\phi}{u^t},\quad
    \iota=-\frac{u^r}{u^t}, 
\end{align}
and $u^\mu$ are the contravariant components of the four-velocity.
In App.~\ref{app:FourVelocity}, we introduce two purely circular flows: a geodesic, Keplerian flow $\mathring{u}$, together with a non-geodesic, sub-Keplerian flow $\hat{u}$ obtained by rescaling the Keplerian angular momentum by a sub-Keplerianity parameter $0<\xi\le1$, such that $\hat{u}\to\mathring{u}$ as $\xi\to1$.
We also introduce the four-velocity $\bar{u}$ of purely radial geodesic infall.
Then, following Refs.~\cite{Pu2016,Vincent2022}, we introduce a flow $\tilde{u}$ given by a linear superposition of purely circular and purely radial motion.
The resulting combined motion, which we present and derive in detail in App.~\ref{app:FourVelocity}, is a non-geodesic family of flows with three parameters: the sub-Keplerianity factor $0<\xi\le1$, and two other parameters $0\le\beta_r\le1$ and $0\le\beta_\phi\le1$ controlling the radial and angular components of the superposition, respectively.
For $\beta_r=\beta_\phi=0$, $\tilde{u}$ reduces to the $\xi$-independent radial inflow $\bar{u}$.
At the other extreme, when $\beta_r=\beta_\phi=1$, $\tilde{u}$ reduces to the sub-Keplerian flow $\hat{u}$, which for $\xi=1$ recovers the circular-equatorial geodesic flow first used by Cunningham \cite{Cunningham1975}.
We summarize these flows in Table~\ref{tbl:Flows}.

\begin{table}[]
	\begin{tabular}{|cc@{\hspace*{15pt}}c@{\hspace*{15pt}}c|}
	\hline 
	\multicolumn{2}{|c}{Motion type} & Four-velocity & Definition \tabularnewline
	\hline 
	\hline 
	\multirow{2}{*}{Geodesic} & Circular Keplerian  & $\mathring{u}$ & App.~\ref{subapp:KeplerianFlow} \tabularnewline
	& Radial infall & $\bar{u}$ & App.~\ref{subapp:RadialInflow} \tabularnewline
	\hline 
	\hline 
	\multirow{2}{*}{Non-geodesic} & Sub-Keplerian  & $\hat{u}$ & App.~\ref{subapp:SubKeplerianFlow}  \tabularnewline
	& General flow  & $\tilde{u}$ & App.~\ref{subapp:GeneralFlow} \tabularnewline
	\hline 
	\end{tabular}
	\caption{Summary of the different accretion flows that we consider.
	The circular Keplerian flow $\mathring{u}$ corresponds to Cunningham's geodesic prescription \cite{Cunningham1975}, while $\bar{u}$ denotes the radial geodesic inflow.
	The non-geodesic, sub-Keplerian flow $\hat{u}$ is the circular motion obtained by rescaling the Keplerian specific angular momentum $\mathring{\ell}=-\mathring{u}_\phi/\mathring{u}_t$ by a factor $0<\xi<1$.
	The non-geodesic flow $\tilde{u}$ is a general linear superposition of these purely circular and radial motions.}
	\label{tbl:Flows}
\end{table}

Once a specific four-velocity \eqref{eq:GeneralFourVelocity} has been prescribed, the observed redshift is then given by [Eq.~\eqref{eq:ObservedRedshift}]
\begin{align}
    \label{eq:ObservedRedshift}
    g=\frac{E}{-p_\mu u^\mu}
    =\br{u^t\pa{1\pm_r\frac{\sqrt{\mathcal{R}(r)}}{\Delta(r)}\iota-\lambda\Omega}}^{-1},
\end{align}
where the conserved quantities $(\lambda,\eta)$ of the photon trajectory are related to its apparent position $(\alpha,\beta)$ via Eqs.~\eqref{eq:BardeenCoordinates}, while $\pm_r=\sign p_{\rm s}^r$ denotes the sign of the photon radial momentum at its source whose computation is described in Sec.~\ref{subsec:SourceMomentum} above.

Given a black hole spin $a$, an observer inclination $\theta_{\rm o}$, some prescribed accretion flow $u$, and an equatorial emission profile $I_{\rm s}$, we now have everything needed to compute an observed image \eqref{eq:ObservedIntensity}.
To summarize, the procedure is the following:
\begin{enumerate}
	\item determine the lensing bands as described in Sec.~\ref{subsec:LensingBands};
	\item in each lensing band, define a regular Cartesian grid of appropriate resolution---the resulting grids (indexed by $n$) form the different image layers described in Sec.~\ref{subsec:GridAdaptiveness};
	\item for each pixel in a given layer, trace the corresponding ray back into the geometry and compute its equatorial crossings $x_{\rm s}^{(n)}$ using the transfer functions in Sec.~\ref{subsec:RayTracing};
	\item use the prescribed flow $u$ to compute the redshift \eqref{eq:ObservedRedshift};
	\item finally, for each pixel in the image, use the equatorial crossings from 3, the redshift from 4, and the prescribed source profile $I_{\rm s}$ to compute the observed intensity \eqref{eq:ObservedIntensity}.
\end{enumerate}
A key observation is that the source profile $I_{\rm s}$ only enters this procedure at the very last step.
In particular, steps 1 through 3 depend only on the black hole spin and observer inclination, and can therefore be computed once and for all for each choice of $(a,\theta_{\rm o})$.
The output can then be reused for each new choice of accretion flow $u$ or source profile $I_{\rm s}$, offering a significant time advantage (particularly when ray tracing \textit{movie} frames).

Some comments are in order about step 2.
First, while there is no reason to favor regular grids in principle, they do offer computational advantages in practice, since many numerical algorithms (such as grid interpolation) are optimized for such grids.
Second, there is no real reason to favor Cartesian grids, and \texttt{AART} only relies on them for simplicity---it may be, however, that for extremely high $n$, regular but non-Cartesian grids become more computationally efficient.
Third, the exact choice of ``appropriate resolution'' adapted to each grid will be described in detail in Sec.~\ref{sec:Requirements} below.

\subsection{Polarimetric images}

In principle, besides the observed intensity (Stokes $I$), we could also ray trace the observed linear polarization (Stokes $P=Q+iU$) to obtain a \textit{polarimetric image}.
In practice, rather than computing the Stokes parameters $Q$ and $U$ separately, we will instead simply express the observed linear polarization as
\begin{align}
	P_{\rm o}=mI_{\rm o}e^{2i\chi},
\end{align}
where $0\le m\le1$ is a fractional degree of polarization and $\chi$ denotes the EVPA \eqref{eq:EVPA}.
In a realistic model, $m$ itself would vary across the image, since the degree of polarization of a photon usually depends on the angle at which the corresponding light ray intersects the local magnetic field at the source.

\begin{figure*}
    \centering
    \includegraphics[width=\textwidth]{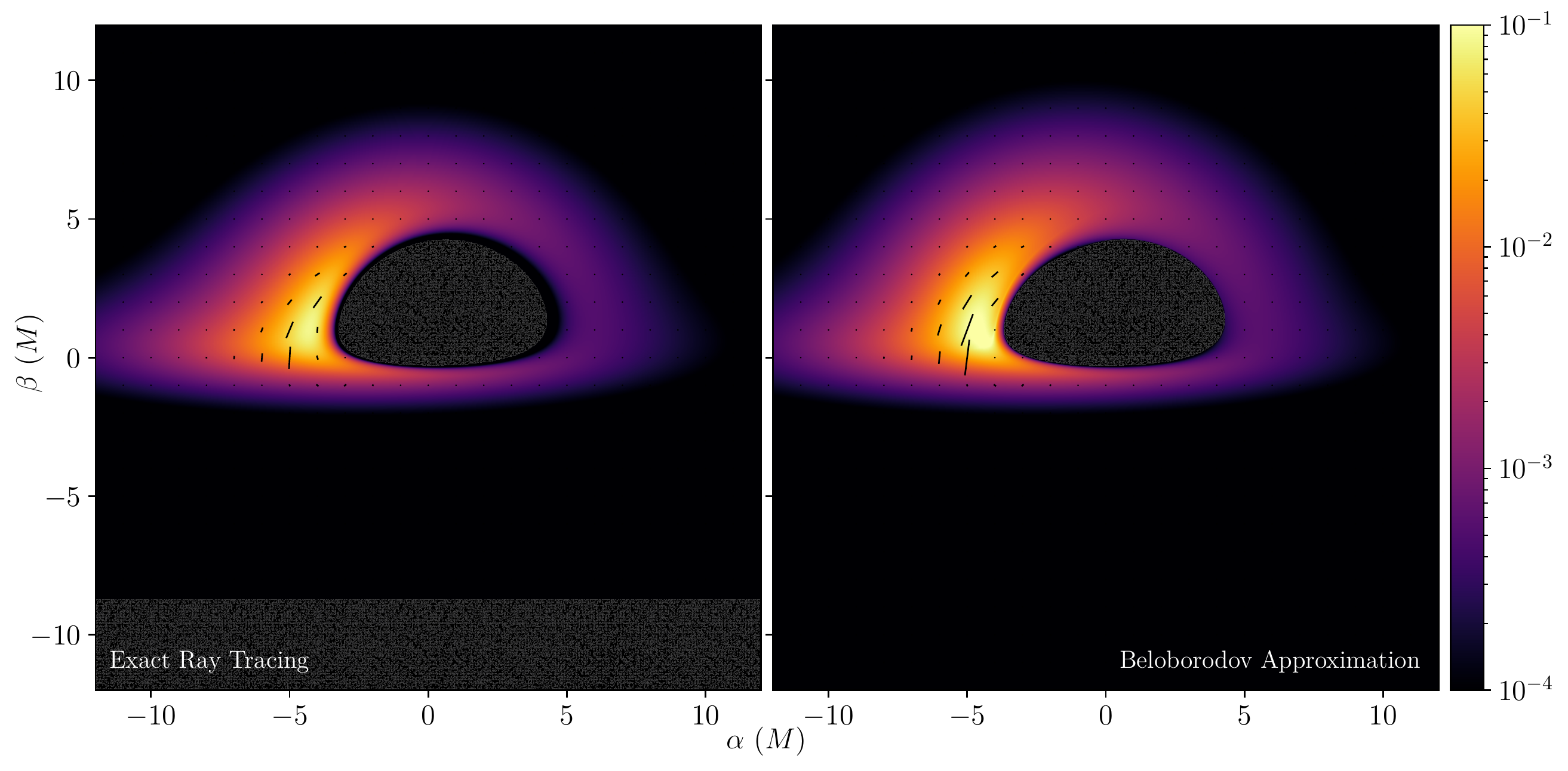}
    \caption{Observed intensity $I_{\rm o}$ (in logarithmic scale) and linear polarization $P_{\rm o}$ corresponding to the stationary and axisymmetric source profile \eqref{eq:JonhnsonSU} with parameters $\mu=r_-$, $\vartheta=M/4$, and $\gamma=-1$.
    The linear polarization is represented by ticks \eqref{eq:PolarizationTicks} that are aligned with the plane of polarization of the light---given by the electric-vector position angle (EVPA) [Eq.~\eqref{eq:EVPA}]---and whose length encodes the magnitude of Stokes $P$.
    Left: Polarimetric $n=0$ image ray traced using the exact analytic transfer functions in Sec.~\ref{subsec:RayTracing}.
    Right: Same image ray traced using the Beloborodov approximation \eqref{eq:BeloborodovApproximation} for the source radius $r_{\rm s}^{(0)}$.
    The black hole spin $a/M=50\%$ and observer inclination $\theta_{\rm o}=80^\circ$ correspond to the case from Fig.~\ref{fig:BeloborodovApproximation} with the largest difference between the exact and approximate isoradial curves, yet the resulting images appear almost identical.
    Here, the accretion flow follows the Cunningham prescription \cite{Cunningham1975} corresponding to geodesic circular motion, i.e., $u=\mathring{u}$ in App.~\ref{subapp:KeplerianFlow}, and we took the source magnetic field to have purely radial spacetime components, i.e., only $B_{\rm s}^r$ is nonzero.}
    \label{fig:BeloborodovComparison}
\end{figure*}

Here, we will be content to set $m$ to a constant, resulting in sufficiently realistic polarimetric images for our purposes.
In effect, this amounts to assuming that our astrophysical source emits isotropically, which is a convenient choice that allows us to continue treating the emission profile $I_{\rm s}$ as a scalar, rather than a directed quantity.
With this choice (redefining the color scale to absorb the proportionality factor $m$), the magnitude $\ab{P_{\rm o}}=mI_{\rm o}$ of the observed linear polarization is identical to the image intensity $I_{\rm o}$, leaving only its direction---the EVPA---as the sole new feature to be displayed in polarimetric images.
We indicate the orientation of the plane of polarization using ``polarimetric ticks'': these are polarization vectors \cite{Roberts1994,Himwich2020}
\begin{align}
	\label{eq:PolarizationTicks}
	f_{\rm o}=mI_{\rm o}\pa{-\sin{\chi}\pd_\alpha+\cos{\chi}\pd_\beta},
\end{align}
where $m$ is chosen to make the ticks legible and $\chi$ is computed from a prescribed source polarization profile (described in the next section) that we parallel transport along light rays using the conservation of the Penrose-Walker constant $\kappa$ defined in Eq.~\eqref{eq:PenroseWalker}.\footnote{A slightly more realistic synchrotron emission model would set $m\propto\sin^2{\zeta}$, where $\zeta$ denotes the angle of emission relative to the local magnetic field, but including this non-isotropy does not produce a large effect \cite{Narayan2021}.}
The length $|f_{\rm o}|=\ab{P_{\rm o}}\propto I_{\rm o}$ of the ticks also encodes the magnitude of the polarization, which improves readability.
We present an example of a polarimetric image in  Fig.~\ref{fig:BeloborodovComparison}.

In the remainder of this section, we describe the equatorial emission profiles implemented in \texttt{AART}.
The modularity of the code makes it easy to include any quantity of interest in the ray tracing, but we limit ourselves to Stokes $I$ and $P$ in this paper.

\subsection{Stationary and axisymmetric source profiles}

State-of-the-art time-averaged GRMHD-simulated images of realistic sources can be mimicked by ray tracing images of stationary, axisymmetric, equatorial emission profiles \cite{Chael2021}.
Such models therefore offer a computationally cheap way to study either time-independent or long-time-averaged sources.

Following \cite{GLM2020,Paugnat2022}, we model such sources by setting
\begin{align}
	\label{eq:RadialProfile}
	I_{\rm s}(r_{\rm s},\phi_{\rm s},t_{\rm s})=J(r_{\rm s}),
\end{align}
where we take the radial profile $J(r)$ to be the analytic function
\begin{align}
	\label{eq:JonhnsonSU}
	J_{\rm{SU}}(r;\mu,\vartheta,\gamma)\equiv\frac{e^{-\frac{1}{2}\br{\gamma+\arcsinh\pa{\frac{r-\mu}{\vartheta}}}^2}}{\sqrt{\pa{r-\mu}^2+\vartheta^2}},
\end{align}
which is derived from Johnson's SU distribution and can be rapidly computed to arbitrary precision everywhere.
The three parameters $\mu$, $\vartheta$, and $\gamma$ respectively control the location of the profile's peak, its width, and the profile asymmetry.
We invite the reader to consult Sec.~5 of Ref.~\cite{Paugnat2022} for a more in-depth description of these parameters and their interpretation. 

To illustrate the computation of a polarimetric image and the validity of Beloborodov's approximation in Sec.~\ref{subsec:Beloborodov}, we now consider a black hole of spin $a/M=50\%$ viewed from an inclination $\theta_{\rm o}=80^\circ$.
This is the case in Fig.~\ref{fig:BeloborodovApproximation} for which the differences between the exact and approximate $r_{\rm s}^{(0)}$ are largest.
We focus on the $n=0$ image layer, which we ray trace with both the exact transfer function \eqref{eq:rs} and its approximation \eqref{eq:BeloborodovApproximation}.
We adopt Cunningham's prescription $u=\mathring{u}$ (described in App.~\ref{subapp:KeplerianFlow}) for the accretion flow, resulting in the redshift $\mathring{g}$ given by Eq.~\eqref{eq:KeplerianRedshiftOutsideISCO} for sources outside the ISCO \eqref{eq:ISCO} and by Eq.~\eqref{eq:KeplerianRedshiftInsideISCO} for sources within.
This completes steps 1 through 4 above.
In step 5, we set $J(r_{\rm s})=J_{\rm{SU}}(r_{\rm s})$ with $\mu=r_-$ [Eq.~\eqref{eq:Horizons}], $\vartheta=M/4$, and $\gamma=-1$, so the emission peaks past the horizon.

Finally, to prescribe the source polarization $f_{\rm s}$, we impose
\begin{align}
	\label{eq:PolarizationConditions}
	f_{\rm s}\cdot p_{\rm s}=0,\quad
	f_{\rm s}\cdot u_{\rm s}=0,\quad
	f_{\rm s}\cdot B_{\rm s}=0.
\end{align}
The first condition states that the polarization is perpendicular to the direction of light propagation---this must by definition be true everywhere, including at the source.
Meanwhile, the second condition implies that the polarization vector is purely spatial in the emitter frame, since
\begin{align}
	f^{(t)}=-f_{(t)}
	=-f_\mu\mathbf{e}_{(t)}^\mu
	=-f_{\rm s}\cdot u_{\rm s},
\end{align}
where Latin indices ${(a)}$ label tensor components in the local orthonormal frame $\mathbf{e}_{(a)}^\mu$ of the emitter, with time leg $\mathbf{e}_{(t)}^\mu=u_{\rm s}^\mu$, while Greek indices $\mu$ label spacetime components.
Imposing this condition does not lead to any loss of generality.
Rather, it is simply a way to fix gauge: under gauge transformations, the polarization undergoes a gauge shift $f_{\rm s}\to f_{\rm s}+cp_{\rm s}$ \cite{Himwich2020} that leaves the first condition invariant but shifts $f_{\rm s}\cdot u_{\rm s}$ by $cp_{\rm s}\cdot u_{\rm s}$.
Thus, setting $c=-(f_{\rm s}\cdot u_{\rm s})/(f_{\rm s}\cdot u_{\rm s})$ results in a gauge-fixed $f_{\rm s}$ that obeys the second condition by construction.
Lastly, the third condition asserts that the polarization is perpendicular to the spacetime vector $B_{\rm s}$.
This is the only physical assumption imposed by the relations \eqref{eq:PolarizationConditions}.
A suitable choice for modeling synchrotron emission is to let 
$B_{\rm s}$ be the local magnetic field.\footnote{As a technical aside, we emphasize here that the condition $f_{\rm s}\cdot u_{\rm s}=0$ is also crucial for the following reason.
Suppose that we start with a polarization $f_{\rm s}'$ which only obeys $f_{\rm s}'\cdot p_{\rm s}=0$, while $f_{\rm s}'\cdot u_{\rm s}\neq0$ and $f_{\rm s}'\cdot B_{\rm s}\neq0$.
Then a gauge shift $f_{\rm s}'\to f_{\rm s}=f_{\rm s}'+cp_{\rm s}$ with $c=-(f_{\rm s}'\cdot B_{\rm s})/(p_{\rm s}\cdot B_{\rm s})$ results in a physically equivalent polarization vector that obeys $f_{\rm s}\cdot B_{\rm s}=0$.
Thus, one can always make the initial polarization perpendicular to the magnetic field by applying a gauge transformation that leaves the observed polarization invariant.
The gauge-fixing condition in Eq.~\eqref{eq:PolarizationConditions} is therefore essential.}

Together, the three conditions \eqref{eq:PolarizationConditions} determine three out of four components $f_{\rm s}^\mu$ of the source polarization, which suffices to fix its spacetime orientation and hence the observed EVPA.
The remaining component essentially controls the magnitude of the observed polarization, which in our model is not fixed at the source, but rather at the observer where we peg it to the observed intensity $I_{\rm o}$.
Mathematically, this prescription fixes the Penrose-Walker constant \eqref{eq:PenroseWalker} up to an overall scale factor that drops out of the formula \eqref{eq:EVPA} for the EVPA $\chi$.
More precisely, solving \eqref{eq:PolarizationConditions} yields, in terms of $b^{\mu\nu}=B_{\rm s}^\mu p_{\rm s}^\nu-B_{\rm s}^\nu p_{\rm s}^\mu$,
\begin{align}
	f_r&=\frac{b^{\theta\phi}-b^{t\theta}\Omega}{b^{\theta\phi}\iota+b^{r\phi}\Omega}f_t,\\
	f_\theta&=\frac{b^{r\phi}+b^{t\phi}\iota+b^{tr}\Omega}{b^{\theta\phi}\iota+b^{r\phi}\Omega}f_t,\\
	f_\phi&=-\frac{b^{r\theta}+b^{t\theta}\iota}{b^{\theta\phi}\iota+b^{r\phi}\Omega}f_t,
\end{align}
where, for clarity, we have temporarily dropped the subscripts `s'.
Lowering indices in Eq.~\eqref{eq:PenroseWalker} and evaluating it at $\theta=\pi/2$ then gives the source Penrose-Walker constant $\kappa_{\rm s}=\kappa_1+i\kappa_2$:
\begin{align}
	r\kappa_1&=a\pa{p_rf_\phi-p_\phi f_r}-\pa{r^2+a^2}\pa{p_tf_r-f_tp_r},\\
	r\kappa_2&=\pa{p_\theta f_\phi-p_\phi f_\theta}-a\pa{p_tf_\theta-p_\theta f_t}.
\end{align}

The computation of the photon momentum $p_{\rm s}$ at the source is described in Sec.~\ref{subsec:SourceMomentum}, while the parameters $(\iota,\Omega)$ of the flow $\mathring{u}$ are given in App.~\ref{subapp:KeplerianFlow}.
The component $f_t$ that controls the magnitude of the polarization remains undetermined, but it factors out of both $\kappa_1$ and $\kappa_2$, and therefore drops out of the ratio in the formula \eqref{eq:EVPA} for the EVPA $\chi$, as mentioned above.
The resulting explicit expressions are implemented in \texttt{AART}.

All that is left to specify are the spacetime components $B_{\rm s}^\mu$ of the local magnetic field.
The decomposition of the electromagnetic field strength $F_{\mu\nu}$ into electric and magnetic fields is frame-dependent.
In a realistic synchrotron emission model, the polarization would be perpendicular to the local magnetic field in the emitter frame, with components $B_{\rm s}^\mu=-{\star F}^{\mu\nu}u_\mu$.

For our example, we simply choose the spacetime field $B_{\rm s}$ to be purely radial, keeping only a nonzero $B_{\rm s}^r$ component whose magnitude also scales out of the EVPA $\chi$.
We display the resulting $n=0$ images, ray traced using both the exact and approximate $r_{\rm s}^{(0)}$, in the left and right panels of Fig.~\ref{fig:BeloborodovComparison}, respectively.
Even though the corresponding isoradial curves (plotted in the right column of Fig.~\ref{fig:BeloborodovApproximation}) differ noticeably, the ray-traced images in Fig.~\ref{fig:BeloborodovComparison} nonetheless look very similar and would be indistinguishable with near-term observations.
The Beloborodov approximation is thus excellent for ray tracing direct images of axisymmetric source configurations.

Having showcased the polarimetric capabilities of \texttt{AART}, we now leave polarization aside and turn to variable sources.

\subsection{Modeling variable sources with Gaussian random fields}
\label{subsec:GRF}

Variable (non-stationary and non-axisymmetric) sources can broadly be divided into two classes.
On the one hand, one can consider specific physical phenomena that are governed by deterministic equations.
A commonly studied example is that of a ``hot spot'': a localized source of enhanced emissivity that usually orbits around the black hole and thereby produces a characteristic pattern of light echoes \citep[see, e.g.,][]{Cunningham1973,Broderick2005,Broderick2006,Hamaus2009,Zamaninasab2010,Gralla2018}.
Adding such a source to our model is straightforward in \texttt{AART}: one merely needs to insert the deterministic source to the RHS of Eq.~\eqref{eq:RadialProfile} before carrying out the usual ray tracing.

On the other hand, one can also consider sources subject to statistical fluctuations.
Hot spots are transient and unlikely to be present in any given observation of M87*, but we certainly expect its surrounding plasma to flare and produce emission ropes (or other photon ring mimickers), which in the absence of a definite physical model we can still represent as random fluctuations in the astrophysical source.
A widely used tool to model such astrophysical noise is the Gaussian random field (GRF); see, e.g., Ref.~\cite{Bardeen1986} for applications to cosmology.
A random field $\mathcal{G}$ is a function on a space $X$ such that $\mathcal{G}(\mathbf{x})$ is a random variable for every $\mathbf{x}\in X$.
A GRF is a random field that is completely determined by its mean $\mu$ and covariance $C$,
\begin{align}
	\mu(\mathbf{x})=\av{\mathcal{G}(\mathbf{x})},\quad
	C(\Delta\mathbf{x})=\av{\mathcal{G}(\mathbf{x})\mathcal{G}(\mathbf{x}+\Delta\mathbf{x})}
	\ge0.
\end{align}
More precisely, we define a random field $\mathcal{G}(\mathbf{x})$ with mean $\mu(\mathbf{x})$ and covariance $C(\Delta\mathbf{x})$ to be a GRF if it satisfies the property
\begin{align}
	\label{eq:GRF}
    \av{e^{i\sum_{\ell=1}^ks_\ell\mathcal{G}(\mathbf{x}_\ell)}}=e^{-\frac{1}{2}\sum_{\ell,j=1}^kC(\Delta\mathbf{x}_{\ell j})s_\ell s_j+i\sum_{\ell=1}^k\mu(\mathbf{x}_\ell)s_\ell},
\end{align}
which states that its joint probability distribution on any set of $k$ points $\mathbf{x}_\ell$, with $\ell\in\cu{1,\ldots,k}$, is a $k$-dimensional multivariate Gaussian distribution with mean vector $\boldsymbol{\mu}=\cu{\mu_1,\ldots,\mu_k}$ and covariance matrix $\mathbf{C}_{\ell j}=C(\Delta\mathbf{x}_{\ell j})$, where $\Delta\mathbf{x}_{\ell j}=\ab{\mathbf{x}_\ell-\mathbf{x}_j}$.

The following discussion is technical and readers interested only in applications may skip down to Sec.~\ref{subsec:VariableProfiles} below.

The mean and covariance are also known as the one-point and two-point (or autocorrelation) function, respectively.
As a result of the defining relation \eqref{eq:GRF}, these low-point correlation functions determine all higher $k$-point functions of a GRF via Wick's theorem (also known as the Isserlis theorem), which may be derived by differentiation with respect to multiple $s_\ell$.

For instance, letting $\mathcal{G}_\ell$ denote $\mathcal{G}(\mathbf{x}_\ell)$ for a zero-mean GRF, it is immediately seen by differentiating Eq.~\eqref{eq:GRF} with respect to $s_1$, $s_2$, and $s_3$ that $\mathcal{G}(\mathbf{x})$ must have a vanishing three-point function, $\av{\mathcal{G}_1\mathcal{G}_2\mathcal{G}_3}=0$.
An additional derivative with respect to $s_4$ implies that the four-point function $\av{\mathcal{G}_1\mathcal{G}_2\mathcal{G}_3\mathcal{G}_4}$ is
\begin{align}
	\label{eq:FourPoints}
	\av{\mathcal{G}_1\mathcal{G}_2}\av{\mathcal{G}_3\mathcal{G}_4}+\av{\mathcal{G}_1\mathcal{G}_3}\av{\mathcal{G}_2\mathcal{G}_4}+\av{\mathcal{G}_1\mathcal{G}_4}\av{\mathcal{G}_2\mathcal{G}_3}.
\end{align}
In general, for a zero-mean GRF, all $k$-point functions vanish for $k$ odd, whereas for $k$ even they are given by all the possible ``contractions'' of the two-point functions $\av{\mathcal{G}_\ell\mathcal{G}_j}=C(\Delta\mathbf{x}_{\ell j})$.
This is Wick's theorem in a nutshell, and it makes precise the idea that a GRF is a ``simple'' random field whose correlation structure is fully determined by two functions $\mu(\mathbf{x})$ and $C(\Delta\mathbf{x})$.

This property is very convenient for our modeling purposes, since only two functions need to be specified to prescribe all of the statistics of our astrophysical source.
For example, Wick's theorem also fixes all the even moments of a zero-mean GRF:
\begin{align}
	\label{eq:Moments}
	\av{\mathcal{G}^{2k}(\mathbf{x})}=(2k-1)!!\br{C(0)}^k,\quad
	\av{\mathcal{G}^{2k+1}(\mathbf{x})}=0,
\end{align}
where the $k=2$ case is compatible with Eq.~\eqref{eq:FourPoints} evaluated at coincident points.
(Of course, the odd moments all vanish.)

We now specialize to GRFs in $d$-dimensional Euclidean space $X=\mathbb{R}^d$.
A GRF is homogeneous if it is invariant under translations, so $\mathcal{G}(\mathbf{x}+\mathbf{a})=\mathcal{G}(\mathbf{x})$ for $\mathbf{a}\in\mathbb{R}^d$, and it is isotropic if it looks the same in all directions, so $\mathcal{G}(\mathbf{x})=\mathcal{G}(x)$ with $x=\ab{\mathbf{x}}$.

A real-space GRF $\mathcal{G}(\mathbf{x})$ defines a (generally complex) GRF $\tilde{\mathcal{G}}(\mathbf{k})$ in momentum space via the Fourier transform
\begin{align}
	\label{eq:FourierConventions}
	\tilde{\mathcal{G}}(\mathbf{k})&=\int\mathcal{G}(\mathbf{x})e^{-i\mathbf{k}\cdot\mathbf{x}}\ed^d\mathbf{x},\quad
	\mathcal{G}(\mathbf{x})=\int\frac{\tilde{\mathcal{G}}(\mathbf{k})}{(2\pi)^d}e^{i\mathbf{k}\cdot\mathbf{x}}\ed^d\mathbf{k}.
\end{align}
If $\mathcal{G}(\mathbf{x})$ is homogeneous, then we let $\mathbf{x}'=\mathbf{x}+\Delta\mathbf{x}$ and Fourier transform its autocorrelation to derive the covariance of $\tilde{\mathcal{G}}(\mathbf{k})$:
\begin{align}
	\av{\tilde{\mathcal{G}}^*(\mathbf{k})\tilde{\mathcal{G}}\pa{\mathbf{k}'}}&=\iint\av{\mathcal{G}(\mathbf{x})\mathcal{G}\pa{\mathbf{x}'}}e^{i\mathbf{k}\cdot\mathbf{x}-i\mathbf{k}'\cdot\mathbf{x}'}\ed^d\mathbf{x}\ed^d\mathbf{x}'\\
	&=\iint C(\Delta\mathbf{x})e^{i\pa{\mathbf{k}-\mathbf{k}'}\cdot\mathbf{x}}e^{-i\mathbf{k}'\cdot\Delta\mathbf{x}}\ed^d\mathbf{x}\ed^d\Delta\mathbf{x}\\
	&=\int e^{i\pa{\mathbf{k}-\mathbf{k}'}\cdot\mathbf{x}}\ed^d\mathbf{x}\int C(\Delta\mathbf{x})e^{-i\mathbf{k}'\cdot\Delta\mathbf{x}}\ed^d\Delta\mathbf{x}\\
	\label{eq:MomentumCovariance}
	&=(2\pi)^d\delta^{(d)}\pa{\mathbf{k}-\mathbf{k}'}\tilde{C}\pa{\mathbf{k}'}.
\end{align}
Similarly, higher-point functions of $\tilde{\mathcal{G}}(\mathbf{k})$ may also be obtained by Fourier transforming those of $\mathcal{G}(\mathbf{x})$.
One finds that $\tilde{\mathcal{G}}(\mathbf{k})$ is a GRF obeying Eq.~\eqref{eq:GRF} with mean 
$\hat{\mu}(\mathbf{k})$ and covariance \eqref{eq:MomentumCovariance}, which is also referred to as the power spectrum $P(\mathbf{k})$ of $\mathcal{G}(\mathbf{x})$.
When $\mathcal{G}(\mathbf{x})$ is also isotropic, so is $P(\mathbf{k})=P(k)$ with $k=\ab{\mathbf{k}}$.
Like covariances, power spectra are always non-negative.

The most ubiquitous GRF is the white noise process $\mathcal{W}(\mathbf{x})$.
It is the standard, homogeneous and isotropic GRF defined by a flat power spectrum with uniform $\tilde{C}\pa{\mathbf{k}'}=1$, so that
\begin{align}
	\av{\tilde{\mathcal{W}}^*(\mathbf{k})\tilde{\mathcal{W}}\pa{\mathbf{k}'}}&=(2\pi)^d\delta^{(d)}\pa{\mathbf{k}-\mathbf{k}'},\\
	\av{\mathcal{W}(\mathbf{x})\mathcal{W}\pa{\mathbf{x}'}}&=\delta^{(d)}\pa{\mathbf{x}-\mathbf{x}'}.
\end{align}
With Eq.~\eqref{eq:GRF}, this delta-function autocorrelation implies that $\mathcal{W}(\mathbf{x})$ consists of independent Gaussian random variables at every $\mathbf{x}\in\mathbb{R}^d$, and likewise in momentum space for $\tilde{\mathcal{W}}\pa{\mathbf{k}}$.
Thus, creating realizations of white noise is trivial: it suffices to draw from independent normal distributions at every point.

In fact, Eq.~\eqref{eq:MomentumCovariance} provides a straightforward way to produce realizations of any homogeneous GRF $\mathcal{G}(\mathbf{x})$: first, one creates a realization $\tilde{\mathcal{W}}\pa{\mathbf{k}}$ of white noise; next, one multiplies it by $\sqrt{\tilde{C}(\mathbf{k})}$ to obtain a realization of $\hat{\mathcal{G}}(\mathbf{k})=\sqrt{\tilde{C}(\mathbf{k})}\tilde{\mathcal{W}}\pa{\mathbf{k}}$; finally, inverse Fourier transforming $\hat{\mathcal{G}}(\mathbf{k})$ yields a realization of $\mathcal{G}(\mathbf{x})$.
(We note that this procedure is well-defined since $\tilde{C}(\mathbf{k})\ge0$.)

Just as a standard Gaussian distribution has zero mean and unit variance, we define a standard GRF as a zero-mean GRF with unit covariance at the origin: $\mu(\mathbf{x})=0$ and $C(0)=1$.

We now consider the Mat\'ern field $\mathcal{F}_\nu(\mathbf{x})$ of order $\nu$, which is another standard, homogeneous, and isotropic GRF that is well-known to statisticians, who use it to model a wide range of processes; see, e.g., Ref.~\cite{Guttorp2006} for its various applications.
This zero-mean GRF obeys the Mat\'ern covariance of order $\nu$,
\begin{align}
	\label{eq:MaternCovariance}
	C_\nu(\mathbf{x})=\frac{1}{2^{\nu-1}\Gamma(\nu)}\pa{\frac{x}{\lambda}}^\nu K_\nu\pa{\frac{x}{\lambda}},
\end{align}
where $\Gamma$ denotes the Gamma function and $K_\nu(x)$ the modified Bessel function of the second kind  of order $\nu$.
Here, $\lambda$ is a correlation length, while $\nu$ is a differentiability parameter that we will always take to be $\nu=n-d/2$ for some positive integer $n$.
At short distances, $C_\nu(\mathbf{x})\stackrel{x\ll\lambda}{\approx}1+c_\nu x^{2\nu}$ for some constant $c_\nu$, so it is best (though not strictly necessary) to require $\nu>0$.

The Mat\'ern field $\mathcal{F}_\nu(\mathbf{x})$ is also ubiquitous in physics:
though its position-space covariance \eqref{eq:MaternCovariance} may seem unfamiliar, in momentum space it becomes\footnote{Equivalently, $\tilde{C}_\nu(\mathbf{k})=\frac{\sigma^{-2}}{\pa{k^2+\lambda^{-2}}^{\nu+\frac{d}{2}}}$ with $
\sigma^2=\frac{\lambda^{2\nu}}{\pa{4\pi}^\frac{d}{2}}\frac{\Gamma(\nu)}{\Gamma\pa{\nu+\frac{d}{2}}}$ as in Ref.~\cite{Lindgren2011}.} (see App.~\ref{app:MaternCovariance} for a derivation)
\begin{align}
	\label{eq:MaternMomentumCovariance}
	\tilde{C}_\nu(\mathbf{k})=\frac{\mathcal{N}^2\lambda^d}{\pa{1+\lambda^2k^2}^{\nu+\frac{d}{2}}},\quad
	\mathcal{N}^2=\pa{4\pi}^\frac{d}{2}\frac{\Gamma\pa{\nu+\frac{d}{2}}}{\Gamma(\nu)},
\end{align}
which for $n=1$ (or $\nu=1-d/2$), we recognize as the quantum propagator for a free scalar field of mass $m=\lambda^{-1}$ in Euclidean signature.
To the best of our knowledge, this link has not been stated explicitly before, and we explore it in detail in App.~\ref{app:MaternCovariance}.

The Mat\'ern field $\mathcal{F}_\nu(\mathbf{x})$ enjoys a special connection to linear stochastic partial differential equations (SPDE).
Realizations of its Fourier transform $\tilde{\mathcal{F}}_\nu(\mathbf{k})$ are related to white noise via
\begin{align}
	\label{eq:FourierTrick}
	\tilde{\mathcal{F}}_\nu(\mathbf{k})=\sqrt{\tilde{C}_\nu(\mathbf{k})}\tilde{\mathcal{W}}(\mathbf{k}).
\end{align}
Such a relation holds for any homogeneous GRF, but it takes a particularly simple form for the Mat\'ern covariance \eqref{eq:MaternMomentumCovariance}:
\begin{align}
	\pa{1+\lambda^2k^2}^{\frac{\nu}{2}+\frac{d}{4}}\tilde{\mathcal{F}}_\nu(\mathbf{k})=\mathcal{N}\lambda^\frac{d}{2}\tilde{\mathcal{W}}(\mathbf{k}).
\end{align}
Returning to position space gives the linear (fractional) SPDE
\begin{align}
	\label{eq:FractionalSPDE}
	\pa{1-\lambda^2\nabla^2}^{\frac{\nu}{2}+\frac{d}{4}}\mathcal{F}_\nu(\mathbf{x})=\mathcal{N}\lambda^\frac{d}{2}\mathcal{W}(\mathbf{x}),
\end{align}
which exactly matches Eq.~(2) of Ref.~\cite{Lindgren2011} since $\mathcal{N}^2\sigma^2=\lambda^{2\nu}$.
As we show in App.~\ref{app:MaternCovariance}, for $n=1$ (or $\nu=1-d/2$), this SPDE is compatible with the identification $\mathcal{F}_\nu(\mathbf{x})\equiv\Phi(\mathbf{x})$, where $\Phi(\mathbf{x})$ is a free Euclidean scalar field of mass $m=\lambda^{-1}$, as expected.

If instead, $n=2$ (or $\nu=2-d/2$), then the resulting Mat\'ern field, which we will simply denote $\mathcal{F}(\mathbf{x})$, obeys a linear SPDE
\begin{align}
	\pa{1-\lambda^2\nabla^2}\mathcal{F}(\mathbf{x})=\mathcal{N}\lambda^\frac{d}{2}\mathcal{W}(\mathbf{x}),
\end{align}
which has no fractional derivative and is thus simpler to solve numerically, as explained in Ref.~\cite{Lee2021} below Eq.~(3) therein.

This connection between the Mat\'ern field $\mathcal{F}(\mathbf{x})$ and a linear SPDE is especially useful once we introduce inhomogeneities.
Inhomogeneous GRFs are hard to generate via other means than solving the associated SPDE; in particular, the trick \eqref{eq:FourierTrick} for generating realizations of $\tilde{\mathcal{F}}(\mathbf{k})$ breaks down because the Fourier modes are no longer delta-correlated as in Eq.~\eqref{eq:MomentumCovariance}.

\subsection{Inhomogeneous and anisotropic Mat\'ern fields}

Before describing how to include inhomogeneity, we first discuss how to model homogeneous anisotropies, following closely the excellent treatment in Ref.~\cite{Lee2021}.

Still in $d$ dimensions, we introduce $d$ orthonormal vectors $\mathbf{u}_\ell$, $d$ correlation lengths $\lambda_\ell>0$, with $\ell\in\cu{1,\ldots,d}$, and define
\begin{align}
	\label{eq:Metric}
	\mathbf{\Lambda}=\sum_{\ell=1}^d\lambda_\ell^2\mathbf{u}_\ell\mathbf{u}_\ell^{\rm T},\quad
	\ab{\mathbf{\Lambda}}\equiv\det{\mathbf{\Lambda}}
	=\prod_{\ell=1}^d\lambda_\ell^2.
\end{align}
We will also allow the unit vectors $\mathbf{u}_\ell$ to not be orthogonal, as long as $\ab{\mathbf{\Lambda}}$ retains the same form.
The matrix $\mathbf{\Lambda}$ is invertible,
and its inverse $\mathbf{\Lambda}^{-1}$ defines a metric on $\mathbb{R}^d$ with line element
\begin{align}
	\ed s^2(\Delta\mathbf{x})=\Delta\mathbf{x}\cdot\mathbf{\Lambda}^{-1}\Delta\mathbf{x}
	=\sum_{\ell=1}^d\pa{\frac{\Delta\mathbf{x}\cdot\mathbf{u}_\ell}{\lambda_\ell}}^2.
\end{align}
We now use this to define the generalized Mat\'ern covariance
\begin{align}
	\label{eq:GeneralizedMaternCovariance}
	C_\nu(\mathbf{x})=\frac{1}{2^{\nu-1}\Gamma(\nu)}\ed s^\nu(\mathbf{x})K_\nu\pa{\ed s(\mathbf{x})}.
\end{align}
When $\lambda_1=\ldots=\lambda_n=\lambda$, so that $\mathbf{\Lambda}=\lambda^2\mathbf{I}$, this reproduces the homogeneous, isotropic Mat\'ern covariance \eqref{eq:MaternCovariance}, as desired.

However, if the constants $\lambda_\ell$ are unequal, then the resulting GRF remains homogeneous but becomes anisotropic along a direction set by the unit vectors $\mathbf{u}_\ell$. An example with $d=2$ is displayed in Fig.~1 of Ref.~\cite{Lee2021}.
Following the steps that led to Eq.~\eqref{eq:MomentumCovariance}, the corresponding momentum-space covariance is
\begin{align}
	\label{eq:GeneralizedMomentumCovariance}
	\tilde{C}_\nu(\mathbf{k})=\frac{\mathcal{N}^2\sqrt{\ab{\mathbf{\Lambda}}}}{\pa{1+\mathbf{k}\cdot\mathbf{\Lambda}\mathbf{k}}^{\nu+\frac{d}{2}}}.\end{align}
In this case, we can still use Eq.~\eqref{eq:FourierTrick} to obtain realizations of the anisotropic Mat\'ern field.
We can also follow the derivation of Eq.~\eqref{eq:FractionalSPDE} to find that the anisotropic field obeys the SPDE
\begin{align}
	\label{eq:SPDE}
	\pa{1-\mathbf{\nabla}\cdot\mathbf{\Lambda}\mathbf{\nabla}}^{\frac{\nu}{2}+\frac{d}{4}}\mathcal{F}_\nu(\mathbf{x})=\mathcal{N}\ab{\mathbf{\Lambda}}^\frac{1}{4}\mathcal{W}(\mathbf{x}).
\end{align}

Finally, to include inhomogeneity, we allow the unit vectors $\mathbf{u}_\ell\to\mathbf{u}_\ell(\mathbf{x})$ to become functions.
Then $\mathbf{\Lambda}\to\mathbf{\Lambda}(\mathbf{x})$ becomes a function too---with the same definition \eqref{eq:Metric}---but the rest of the previous discussion breaks down.
Instead, we must reverse the logic and \textit{define} the inhomogeneous, anisotropic Mat\'ern field $\mathcal{F}_\nu(\mathbf{x})$ by the SPDE \eqref{eq:SPDE} with variable coefficients $\mathbf{\Lambda}(\mathbf{x})$.

We emphasize that the resulting field $\mathcal{F}_\nu(\mathbf{x})$ no longer obeys the covariance \eqref{eq:GeneralizedMomentumCovariance}, which is not a function on momentum space once $\mathbf{\Lambda}(\mathbf{x})$ is position-dependent, but it is a well-defined GRF arising as a solution to the SPDE \eqref{eq:SPDE}; see also Ref.~\cite{Fuglstad2013}.

As explained below Eq.~\eqref{eq:FractionalSPDE}, in practice, we will only use the field $\mathcal{F}(\mathbf{x})$ with $n=2$ (or $\nu=2-d/2$) as it obeys the SPDE
\begin{align}
	\label{eq:inoisy}
	\br{1-\mathbf{\nabla}\cdot\mathbf{\Lambda}(\mathbf{x})\mathbf{\nabla}}\mathcal{F}(\mathbf{x})=\mathcal{N}\ab{\mathbf{\Lambda}(\mathbf{x})}^\frac{1}{4}\mathcal{W}(\mathbf{x}),
\end{align}
which has no fractional derivatives and is therefore more tractable numerically.
Since we took our white noise to have unit variance, this SPDE exactly matches Eq.~(5) of Ref.~\cite{Lee2021} (after setting $\sigma=1$ therein), which is what \texttt{inoisy} solves.

One last comment is in order: in the inhomogeneous case, the Mat\'ern field $\mathcal{F}(\mathbf{x})$ that \texttt{inoisy} produces by numerically integrating Eq.~\eqref{eq:inoisy} does not follow the covariance \eqref{eq:GeneralizedMaternCovariance}; in particular, it may not be standard with $\mu(\mathbf{x})=0$ and $C(0)=1$.
To ``standardize'' it, we will always work with a rescaled field
\begin{align}
	\label{eq:MaternField}
	\hat{\mathcal{F}}(\mathbf{x})=\frac{\mathcal{F}(\mathbf{x})-\av{\mathcal{F}(\mathbf{x})}}{\sqrt{\av{\mathcal{F}^2(\mathbf{x})}-\av{\mathcal{F}(\mathbf{x})}^2}},
\end{align}
which also happens to be precisely what \texttt{inoisy} implements.
For a homogeneous field, this makes no statistical difference.

\subsection{Non-stationary and non-axisymmetric source profiles}

With the inhomogenous, anisotropic Mat\'ern field $\hat{\mathcal{F}}(\mathbf{x})$ in hand, we are now in a position to model the fluctuations of an equatorial accretion disk and produce \texttt{inoisy} simulations.

Before doing so, we first describe what we want to achieve intuitively, without being too rigorous.
We let $\mathbf{x}_{\rm s}=(r_{\rm s},\phi_{\rm s},t_{\rm s})$.
As in Ref.~\citep{Hadar2021}, we want the total intensity \eqref{eq:RadialProfile} to consist of a background radial profile $J(r_{\rm s})$ with some fluctuations $\Delta J(\mathbf{x}_{\rm s})$:
\begin{align}
	\label{eq:IntuitiveDefinition}
	I_{\rm s}(\mathbf{x}_{\rm s})=J(r_{\rm s})+\Delta J(\mathbf{x}_{\rm s})
	=J(r_{\rm s})\br{1+F(\mathbf{x}_{\rm s})}.
\end{align}
Here, $F\equiv\Delta J/J$ is the fractional variation in the source, which at first we assume to be fractionally small: $0<\ab{F(\mathbf{x}_{\rm s})}\ll1$.
We also want these fluctuations to wash out under averaging,
\begin{align}
	\label{eq:FluctuationAverage}
	\av{F(\mathbf{x}_{\rm s})}=0,
\end{align}
so that after averaging over fluctuations using $\av{J(r_{\rm s})}=J(r_{\rm s})$, we recover the stationary, axisymmetric profile \eqref{eq:RadialProfile}:
\begin{align}
	\label{eq:DiskFluctuations}
	\av{I_{\rm s}(\mathbf{x}_{\rm s})}=\av{J(r_{\rm s})}+\av{J(r_{\rm s})F(\mathbf{x}_{\rm s})}
	=J(r_{\rm s}).
\end{align}

Eventually, we will also want to consider large fluctuations with $\Delta J\sim J$, or $F\sim\bigO(1)$, but this could pose a problem: if at some point $\mathbf{x}_{\rm s}$, a fluctuation grew so large and negative that $F(\mathbf{x}_{\rm s})<-1$, then we would have $\Delta J(\mathbf{x}_{\rm s})<-J(r_{\rm s})$, and hence $I_{\rm s}(\mathbf{x}_{\rm s})<0$ at that point, which is nonsensical because an intensity is a (necessarily positive) count of photons---we must therefore prevent this from happening.

Of course, we could demand that the fluctuations always remain small, but there is a way to allow them to grow large while still ensuring they do not result in a negative intensity.
The idea is to note that for small fluctuations $0<\ab{F(\mathbf{x}_{\rm s})}\ll1$, $1+F\approx 1+F+\frac{1}{2!}F^2+\frac{1}{3!}F^3+\ldots=e^F$, so we may well replace the RHS of Eq.~\eqref{eq:IntuitiveDefinition} by a (necessarily positive) exponential:
\begin{align}
	\label{eq:MoralDefinition}
	I_{\rm s}(\mathbf{x}_{\rm s})\cong J(r_{\rm s})e^{F(\mathbf{x}_{\rm s})}.
\end{align}
For small fluctuations, this seems completely equivalent, since
\begin{align}
	\frac{\Delta J(\mathbf{x}_{\rm s})}{J(r_{\rm s})}=\frac{I_{\rm s}(\mathbf{x}_{\rm s})-J(r_{\rm s})}{J(r_{\rm s})}
	\cong e^{F(\mathbf{x}_{\rm s})}-1
	\stackrel{\ab{F}\ll1}{\approx}F(\mathbf{x}_{\rm s}),
\end{align}
reproducing Eq.~\eqref{eq:IntuitiveDefinition}.
The advantage of this definition is that it can now be freely extended to arbitrarily large $\ab{F}$, even in the negative direction: a fluctuation $F(\mathbf{x}_{\rm s})\ll-1$ would just result in a small but still positive (and hence still physical) intensity, rather than an unphysical, negative emissivity.
Hence, our new definition \eqref{eq:MoralDefinition} for $I_{\rm s}(\mathbf{x}_{\rm s})$ remains sensible even for large and negative field excursions, unlike our first attempt \eqref{eq:IntuitiveDefinition}.

However, as we are about to see, this new definition suffers from a subtle issue: it does not actually satisfy Eq.~\eqref{eq:DiskFluctuations}. For now, we use `$\cong$' to remind us of this, and proceed naively.

The preceding intuitive discussion hinges on being able to produce a fluctuation field $F(\mathbf{x}_{\rm s})$ with some desired properties, such as Eq.~\eqref{eq:FluctuationAverage}.
We now make this mathematically precise.

We replace $F(\mathbf{x}_{\rm s})\to\sigma\hat{\mathcal{F}}(\mathbf{x}_{\rm s})$, where $\hat{\mathcal{F}}(\mathbf{x}_{\rm s})$ is the standard, zero-mean Mat\'ern field defined in Eq.~\eqref{eq:MaternField} as the solution to the SPDE \eqref{eq:SPDE}.
We then define an emissivity field \cite{Lee2021}
\begin{align}
	I_{\rm s}(\mathbf{x}_{\rm s})\equiv\mathcal{J}(\mathbf{x}_{\rm s})
	\cong J(r_{\rm s})e^{\sigma\hat{\mathcal{F}}(\mathbf{x}_{\rm s})},
\end{align}
where the parameter $\sigma$ controls the scale of the fluctuations, since $\hat{\mathcal{F}}(\mathbf{x}_{\rm s})$ has unit covariance $C(0)=1$.
Because $\hat{\mathcal{F}}(\mathbf{x}_{\rm s})$ has zero mean, Eq.~\eqref{eq:FluctuationAverage} is satisfied.
Let us now check Eq.~\eqref{eq:DiskFluctuations}:
\begin{align}
	\av{\mathcal{J}(\mathbf{x}_{\rm s})}\cong J(r_{\rm s})\av{e^{\sigma\hat{\mathcal{F}}(\mathbf{x}_{\rm s})}}
	=J(r_{\rm s})\sum_{n=0}^\infty\frac{\big\langle\sigma^n\hat{\mathcal{F}}^n(\mathbf{x}_{\rm s})\big\rangle}{n!},
\end{align}
where we series-expanded the exponential and used linearity of the expectation value.
Then by Eq.~\eqref{eq:Moments}, relabeling $k=2n$,
\begin{align}
	\frac{\av{\mathcal{J}(\mathbf{x}_{\rm s})}}{J(r_{\rm s})}\cong\sum_{k=0}^\infty\frac{\sigma^{2k}\big\langle\hat{\mathcal{F}}^{2k}(\mathbf{x}_{\rm s})\big\rangle}{
(2k)!}
	=\sum_{k=0}^\infty\frac{(2k-1)!!}{
(2k)!}\sigma^{2k}.
\end{align}
Evaluating the sum results in
\begin{align}
	\av{\mathcal{J}(\mathbf{x}_{\rm s})}\cong J(r_{\rm s})e^{\frac{1}{2}\sigma^2},
\end{align}
and we now see why the definition \eqref{eq:MoralDefinition} is not what we wanted,
since it does not satisfy the requirement \eqref{eq:DiskFluctuations}.
Instead, we let
\begin{align}
	\label{eq:VariableSource}
	I_{\rm s}(\mathbf{x}_{\rm s})=\mathcal{J}(\mathbf{x}_{\rm s})
	\equiv J(r_{\rm s})e^{\sigma\hat{\mathcal{F}}(\mathbf{x}_{\rm s})-\frac{1}{2}\sigma^2},
\end{align}
which does have the desired property \eqref{eq:DiskFluctuations}, and hence recovers the background profile \eqref{eq:RadialProfile} after averaging.
We note that this can also be checked directly from the property \eqref{eq:GRF} with $k=1$,
\begin{align}
	\av{e^{is\mathcal{G}(\mathbf{x})}}=e^{-\frac{1}{2}C(0)s^2},
\end{align}
by taking $s=-i\sigma$ and $\mathcal{G}=\hat{\mathcal{F}}$ with $\mu=0$ and $C(0)=1$.
This also shows that $\sigma^2$ effectively rescales the covariance $C(0)$.
In particular, the fluctuations disappear as $s\to0$ and we recover the pure envelope $J(r_{\rm s})$ of the background surface brightness.

\subsection{Complete statistics of the variable source}

We can also use the GRF property to explicitly write down the complete correlation structure of the random field $\mathcal{J}(\mathbf{x}_{\rm s})$.
With $s_\ell=-i\sigma$, $\mathcal{G}=\hat{\mathcal{F}}$, $\mu=0$ and $C(0)=1$, Eq.~\eqref{eq:GRF} becomes
\begin{align}
    \av{e^{\sigma\sum_{\ell=1}^k\hat{\mathcal{F}}(\mathbf{x}_\ell)}}=e^{\frac{1}{2}\sigma^2\sum_{\ell,j=1}^kC(\Delta\mathbf{x}_{\ell j})}.
\end{align}
Since $\av{\mathcal{J}(\mathbf{x}_1)\ldots\mathcal{J}(\mathbf{x}_k)}=J(r_1)\ldots J(r_k)\av{e^{\sigma\sum_{\ell=1}^k\hat{\mathcal{F}}(\mathbf{x}_\ell)-\frac{k}{2}\sigma^2}}$,
\begin{align}
	\label{eq:CorrelationStructure}
    \frac{\av{\mathcal{J}(\mathbf{x}_1)\ldots\mathcal{J}(\mathbf{x}_k)}}{J(r_1)\ldots J(r_k)}=\exp\br{-\frac{k\sigma^2}{2}+\frac{\sigma^2}{2}\sum_{\ell,j=1}^kC(\Delta\mathbf{x}_{\ell j})}.
\end{align}
This result characterizes all the statistics of the variable source $\mathcal{J}(\mathbf{x}_{\rm s})$ in terms of a single covariance function, as advertised.

Of course, this formula with $k=1$ confirms that Eq.~\eqref{eq:DiskFluctuations} holds.
Meanwhile, for $k=2$ the double sum in the exponential is $\sum_{\ell,j=1}^2C(\Delta\mathbf{x}_{\ell j})=C(\Delta\mathbf{x}_{11})+C(\Delta\mathbf{x}_{12})+C(\Delta\mathbf{x}_{21})+C(\Delta\mathbf{x}_{22})$ so we obtain the two-point function
\begin{align}
	\frac{\av{\mathcal{J}(\mathbf{x}_1)\mathcal{J}(\mathbf{x}_2)}}{J(r_1)J(r_2)}=e^{-\sigma^2+\frac{\sigma^2}{2}\br{2C(0)+2C(\Delta\mathbf{x}_{12})}}
	=e^{\sigma^2C(\Delta\mathbf{x}_{12})},
\end{align}
which is the exponential of the Mat\'ern field's autocorrelation.
Going further, for $k=3$ the double sum in the exponential is $\sum_{\ell,j=1}^2C(\Delta\mathbf{x}_{\ell j})=3C(0)+2C(\Delta\mathbf{x}_{12})+2C(\Delta\mathbf{x}_{13})+2C(\Delta\mathbf{x}_{23})$ so
\begin{align}
	\frac{\av{\mathcal{J}(\mathbf{x}_1)\mathcal{J}(\mathbf{x}_2)\mathcal{J}(\mathbf{x}_3)}}{J(r_1)J(r_2)J(r_3)}
	=e^{\sigma^2\br{C(\Delta\mathbf{x}_{12})+C(\Delta\mathbf{x}_{13})+C(\Delta\mathbf{x}_{23})}},
\end{align}
and one can check that $\mathcal{J}(\mathbf{x}_{\rm s})$ is a non-Gaussian random field.
That is, it does not quite obey the GRF property \eqref{eq:GRF}, though its statistics \eqref{eq:CorrelationStructure} are still governed by a single covariance. 

For some intuition, let us reexamine the fractional variation $F\equiv\Delta J/J$, whose statistics are encoded in the random field
\begin{align}
	\hat{\mathcal{J}}(\mathbf{x}_{\rm s})\equiv\frac{\mathcal{J}(\mathbf{x}_{\rm s})-J(r_{\rm s})}{J(r_{\rm s})}
	=e^{\sigma\hat{\mathcal{F}}(\mathbf{x}_{\rm s})-\frac{1}{2}\sigma^2}-1.
\end{align}
Its one-point and two-point functions are
\begin{align}
	\av{\hat{\mathcal{J}}(\mathbf{x}_{\rm s})}=0,\quad
	\av{\hat{\mathcal{J}}(\mathbf{x}_{\rm s})\hat{\mathcal{J}}\pa{\mathbf{x}_{\rm s}'}}=e^{\sigma^2C(\Delta\mathbf{x})}-1.
\end{align}
If the fluctuations are very small, then this is approximately
\begin{align}
    \av{\hat{\mathcal{J}}(\mathbf{x}_{\rm s})\hat{\mathcal{J}}\pa{\mathbf{x}_{\rm s}'}}\stackrel{\sigma^2\ll1}{\approx}\sigma^2C(\Delta\mathbf{x}),
\end{align}
which is the autocorrelation of a GRF with covariance $\sigma^2$.
This is also true for higher-point functions, using Eq.~\eqref{eq:CorrelationStructure}, we can show that the field $\hat{\mathcal{J}}(\mathbf{x}_{\rm s})$ is approximately Gaussian at leading order in $\sigma$.
Thus, small fluctuations are Gaussian, but larger fluctuations introduce non-Gaussianities. 

\subsection{Generating realizations of an accretion flow using \texttt{inoisy}}
\label{subsec:VariableProfiles}

To summarize, we model variable sources via the emissivity \eqref{eq:VariableSource}, which is defined in terms of the Mat\'ern field \eqref{eq:MaternField} and displays the correlation structure \eqref{eq:CorrelationStructure}.
The zero-mean field $\hat{\mathcal{F}}(\mathbf{x}_{\rm s})$ and its statistics are fully determined by its covariance, which is controlled by a metric $\mathbf{\Lambda}(\mathbf{x_{\rm s}})$ of the form \eqref{eq:Metric}.
Given such a metric, a realization of $\hat{\mathcal{F}}(\mathbf{x}_{\rm s})$ is obtained by solving the linear SPDE \eqref{eq:inoisy} with some realization $\mathcal{W}(\mathbf{x}_{\rm s})$ of white noise.
This is essentially what \texttt{inoisy} does, though in practice, the continuous SPDE \eqref{eq:inoisy} is actually discretized and its solution is a Gaussian Markov random field approximating $\hat{\mathcal{F}}(\mathbf{x}_{\rm s})$ \cite{Lee2021}.
As for the envelope $J(r_{\rm s})$, we use the radial profile \eqref{eq:JonhnsonSU}.

Therefore, all we need to do to specify our model is to fix a choice of unit vectors $\mathbf{u}_\ell(\mathbf{x_{\rm s}})$ and correlation lengths $\lambda_\ell$, where now $\ell\in\cu{0,1,2}$.
Such a choice will define $\mathbf{\Lambda}(\mathbf{x_{\rm s}})$ and hence our non-stationary, non-axisymetric stochastic source profile.

Since we want to produce ``realistic'' images and visibility amplitudes, we will choose $\hat{\mathcal{F}}(\mathbf{x}_{\rm s})$ to be an inhomogeneous, anisotropic Mat\'ern field whose power spectrum is comparable to that observed in GRMHD simulations \cite{Lee2021}, and therefore serves as a good phenomenological model.

\begin{table}
	\begin{tabular}{|c@{\hspace*{10pt}}c@{\hspace*{10pt}}c|}
	\hline 
	Parameter & Default value & Description \\
	\hline\hline 
	$\psi$ & 1.07 & BH mass-to-distance ratio \\
	$a/M$ & 94$\%$ & BH spin \\
	$\theta_{\rm o}$ & $17^\circ$ & Observer inclination \\
	\hline\hline
	$\zeta$ & 1.5 & Geometrical factor \\
	$\sigma$ & 0.4 & Fluctuation scale \\
	$\xi$ & $0.95$ & Sub-Keplerian factor \\
	$\beta_r$ & $0.95$ & Radial velocity factor\\
	$\beta_\phi$ & $0.95$ & Angular velocity factor \\
	$\theta_\angle$ & $20^\circ$ & Anisotropy direction \\
	$\lambda_0$ & $2\pi/\Omega$ & Temporal correlation \\
	$\lambda_1$ & $5r_{\rm s}$ & Spatial correlation in the $x_{\rm s}$ direction \\
	$\lambda_2$ & $0.1\lambda_1$ & Spatial correlation in the $y_{\rm s}$ direction \\
	$\mu$ &$r_{-}$ & Controls location of profile peak \\
	$\vartheta$ & $1/2$ & Controls profile width \\
	$\gamma$ & $-3/2$ & Controls profile asymmetry \\
	\hline 
	\end{tabular}
	\caption{Summary of the fifteen parameters in our non-stationary and non-axisymmetric stochastic source model, together with their default values used for the majority of our examples.
	The first three pertain to the geometry, while the other twelve prescribe the statistics of the equatorial surface brightness.}
	\label{tbl:Parameters}
\end{table}

Instead of $\mathbf{x}_{\rm s}=(r_{\rm s},\phi_{\rm s},t_{\rm s})$, we use a regular Cartesian grid $\mathbf{x}_{\rm s}=(t_{\rm s},x_{\rm s},y_{\rm s})$ in the Kerr equatorial plane, and \texttt{inoisy} solves the SPDE \eqref{eq:inoisy} on this grid with periodic boundary conditions.
Following Eq.~\eqref{eq:Metric}, we pick a position-dependent anisotropy
\begin{align}
	\mathbf{\Lambda}(\mathbf{x}_{\rm s})=\sum_{\ell=0}^2\lambda_\ell^2(r_{\rm s})\mathbf{u}_\ell(\mathbf{x}_{\rm s})\mathbf{u}_\ell^{\rm T}(\mathbf{x}_{\rm s}).
\end{align}
Under this prescription, $\mathbf{u}_0(\mathbf{x}_{\rm s})$ sets the  temporal correlation of the flow, with characteristic correlation time $\lambda_0(r_{\rm s})$, while $\mathbf{u}_1(\mathbf{x}_{\rm s})$ and $\mathbf{u}_2(\mathbf{x}_{\rm s})$ determine its spatial structure, which at any given time exhibits correlations of characteristic lengths $\lambda_1(r_{\rm s})$ and $\lambda_2(r_{\rm s})$, respectively.
Following Ref.~\cite{Lee2021}, we take these 3D unit vectors to have $(t_{\rm s},x_{\rm s},y_{\rm s})$ components
\begin{align}
	\mathbf{u}_0(\mathbf{x}_{\rm s})&=\pa{1,v_x(\mathbf{x}_{\rm s}),v_y(\mathbf{x}_{\rm s})},\\
	\mathbf{u}_1(\mathbf{x}_{\rm s})&=\pa{0,\cos{\theta(\mathbf{x}_{\rm s})},\sin{\theta(\mathbf{x}_{\rm s})}},\\
	\mathbf{u}_2(\mathbf{x}_{\rm s})&=\pa{0,-\sin{\theta(\mathbf{x}_{\rm s})},\cos{\theta(\mathbf{x}_{\rm s})}},
\end{align}
where we have yet to specify $v_x$, $v_y$, and $\theta$.
In practice, we will only let these functions depend on the spatial position $(x_{\rm s},y_{\rm s})$.

We note that $\mathbf{u}_1(\mathbf{x}_{\rm s})\cdot\mathbf{u}_2(\mathbf{x}_{\rm s})=0$ are always orthogonal to each other, but not to $\mathbf{u}_0(\mathbf{x}_{\rm s})$.
Nonetheless, the resulting $\mathbf{\Lambda}(\mathbf{x}_{\rm s})$ still has a determinant of the form required by Eq.~\eqref{eq:Metric}, since $\ab{\mathbf{\Lambda}(\mathbf{x}_{\rm s})}=\lambda_0^2\lambda_1^2\lambda_2^2$.
Also, the sign of $\mathbf{u}_1$ and $\mathbf{u}_2$ is arbitrary, as it sets only the spatial correlation; the flow in time is set by $\mathbf{u}_0$. 

To be consistent with our choice of accretion flow \eqref{eq:GeneralFourVelocity}, we must take the temporal correlations to have the same velocity:
\begin{align}
    \vec{v}\equiv\frac{dr}{dt}\pd_r+\frac{d\phi}{dt}\pd_\phi
    =-\iota\pd_r+\Omega\pd_\phi,
\end{align}
where $\Omega$ and $\iota$ respectively denote the angular and radial-infall velocities \eqref{eq:OmegaAndIota}. In Cartesian coordinates this velocity becomes $\vec{v}=v_x\pd_x+v_y\pd_y$ with
\begin{align}
    v_x(\mathbf{x}_{\rm s})=-\frac{x_{\rm s}}{r_{\rm s}}\iota-y_{\rm s}\Omega,\quad
    v_y(\mathbf{x}_{\rm s})=-\frac{y_{\rm s}}{r_{\rm s}}\iota+x_{\rm s}\Omega.
\end{align}
As for the spatial correlations, we adopt the prescription \cite{Lee2021}
\begin{align}
    \theta(\mathbf{x}_{\rm s})=\arctan(y_{\rm s},-x_{\rm s})+\theta_\angle.
\end{align}
This sets the major axis $\mathbf{u}_1(\mathbf{x}_{\rm s})$ of the spatial correlation tensor to lie at a constant angle $\theta_\angle$ relative to the equatorial circles of constant radius $r_{\rm s}$.
The resulting flow displays spiral features with opening angle $\theta_\angle$.
The choice $\theta_\angle\approx20^\circ$ produces spiral arms broadly consistent with GRMHD simulations \cite{Guan2009,Lee2021}.

This completes the specification of our stochastic surface brightness.
To summarize, the accretion flow model takes as input twelve parameters, each of which is listed in Table~\ref{tbl:Parameters}, along with a short description and the default value used for the main examples in this paper.
To compute images, we must also specify three more parameters: the black hole spin $a$, the observer inclination $\theta_{\rm o}$, and the mass-to-distance ratio
\begin{align}
	\psi=\frac{1}{3.62\,\mu\rm{as}}\frac{M}{r_{\rm o}},
\end{align}
which combines the black hole mass $M$ and observer distance $r_{\rm o}$, and which is defined relative to a fiducial value for M87*.

\begin{figure*}
    \centering
    \includegraphics[width=\textwidth]{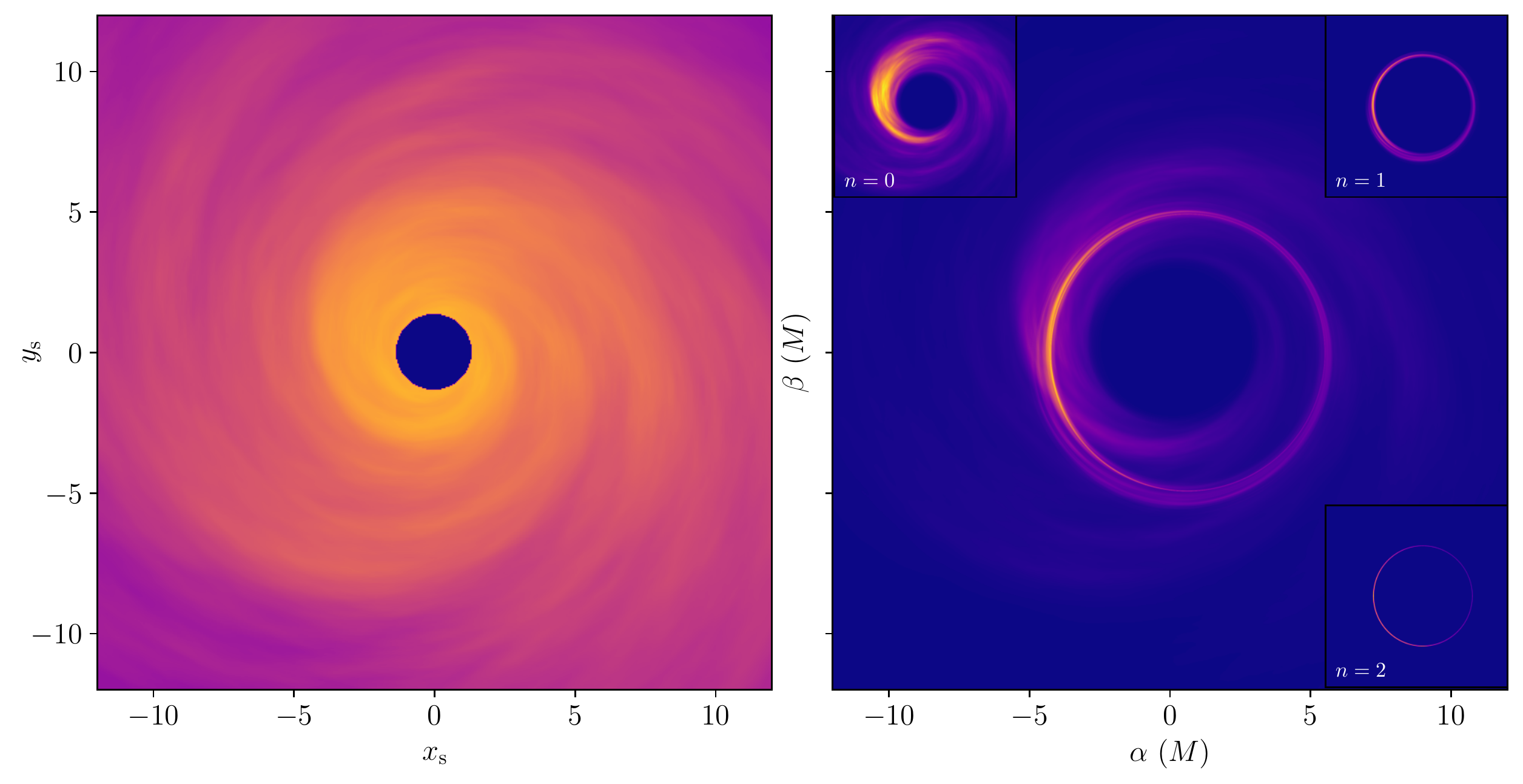}
    \caption{Left: Intensity profile (in logarithmic scale) of a single snapshot from an \texttt{inoisy} simulation with the parameters in Table.~\ref{tbl:Parameters}.
    Right: Ray-traced image corresponding to this equatorial source profile, treating the realization of the random field on the left as static (the ``fast-light'' approximation).
    The inset panels decompose the image into layers: the direct $n=0$ image and the first two ($n=1$ and $n=2$) photon rings.}
    \label{fig:inoisySnapshotM87}
\end{figure*}

\begin{figure}[b!]
    \centering
    \includegraphics[width=\linewidth]{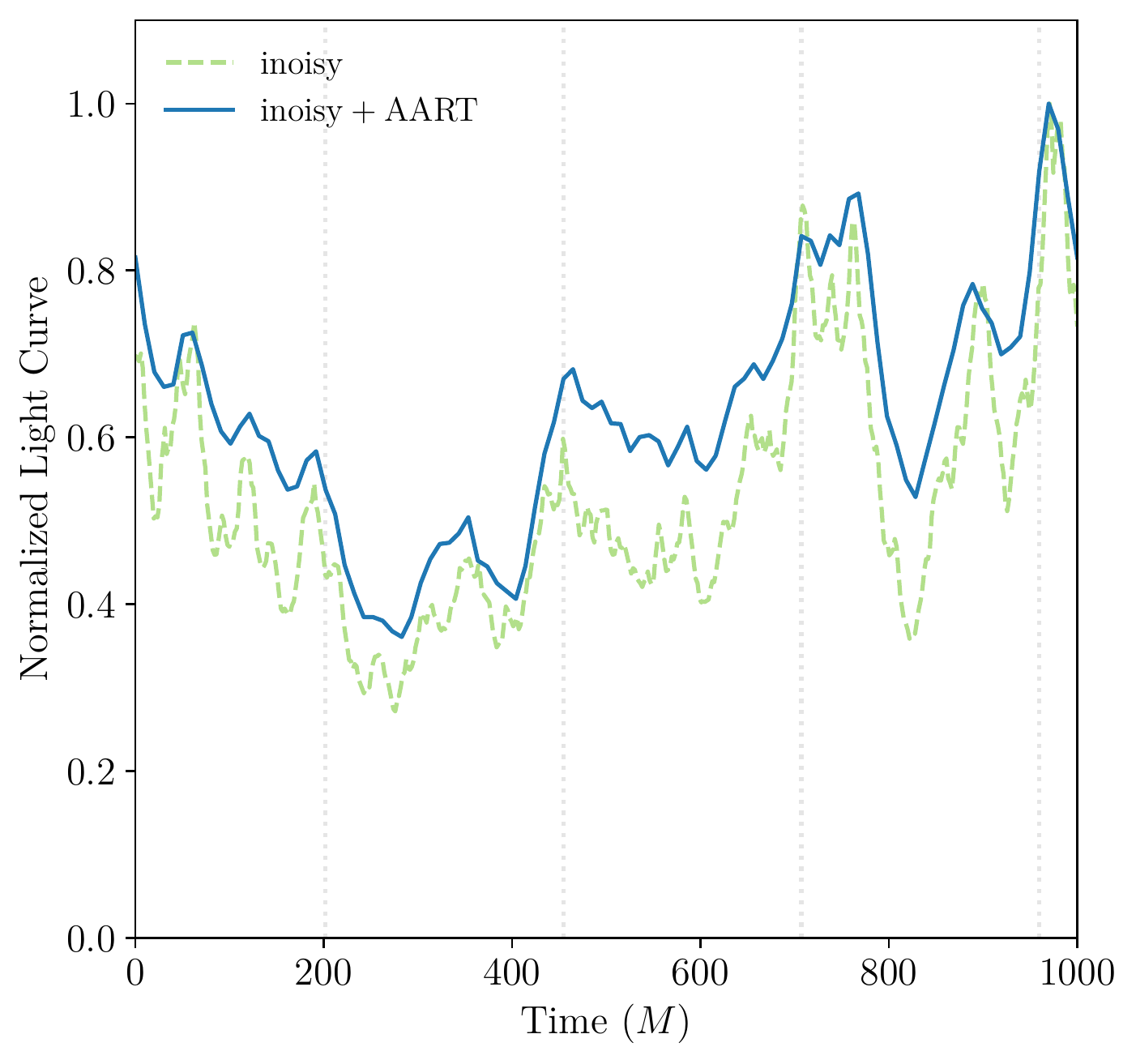} 
    \caption{Normalized light curves of the stochastic \texttt{inoisy} source with parameters in Table~\ref{tbl:Parameters}.
    Green: total emitted flux measured from \texttt{inoisy} snapshots.
    Blue: total observed flux measured from ray-traced images.
    Relativistic effects (gravitational redshift, Doppler boosting, and light bending) make the observed flux smoother and less variable than the emitted flux.
    The vertical dotted lines indicate the times corresponding to the four snapshots displayed in Fig.~\ref{fig:Snapshots}.}
    \label{fig:LightCurves}
\end{figure}

\section{Applications}
\label{sec:Applications}

Having fully described our stochastic source model, we can now simulate it with \texttt{inoisy} and then use \texttt{AART} to ray trace its appearance and compute its visibility.
We will present a whole suite of simulations in a follow-up paper, but for now we limit ourselves to one complete example from which Figs.~\ref{fig:Snapshots}, \ref{fig:TimeAverage}, and \ref{fig:ProjectedDiameter} are derived.
We first explain in detail how to obtain Figs.~\ref{fig:Snapshots} and \ref{fig:TimeAverage}, while Fig.~\ref{fig:ProjectedDiameter} will be the focus of Sec.~\ref{sec:Forecast} below.

For this simulation, we ran \texttt{inoisy} on a regular Cartesian grid $(x_{\rm s},y_{\rm s},t_{\rm s})$ of size $2048\times2048\times512$.
For each of the spatial coordinates $(x_{\rm s},y_{\rm s})$, we uniformly placed 2048 pixels within the range $[-50,50]M$, resulting in a spacing of $0.05M$ between pixels, whereas for the time coordinate $t_{\rm s}$, we placed 512 pixels uniformly within the range $[0,1000]M$, resulting in a cadence of one frame every $1.95M$.
We used the default values listed in Table~\ref{tbl:Parameters} for the fifteen parameters in the model.

The left panel in Fig.~\ref{fig:inoisySnapshotM87} shows an example of a snapshot extracted from the resulting \texttt{inoisy} run.
The accretion flow appears qualitatively similar to GRMHD-simulated flows.
In particular, we tuned the scale $\sigma$ of fluctuations to ensure that the total observed flux (the blue light curve in Fig.~\ref{fig:LightCurves}) varies by a factor of $\sim\!3$ between its minimum and maximum.
This level of time-variability mimics the light curves obtained from GRMHD simulations: see, for example, Figs.~3 and 4 of Ref.~\cite{Noble2009}, Fig.~8 of Ref.~\cite{Chatterjee2020} or Fig.~5 of Ref.~\cite{Wong2022a}.
Ensuring that this agreement also holds quantitatively is an interesting challenge that we hope to tackle in the future.

\subsection{Fast-light: stationary non-axisymmetric source}

After producing our \texttt{inoisy} simulation, we use \texttt{AART} to ray trace it.
A commonly adopted simplification in carrying out ray tracing is the so-called ``fast-light'' approximation.
When staring at a variable source, a single frame $I_{\rm o}(t_{\rm o},\alpha,\beta)$ of its observational appearance at a fixed moment in time is formed by photons that were emitted at different times $t_{\rm s}^{(n)}(\alpha,\beta)$ in the history of the source.
The reason is that photons that appear at different positions $(\alpha,\beta)$ in the image plane follow different paths in the geometry, thereby incurring different time delays $\Delta t=t_{\rm o}-t_{\rm s}^{(n)}$ on their way from source to observer, as shown in Figs.~\ref{fig:n0Time} and~\ref{fig:n1Time}.
The fast-light approximation simply ignores this variable time delay and maps the image plane $(\alpha,\beta)$ onto a single snapshot of the equatorial plane, using only the transfer functions $r_{\rm s}^{(n)}$ and $\phi_{\rm s}^{(n)}$ from Sec.~\ref{subsec:RayTracing} and replacing Eq.~\eqref{eq:ObservedIntensity} by
\begin{align}
	I_{\rm o}(\alpha,\beta)=\sum_{n=0}^{N(\alpha,\beta)-1}\zeta_ng^3\pa{r_{\rm s}^{(n)},\alpha,\beta}I_{\rm s}\pa{r_{\rm s}^{(n)},\phi_{\rm s}^{(n)}}.
\end{align}
Mathematically, this is equivalent to keeping the emission time $t_{\rm s}$ fixed while letting the observation time $t_{\rm o}$ vary across the image.
Naturally, if the source is stationary, then this is a moot distinction since the time-dependence drops out anyway, so this amounts to treating each individual \texttt{inoisy} snapshot as a stationary (though non-axisymmetric) source.
Thus, the fast-light approximation is exact for a stationary source, and it can offer a decent approximation as long as the source varies slowly relative to the ``fast'' light being traced, hence the name.

As an example, we take the first ($t_{\rm s}=0$) snapshot from our \texttt{inoisy} simulation and ray trace it as described in Sec.~\ref{subsec:RayTracing}, assuming a black hole spin of $a/M=94\%$ and an observer inclination of $\theta_{\rm o}=17^\circ$.
In the $n=0$ layer, we ray trace at the same spatial resolution as the underlying \texttt{inoisy} simulation; that is, we use a spacing of $\delta x^{(0)}=\delta x_\texttt{inoisy}=0.05M$ between grid pixels.
In the higher-$n$ layers, we adaptively increase the resolution by a factor of two in each successive band, and ray trace on grids with spacings of $\delta x^{(n)}=2^n\delta x^{(0)}$ between pixels.

The right panel of Fig.~\ref{fig:inoisySnapshotM87} displays the resulting image (in a logarithmic color scale to highlight the accretion flow), which looks qualitatively similar to the single snapshots produced with state-of-the-art GRMHD simulations \cite{Chael2021,Wong2022a}.
The figure also displays a decomposition of the image into its $n\in\cu{0,1,2}$ layers, which are shown individually in the inset panels.
Each separate layer appears well-resolved, suggesting that our grid resolutions are sufficient to resolve the photon rings.
Indeed, we have verified (using a convergence test detailed in Sec.~\ref{sec:Requirements}) that the results derived from the observable of interest (which in our case is the visibility amplitude, presented in Sec.~\ref{subsec:Visibility} below) do not change when the resolutions are doubled.

\subsection{Slow-light: non-stationary non-axisymmetric source}

With the transfer functions \eqref{eq:rs}--\eqref{eq:ts} implemented in \texttt{AART}, it is also easy to perform a ``slow-light'' ray-tracing that takes into account varying photon emission times across the image.
We simply use Eq.~\eqref{eq:ObservedIntensity} and plug in all the transfer functions defined in Sec.~\ref{subsec:RayTracing}.
This requires us to use the entire series of snapshots generated by \texttt{inoisy}; that is, we must now use the full time evolution of the accretion flow.

Slow-light tracing does introduce two new complications.
First, since the pixels in a single frame of a movie of the source can depend on a very wide range of emission times $t_{\rm s}^{(n)}(\alpha,\beta)$, in order to produce a movie of some duration $T_{\rm o}$, we typically need to simulate the source over a longer duration $T_{\rm s}>T_{\rm o}$.
In practice, we can estimate how much longer $T_{\rm s}$ needs to be by examining isochronal curves in the $n=0$ layer, such as those displayed in Fig.~\ref{fig:n0Time}.
At low inclinations, the range of $t_{\rm s}^{(0)}(\alpha,\beta)$ over sampled pixels is about $50M$, but this range grows much larger at higher inclinations.
In addition, the higher-$n$ images are composed of photons emitted roughly a time $n\tau$ earlier, so in order to ray trace a movie containing $n$ layers, we expect to need $T_{\rm s}-T_{\rm o}\gtrsim50M+n\tau\approx80M$.
Some rays may occasionally require sampling the source even earlier than the start of the simulation, but this is not really an issue with \texttt{inoisy} thanks to its use of periodic boundary conditions (including in time).

Second, because $t_{\rm s}^{(n)}(\alpha,\beta)$ varies smoothly across the image plane, the code must be able to sample the source $I_{\rm s}(r_{\rm s},\phi_{\rm s},t_{\rm s})$ for continuous values of $t_{\rm s}$, including at times in between the frames computed by \texttt{inoisy}.
This problem can be dealt with using interpolation, but only so long as the underlying \texttt{inoisy} simulation has sufficiently high resolution to smoothly resolve the flow's motion.
This is the case in our example simulation, as the  \texttt{inoisy} movie of the equatorial source has high enough cadence to be smoothly interpolated in time.

Thus, \texttt{AART} readily produces a movie of the source (whose parameters are listed in Table~\ref{tbl:Parameters}).
We display four snapshots from this movie, taken at intervals of $250M$, in the top row of Fig.~\ref{fig:Snapshots}.
We also plot its light curves in Fig.~\ref{fig:LightCurves}: the total observed flux (measured in the image plane) and total emitted flux (measured in the equatorial plane) are shown in blue and green, respectively, with the former appearing smoother and less variable than the latter due to relativistic effects.

As in the stationary (fast-light) case, the snapshot images in Fig.~\ref{fig:Snapshots} present bright transient features that are qualitatively similar to those seen in state-of-the-art snapshots of GRMHD simulations \cite{Chael2021,Wong2022a}.
These transient features can obscure the photon ring in instantaneous snapshots, but they wash out of the average over 100 snapshots in Fig.~\ref{fig:TimeAverage}, leaving the photon ring as the only prominent feature in the time-averaged image shown in the left panel.
This image can be directly compared the one shown in Fig.~1 of Ref.~\cite{Johnson2020}, which was also obtained by time-averaging over 100 uniformly spaced snapshots taken from a GRMHD simulation with a time range of $1000M$.

\subsection{Visibility on long baselines}
\label{subsec:Visibility}

A radio interferometer like the EHT samples a (complex) radio visibility $V(\mathbf{u})$.
By the van Cittert-Zernike theorem, this is related to a sky brightness $I_{\rm o}(\mathbf{x}_{\rm o})$ via a 2D Fourier transform
\begin{align}
	\label{eq:ComplexVisibility}
	V(\mathbf{u})=\int I_{\rm o}(\mathbf{x}_{\rm o})e^{-2\pi i\mathbf{u}\cdot\mathbf{x}_{\rm o}}\ed^2\mathbf{x}_{\rm o}.
\end{align}
Here, $\mathbf{x}_{\rm o}$ are dimensionless image-plane coordinates measured in radians, such as $(\alpha,\beta)/r_{\rm o}$, while the dimensionless vector $\mathbf{u}$ is a ``baseline'' projected onto the plane perpendicular to the line of sight and measured in units of the observation wavelength \cite{Roberts1994}.
We use radio-astronomy conventions to make this Fourier transform $2\pi$-symmetric, unlike the definition \eqref{eq:FourierConventions}.

Therefore, to relate simulated images to actual observables, we must Fourier transform them to compute their visibility.
In particular, the narrow image features---like the photon rings---that encode precise information about the spacetime dominate the interferometric signal on very long baselines $u=\ab{\mathbf{u}}\gg1$ \cite{Johnson2020,GLM2020}, requiring us to compute Fourier transforms up to very high frequencies.
To do so, we again exploit the layered image structure: since the observed intensity \eqref{eq:ObservedIntensity} decomposes into
\begin{align}
	I_{\rm o}(\alpha,\beta)=\sum_{n=0}^{N(\alpha,\beta)-1}I_n(\alpha,\beta),
\end{align}
with $I_n(\alpha,\beta)$ denoting the $n^\text{th}$ image layer, we can likewise use the linearity of the Fourier transform \eqref{eq:ComplexVisibility} to decompose the total complex visibility into individual subring components:
\begin{align}
	\label{eq:VisibilityLayers}
	V(\mathbf{u})=\sum_{n=0}^\infty V_n(\mathbf{u}),\quad
	V_n(\mathbf{u})=\int I_n(\mathbf{x}_{\rm o})e^{-2\pi i\mathbf{u}\cdot\mathbf{x}_{\rm o}}\ed^2\mathbf{x}_{\rm o}.
\end{align}
The subring images $I_n(\mathbf{x}_{\rm o})$ with $n>0$ consist of narrow rings of characteristic widths $e^{-\gamma n}$ (where $\gamma$ is a Lyapunov exponent) and roughly equal intensities.
Hence the total flux $V_n(0)$ of the $n^\text{th}$ image layer also scales like $e^{-\gamma n}$ \cite{Johnson2020}.
Moreover, a narrow ring produces a characteristic, weak $u^{-1/2}$ power-law falloff in its Fourier transform: indeed one can show that in the regime
\begin{align}
	\label{eq:UniversalRegime}
	\frac{1}{d}\ll u\ll\frac{1}{w},
\end{align}
the visibility \eqref{eq:ComplexVisibility} in polar coordinates $\mathbf{u}=(u,\varphi)$ of a ring of diameter $\sim\!d$ and width $\sim\!w$ takes the ``universal'' form \cite{Gralla2020,GrallaLupsasca2020c}
\begin{align}
	\label{eq:UniversalVisibility}
	V(\mathbf{u})=\frac{e^{-2\pi iC_\varphi u}}{\sqrt{u}}\br{\alpha_\varphi^{\rm L}e^{-\frac{i\pi}{4}}e^{i\pi d_\varphi u}+\alpha_\varphi^{\rm R}e^{\frac{i\pi}{4}}e^{-i\pi d_\varphi u}},
\end{align}
where $\alpha_\varphi^{\rm L,R}=\alpha_{\varphi+\pi}^{\rm R,L}>0$ encodes the polar intensity profile around the ring, while $d_\varphi$ and $C_\varphi$ are its projected diameter and centroid displacement at angle $\varphi$ in the image, respectively.

As first pointed out in Ref.~\cite{Johnson2020}, this suggests that the $n^\text{th}$ photon ring dominates the interferometric signal in the regime
\begin{align}
	\label{eq:RingCascade}
	\frac{1}{w_{n-1}}\ll u\ll\frac{1}{w_n},
\end{align}
where $w_n\approx e^{-\gamma n}w_0$ is the width of the $n^\text{th}$ subring, producing a characteristic cascading structure of damped oscillations with periodicity encoding the diameter of successive subrings; see also Secs.~2 and 4 of Ref.~\cite{Paugnat2022} for a more detailed discussion.

Since we compute images layer-by-layer---as described in Sec.~\ref{subsec:GridAdaptiveness}---with a different grid and resolution in each layer $I_n$, it is convenient to also compute the visibility layer-by-layer as well, and obtain each $V_n$ as the Fourier transform \eqref{eq:VisibilityLayers} of $I_n$.

Instead of taking this 2D Fourier transform directly, as in Refs.~\cite{GLM2020,Paugnat2022}, we make use of the projection-slice theorem to compute $V_n(u,\varphi)$ along slices of fixed polar angle $\varphi$ in the Fourier plane.
The procedure is as follows.
For each angle $\varphi$, we first compute the Radon transform along the cut at angle $\varphi$ across the image; that is, in each lensing band, we integrate the observed intensity $I_n$ along lines perpendicular to the slice of constant $\varphi$.
Then, we interpolate each Radon transform to the resolution of the highest-order lensing band.
Lastly, we sum up the contributions from all the image layers $I_n$, and perform a one-dimensional Fourier transform to finally obtain $V(u,\varphi)$.

The bottom row of Fig.~\ref{fig:Snapshots} presents the visibility amplitudes $|V(u,\varphi)|$ along slices of constant $\varphi=0^\circ$ (in red) and $\varphi=90^\circ$ (in blue) of the corresponding snapshots in the top row.
Since these individual snapshots display a strong dependence on the variable emission profile, the visibility amplitudes also exhibit a large variability.
Nevertheless, as we mentioned previously, these fluctuations wash out under time-averaging, leaving the characteristic ringing signature of the photon ring as the main persistent feature that dominates the visibility.

We also display the visibility amplitudes at baseline angles $\varphi=0^\circ$ and $\varphi=90^\circ$ (again, in red and blue, respectively) for our 100 snapshots in the right panel of Fig.~\ref{fig:TimeAverage}.
Their incoherent time-average is also shown with bold solid lines.
``Incoherent'' here means that the averaging is performed at the level of the visibility \textit{amplitudes} $|V(u,\varphi)|$, as in a realistic experiment \cite{GLM2020}, rather than at the level of the complex visibilities themselves, before taking the amplitude---these operations (averaging and taking amplitudes) do not commute for complex quantities.
(Experimentally, the latter type of ``coherent'' time-averaging would require tracking the visibility phase over the course of all observations, which is beyond our near-term capabilities.)
These time-averaged amplitudes can be directly compared to those of a radial profile; see, e.g., Figs.~4 and 5 of Ref.~\cite{GLM2020}.

\section{Ray tracing requirements}
\label{sec:Requirements}

As described in Sec.~\ref{subsec:GridAdaptiveness}, we decompose images into layers labeled by photon half-orbit number $n$ and ray trace each layer separately.
In principle, one can choose any grid resolution in each lensing band.
In practice, however, these resolutions must be sufficiently high to resolve the image features present in each layer.
Otherwise, the output is not a faithful image of the source, and the resulting visibility is likewise inaccurate.
In addition, the $n=0$ lensing band occupies all of the image plane, whereas the $n=0$ grid must of course have a finite size.
Hence, the direct image of the source may need truncatation, which could in turn introduce errors in the visibility.
In this section, we describe the requirements on grid resolution and size that must be met in order to ensure that both the image and the visibility of a source are correctly computed.
We also strive to explain the types of errors encountered when these requirements are not met.

We first discuss grid resolution and then grid size.
A finite resolution erases small-scale features and a finite field of view cuts off large-scale features, but these effects can be remedied with the use of interpolation and extrapolation, respectively.

\subsection{Resolution requirements}
\label{subsec:Resolution}

\subsubsection{Requirements on the underlying simulation of the source}

When computing images of some source model $I_{\rm s}(\mathbf{x}_{\rm s})$, we must be able to sample the source intensity $I_{\rm s}$ at arbitrary $\mathbf{x}_{\rm s}$ within the range of the simulation.
This is because the transfer functions \eqref{eq:rs}--\eqref{eq:ts} vary smoothly across the image plane, as discussed in Sec.~\ref{subsec:Visualization} and as illustrated in Figs.~\ref{fig:n0TransferFunctions} through \ref{fig:n2TransferFunctions}.

For a stationary and axisymmetric source profile of the form \eqref{eq:RadialProfile}, this requirement is trivial, since we typically specify the radial profile $J(r_{\rm s})$ analytically, as we did in Eq.~\eqref{eq:JonhnsonSU} with Johnson's SU distribution (which can be computed anywhere to arbitrary precision).
On the other hand, if the source profile $I_{\rm s}(\mathbf{x}_{\rm s})$ is not known exactly, then we must define it everywhere in the range where it is to be sampled via interpolation.

In that case, it is important that the interpolation be done on a dense grid, so that we do not miss any features of the source.
In practice, this means that if its smallest features are of size $\sim\!w$, then before interpolating the source, we must first sample it on a grid with pixel spacing $\delta x\lesssim w$.
Otherwise, if the grid spacing is some $\delta x\gtrsim w$, then we may miss a feature of size $w$ in between two successive grid points $x_i$ and $x_i+\delta x$, leading to an error in both the computed image and its visibility.
 
Moreover, the grid density must be high enough to resolve all the features in the source with many points, so as to ensure that the resulting interpolation is smooth: if some features are underresolved (for instance, if a bump in a radial profile has only one point underneath it), then interpolation will produce spurious sharp features in the image that will have a significant impact on its Fourier transform, and hence the visibility.
Thus, $\delta x\lesssim w$ is not enough---we must in fact require that $\delta x\ll w$.\footnote{This smooth interpolation effectively defines the source down to arbitrarily small length scales by decreeing that no new features appear below size $w$, which is the minimal (most conservative) ``ultraviolet completion'' of the model.
Any other choice would modify the visibility on baselines $u\gtrsim1/w$ and should then be specified as part of the model itself, rather than by fiat.}

For the GRFs that we introduced in Sec.~\ref{sec:Model} to model astrophysical fluctuations, this means that the spacing between grid points $\mathbf{x}_{\rm s}=(t_{\rm s},x_{\rm s},y_{\rm s})$ must be smaller than the characteristic size of fluctuations, which is set by the correlation lengths $\lambda_\ell$ with $\ell\in\cu{0,1,2}$.
That is, we require a grid spacing $\delta x\ll\lambda_\ell$.

We reiterate that interpolation is unavoidable here because the \texttt{inoisy} simulation grid does not coincide (barring some incredible fine-tuning) with the equatorial crossings of rays traced backwards from points in the image-plane grids.

\subsubsection{Requirements on the image grids used in ray tracing}

We now describe the resolutions needed in the grids used to ray trace within each lensing band, as described in Sec.~\ref{subsec:GridAdaptiveness}.

For the direct $n=0$ image, we must ray trace on a grid with spacing $\delta x^{(0)}\lesssim\delta x^\texttt{model}$ for precisely the same reasons as listed above for the underlying model---provided we do so, we will not miss any image features and they will all be well resolved.

Likewise, in the higher-$n$ layers, we must ray trace on grids with exponentially fine spacings $\delta x^{(n)}\approx e^{-\gamma n}\delta x^{(0)}$.
The reason is simply that if the smallest feature in the $n=0$ image is of size $w_0\approx w$, then its $n^\text{th}$ mirror image in the $n^\text{th}$ lensing band will be roughly of size $w_n\approx e^{-\gamma n}w_0$.
To avoid missing these features (or underresolving them), we must demand that $\delta x^{(n)}\ll w_n$, a condition that is automatically satisfied with this adaptive grid scaling tailored to the Kerr lensing behavior.

Provided that each image layer is ray traced at a sufficiently high resolution to ensure that no sub-grid structure is omitted, we can safely interpolate the image layers $I_n(\mathbf{x}_{\rm o})$.
In turn, we can evaluate their Radon transforms along any desired slice, and then take a 1D Fourier transform to obtain their correctly computed visibilities $V_n(\mathbf{u})$, as described in Sec.~\ref{subsec:Visibility}.

Conversely, we note that if the resolution is not increased in each successive layer, then eventually some image layer $I_n(\mathbf{x}_{\rm o})$ will have lensed features smaller than its grid spacing.
This unresolved sub-grid structure will render the visibility $V_n(\mathbf{u})$, and thus the full visibility $V(\mathbf{u})$, inaccurate.
Hence, adaptive ray tracing is necessary to guarantee accurate visibilities.

\subsubsection{Convergence tests}

To summarize, for both the emission model $I_{\rm s}(\mathbf{x}_{\rm s})$ and its ray-traced image layers $I_n(\mathbf{x}_{\rm o})$, we must use sufficiently high resolutions to ensure that we do not truncate any sub-grid structure (that is, structure on length scales smaller than the resolution).
This then allows us to safely interpolate $I_{\rm s}(\mathbf{x}_{\rm s})$ and $I_n(\mathbf{x}_{\rm o})$, without missing any features or introducing any spurious new ones in the interpolation procedure.

Interpolating $I_{\rm s}(\mathbf{x}_{\rm s})$ is needed to obtain the layers $I_n(\mathbf{x}_{\rm o})$, as ray tracing requires us to compute the intensity loaded on rays that intersect the equatorial source at generic crossing points.
Interpolating the image layers $I_n(\mathbf{x}_{\rm o})$ is also needed to compute their Fourier transforms $V_n(\mathbf{u})$, and hence the visibility \eqref{eq:VisibilityLayers}.
Any errors in either of these steps introduce errors in the final visibility $V(\mathbf{u})$, so we must carefully avoid them by using the appropriate resolutions, as explained above.

In practice, however, these requirements are only heuristic.
For instance, the Lyapunov exponent is an angle-dependent function $\gamma(\varphi)$ around the image plane, so we cannot precisely realize the exponential scaling $\delta x^{(n)}\sim e^{-\gamma n}\delta x^{(0)}$ on a Cartesian grid.
In addition, $e^{\gamma(\varphi)}$ can be as large as $\sim\!20$, but such an increase in pixel density is not always necessary.
In fact, recall that for the example in Sec.~\ref{sec:Applications}, we simply set $\delta x^{(n)}=2^n\delta x^{(0)}$, a much more modest but still adequate scaling of pixel density.

Inversely, a realization of an  \texttt{inoisy} source realization may occasionally make a field excursion that creates a far smaller fluctuation than the minimum size $w$ typically expected from its correlation matrix $\mathbf{\Lambda}(\mathbf{x}_{\rm s})$.
While statistically improbable, such events cannot be precluded from occurring, and could result in a too-large grid spacing $\delta x^\texttt{model}\gtrsim w$.
Unfortunately, it is intractable to directly check that this does not happen.

\begin{figure}
    \centering
    \includegraphics[width=\columnwidth]{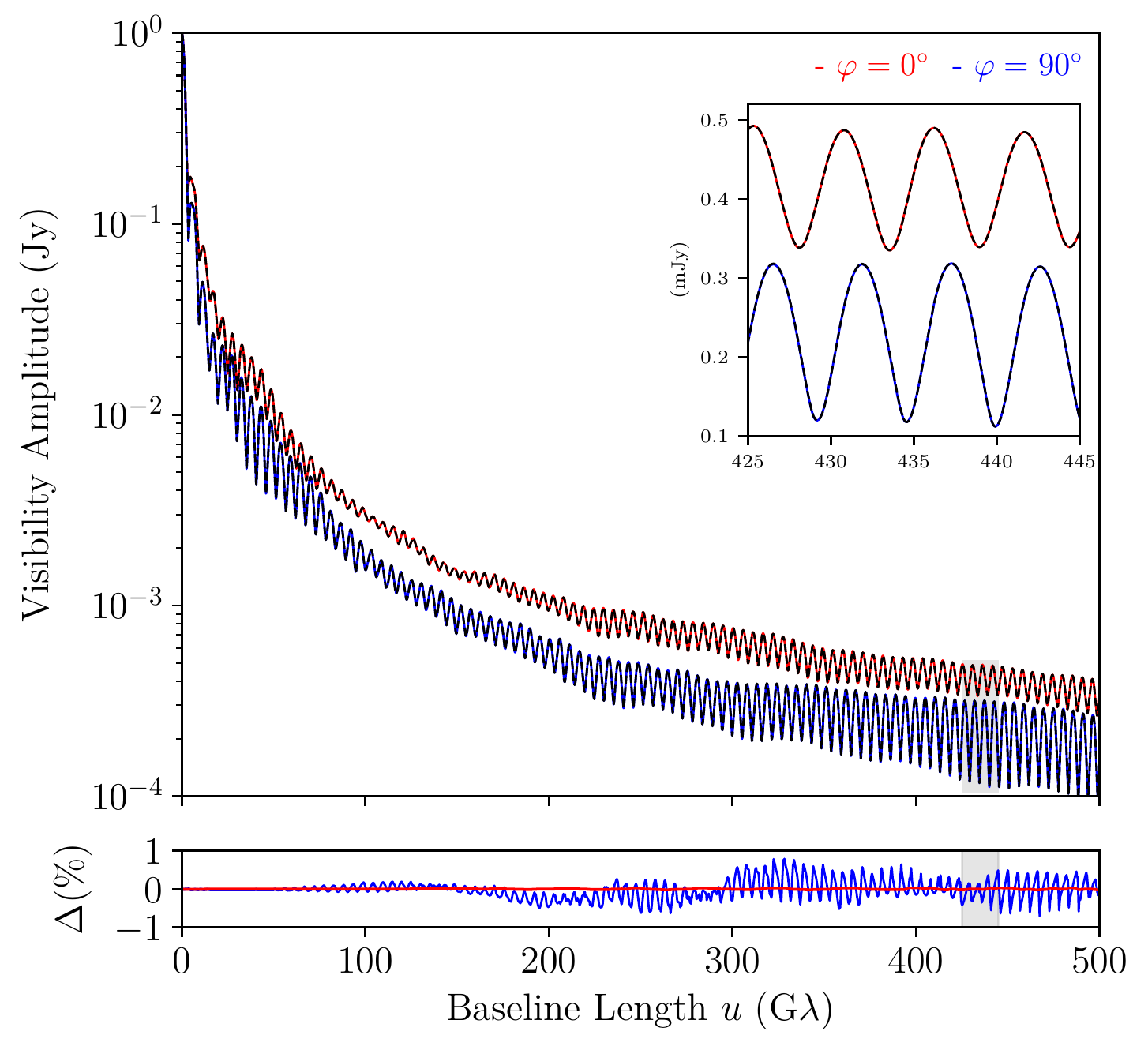}
    \caption{Convergence test of the resolution needed to ray trace an \texttt{inoisy} snapshot.
    We plot its visibility amplitude at baseline angles $\varphi=0^\circ$ (red) and $\varphi=90^\circ$ (blue) computed using the grid resolutions described in the main text.
    The black dashed lines correspond to the visibility amplitudes obtained after doubling the nominal resolution.
    The inset panel zooms into the region were the fits were performed to infer the projected ring diameter as described in Sec.~\ref{sec:Forecast}.
    Since the differences are consistently below 1\%, we can validate the resolution.}
    \label{fig:ResolutionCheck}
\end{figure}

Instead, it is best to numerically check for the absence of errors using a simple convergence test, in which one doubles all the resolutions (of the simulated profile and image grids) and computes again the observables (images and visibilities).
In Fig.~\ref{fig:ResolutionCheck}, we show an example of a convergence test for the example from Sec.~\ref{sec:Applications}: after halving the grid resolutions, the changes in the visibility amplitude remain small, confirming that the resolution scaling $\delta x^{(n)}=2^n\delta x^{(0)}$ is adequate.

\subsection{Field of view requirements}

\begin{figure}
    \centering
    \includegraphics[width=\linewidth]{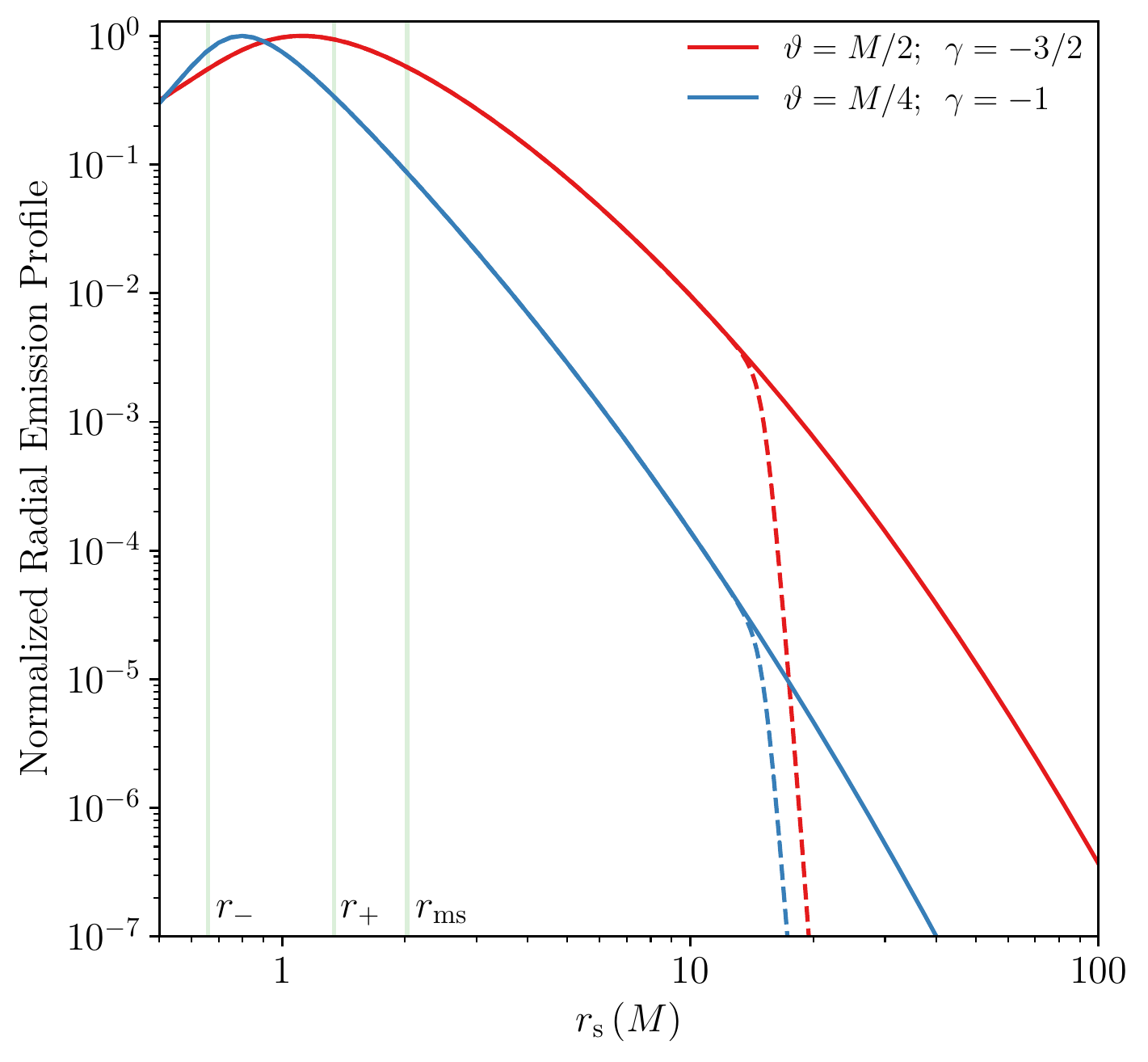} 
    \caption{Radial emission profile $J_{\rm{SU}}(r;\mu,\vartheta,\gamma)$ [Eq.~\eqref{eq:JonhnsonSU}] with $\mu=r_{-}$ to ensure that the profile peaks past the outer horizon at $r=r_+$.
    The values of the other parameters $\vartheta$ and $\gamma$ are shown in the legend.
    The dashed lines depict the profile multiplied by the cutoff function \eqref{eq:Cutoff} with $s_\neg=M$ and $r_{\neg}=15M$.
    From left to right, the vertical lines indicate the inner and outer horizons $r=r_\pm$ and the ISCO radius $r_{\rm ms}$.}
    \label{fig:Profiles}
\end{figure}

\begin{figure*}
    \includegraphics[width=\columnwidth]{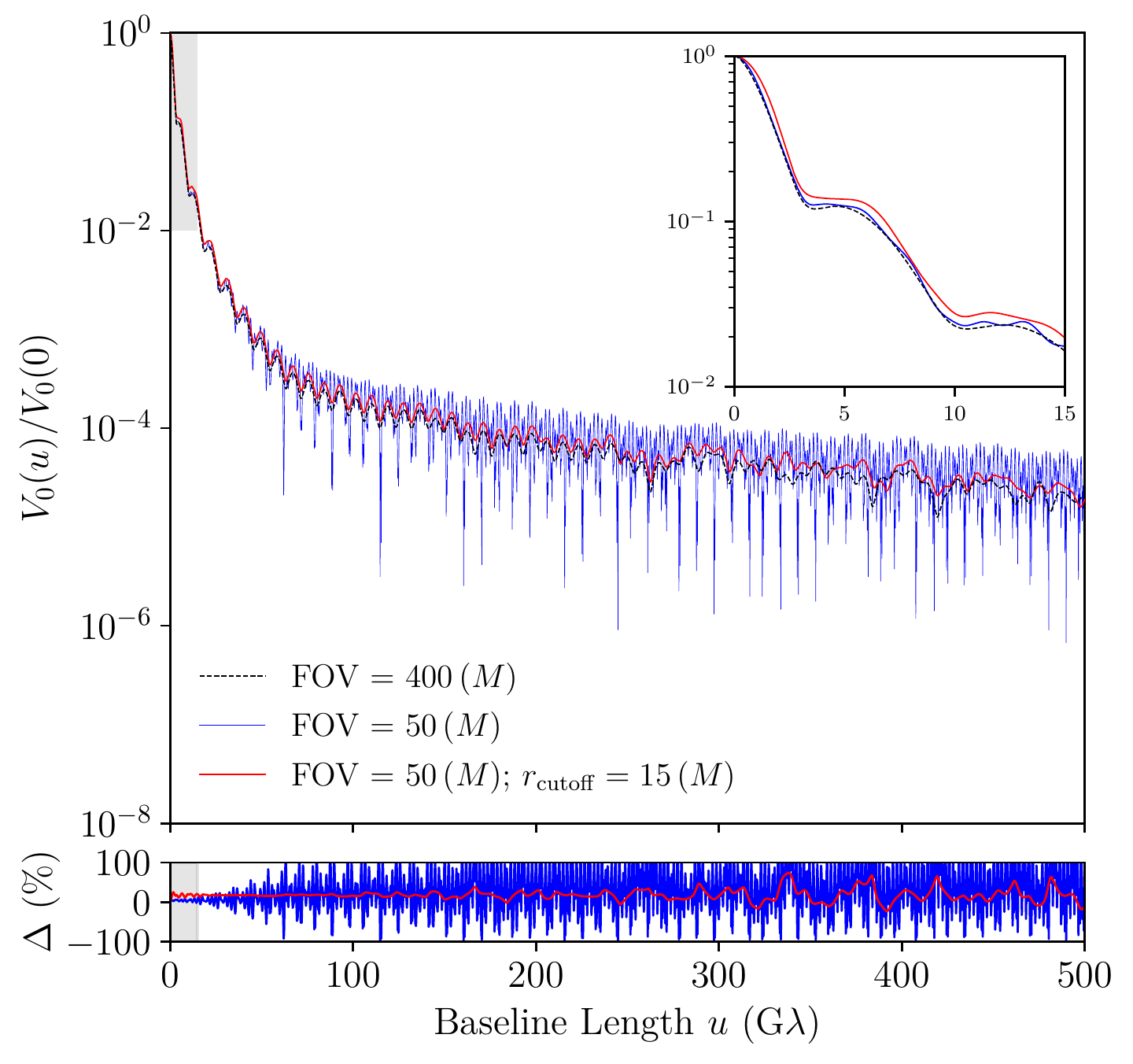}
    \includegraphics[width=\columnwidth]{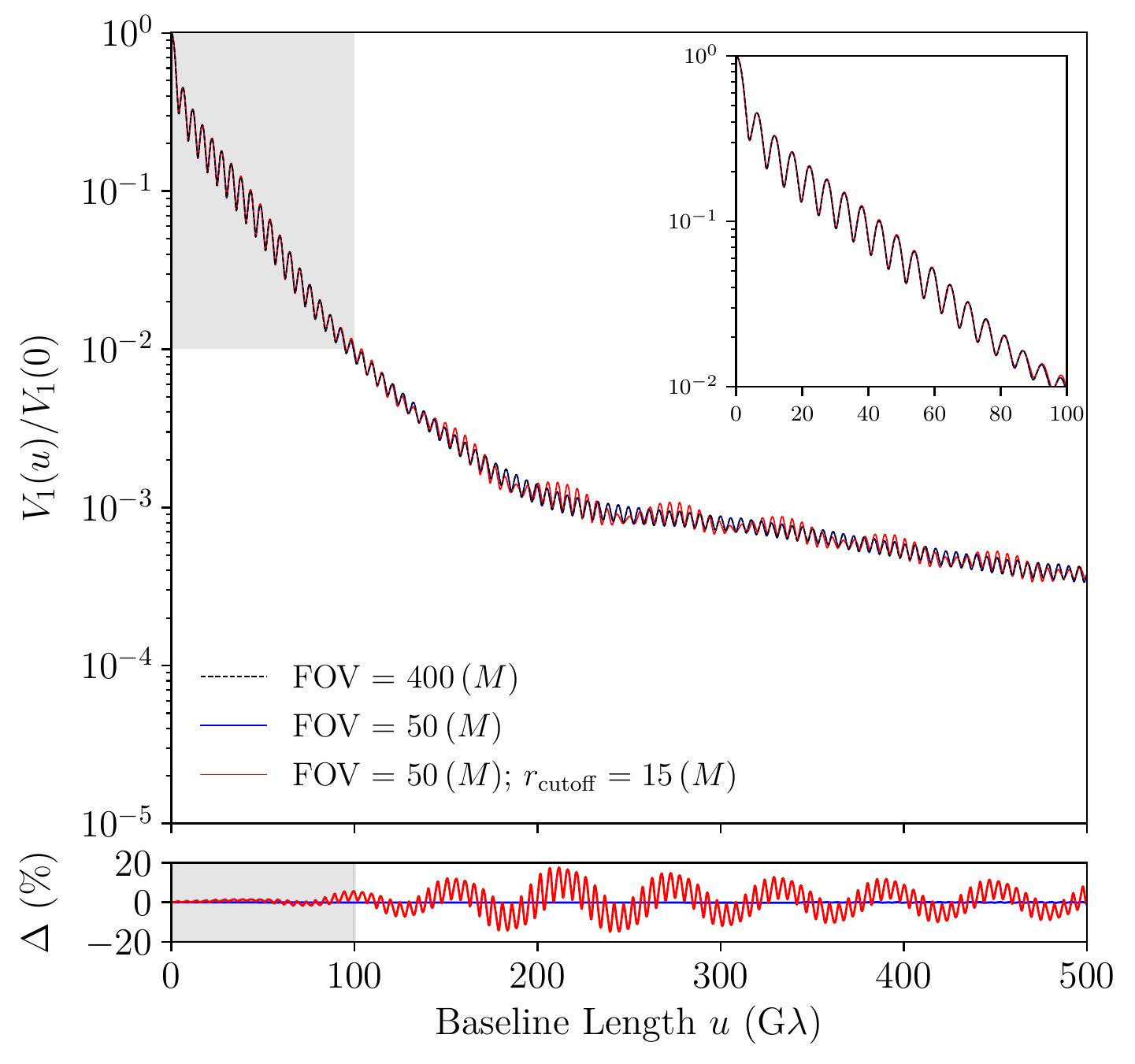}\\
    \includegraphics[width=\columnwidth]{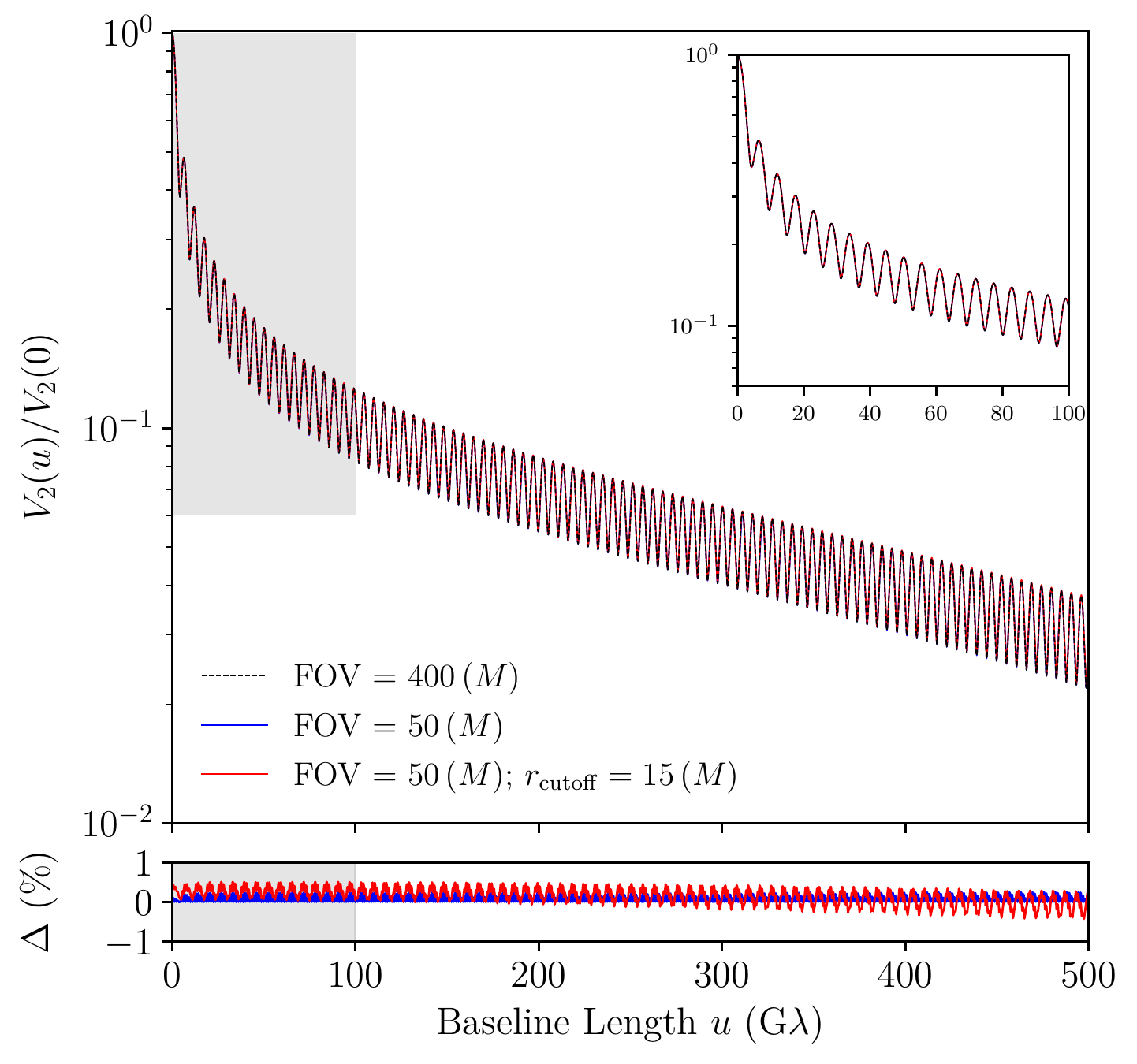}
    \includegraphics[width=\columnwidth]{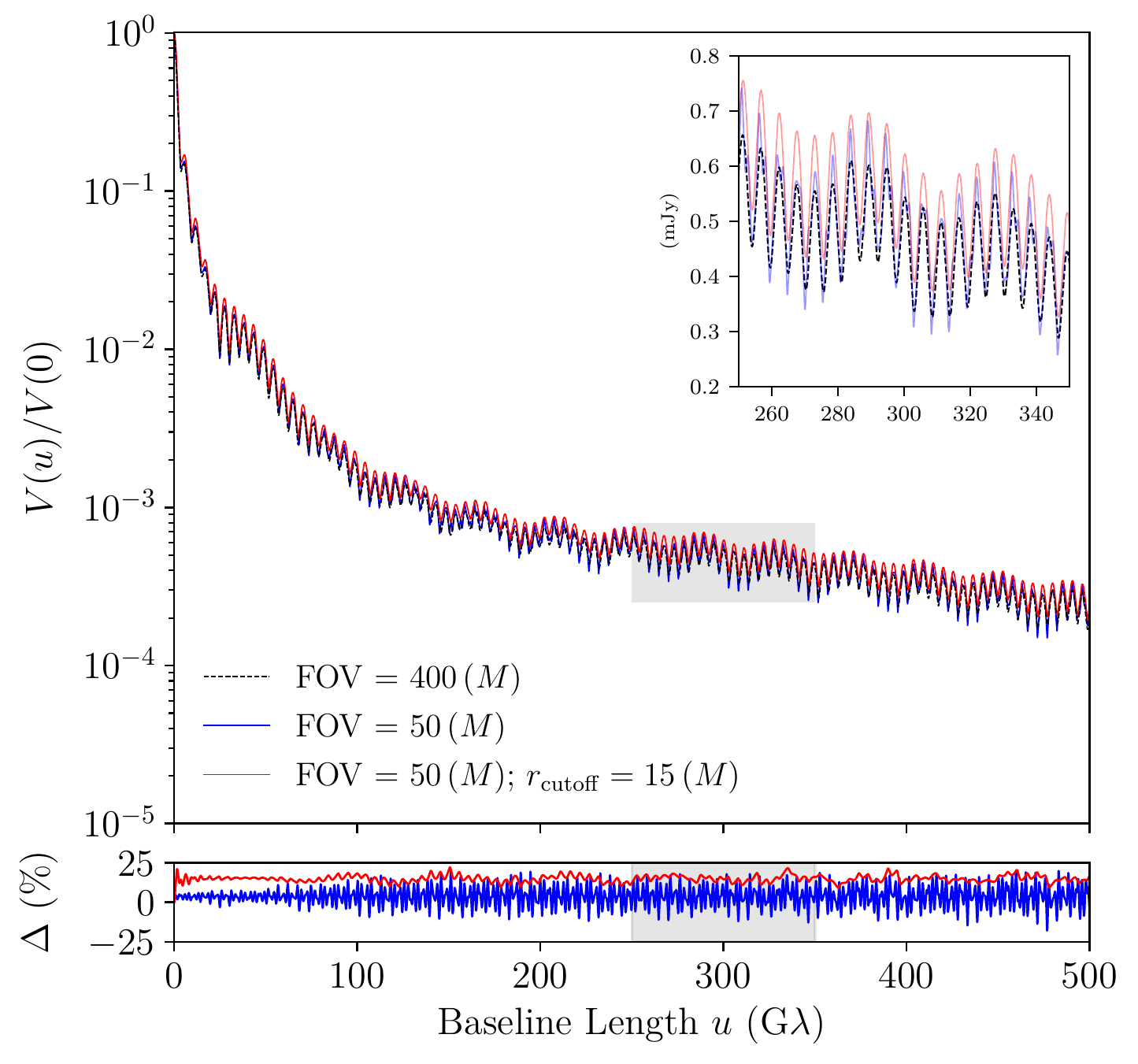}
    \caption{Normalized visibility amplitudes $\ab{V_n\pa{u,0^\circ}}$ corresponding to the image layers $I_n(\alpha,\beta)$ produced by the red radial emission profile in Fig.~\ref{fig:Profiles} (Johnson's SU distribution with $\mu=r_-$, $\vartheta=M/2$ and $\gamma=-3/2$), for $n=0$ (top left), $n=1$ (top right), and $n=2$ (bottom left).
    Bottom right: Normalized visibility amplitude $|V\pa{u,0^\circ}|$ corresponding to the full image $I_{\rm o}(\alpha,\beta)$ obtained by summing over the preceding three layers.
    In each quadrant, the main panel shows the visibility amplitudes obtained using different values for the field of view (FOV): dashed black for a large $400M$ FOV, blue for a smaller  $50M$ FOV, and red for the $50M$ FOV with a cutoff applied [Eq.~\eqref{eq:Cutoff} with $s_\neg=M$ and $r_{\neg}=15M$].
    Inset panels zoom into the shaded regions of each main plot to highlight the differences between the visibility amplitudes, while the smaller panels beneath the main ones plot percentage differences relative to the amplitude with large $400M$ FOV.
    In all these examples, we have assumed that the flow follows the Cunningham prescription \cite{Cunningham1975}, which in our notation corresponds to the Keplerian circular motion described in Sec.~\ref{subapp:KeplerianFlow}.}
    \label{fig:FieldOfView}
\end{figure*}

Being able to completely resolve an emission profile $I_{\rm s}(\mathbf{x}_{\rm s})$ and a (central) part of its image $I_{\rm o}(\mathbf{x}_{\rm o})$ is a necessary but not sufficient condition to correctly compute its visibility $V(\mathbf{u})$.

Another requirement for the accurate computation of a source model's visibility is related to the size of its image's field of view (FOV), and how fast the emission profile decays past it.
Even when all the photon rings are fully resolved (that is, the resolution in each lensing band is sufficiently high), it is still imperative to check that the FOV is sufficiently large for the image to include all of the features with significant flux.

If the FOV is too small and cuts out a portion of the source, then this truncation may create a sharp drop in the observed intensity at the edges of the image.
Such an artificial transition would effectively introduce high-frequency components to the image, polluting its Fourier transform, and hence its visibility.
This is the well-known Gibbs phenomenon.

Since we are interested in narrow features of the image, and in particular the interferometric signature of its $n=2$ photon ring that is encoded in the visibility on very long  baselines, we cannot avoid dealing with this potential effect.

\subsubsection{Field of view of the direct image}

We first focus on the direct image, as it is the most prone to the Gibbs phenomenon.
This is because it occupies the $n=0$ lensing band, which has noncompact support: as described in Sec.~\ref{subsec:ApparentHorizon}, it fills the entire image plane (except for the interior of the apparent equatorial horizon).

If the emission profile has compact support in the equatorial plane $(r_{\rm s},\phi_{\rm s})$, however, its direct image $I_0(\mathbf{x}_{\rm o})$ will also have compact support within the $n=0$ layer, and we can terminate our FOV just past the edge of this image without introducing any truncation errors or triggering the Gibbs phenomenon.

On the other hand, if the emission profile extends infinitely far in the equatorial plane (remaining nonzero as $r_{\rm s}\to\infty$), or else if its support is too large to fully ray trace over in practice, then we must necessarily cut off its $n=0$ image.\footnote{We saw in Sec.~\ref{subsec:Resolution} that a finite grid resolution introduced an ``ultraviolet'' cutoff; now we see a finite grid size as the corresponding ``infrared'' cutoff.}

For instance, consider the simple, stationary, axisymmetric source \eqref{eq:RadialProfile} with radial profile $J(r_{\rm s})$ given by Johnson's SU distribution \eqref{eq:JonhnsonSU}.
While this emission profile does decay very fast as $r_{\rm s}\to\infty$, it always remains finite.
In Fig.~\ref{fig:Profiles}, we plot it in logarithmic scale for two sets of values of its parameters.
From now on, we focus on the red profile, with parameters $\mu=r_-$, $\vartheta=M/2$, and $\gamma=-3/2$.
We first ray trace its direct $n=0$ image on a grid of size $400M$, and then we compute the corresponding visibility amplitude $\ab{V_0(u,0^\circ)}$, which we plot as a black dashed curve in the top left panel of Fig.~\ref{fig:FieldOfView}.

At the edge of our FOV, $J(r_{\rm s})$ has dropped to $\lesssim 10^{-11}$ times its maximum intensity.
As such, the sudden drop introduced by our truncation is too small to produce a significant ringing effect on long baselines, and so it cannot meaningfully affect the signal.
We will consider this large FOV and its associated visibility amplitude as the ``correct'' one.
Next, we ray trace the same image with its FOV truncated at $r_{\rm s}\approx15M$, a cutoff past which $J(r_{\rm s})$ drops to $\lesssim 10^{-4}$ times its maximum intensity.
We then plot the resulting visibility amplitude as the solid blue curve in the top left panel of Fig.~\ref{fig:FieldOfView}.
Perhaps surprisingly, it differs quite significantly from the ``correct'' one!
In fact, their relative difference---which we plot as a blue curve in the small panel below the main---shows a large and growing percentage error that exceeds 100\% before even reaching $100\,$G$\lambda$.
Even though the total flux that we ignored is tiny, the sudden drop in intensity at the FOV's edge triggered the Gibbs phenomenon.

The resulting error can affect the interferometric signature of the $n=2$ photon ring and is therefore an obstacle that we need to overcome, especially since this very profile is roughly consistent with the 2017 EHT observations of M87* \cite{GLM2020}.

There are two approaches to mitigating this problem.
The first is to increase the FOV until the profile decays enough  at the edges that it ``effectively vanishes''---in that case, these spurious ringing effects will not be noticeable.
In effect, this is what we did when we chose an FOV of $400M$.
However, it may not always be feasible to extend the FOV this much.
For instance, if the emission profile is an \texttt{inoisy} source simulated on a grid that stretches out to some maximal $r_{\rm s}$, then extending the source past this cutoff would require extrapolation.
This is already subtle for smooth profiles, as it amounts to introducing ad hoc data to the model.

The second approach---referred to as ``apodization''---has already been used to deal with this problem in Ref.~\cite{Paugnat2022}.
This method involves the multiplication of the image by a suitable window (also known as tapering or apodization function) that smoothly tapers the intensity to precisely zero before the edge of the FOV.
For our example, we will use the window function
\begin{align}
	\label{eq:Cutoff}
	C(s_{\neg},r_{\neg})=\frac{1-\tanh\left[s_{\neg}\left(r-r_{\neg}\right)\right]}{2},
\end{align}
with a cutoff value of $r_\neg=15M$ (well within the ``small'' FOV of $50M$) and $s_\neg=M$, which results in the smooth but rapidly truncated profile depicted with a dashed red line in Fig.~\ref{fig:Profiles}.

Ray tracing the direct image of this profile and computing its visibility amplitude now results in the solid red curve that we plot in the top left panel of Fig.~\ref{fig:FieldOfView}.
Its difference relative to the ``correct'' amplitude is much smaller, and the percentage difference remains fairly constant well past $100\,$G$\lambda$.
This can therefore be an effective method.

In practice, we prefer to always use emission profiles that smoothly terminate within the FOV of our direct image, which we can easily take to be $50M$ wide.
In case our source profile extends beyond this FOV, we apply the window function \eqref{eq:Cutoff} to it and take the resulting profile as our ``true'' source.

\subsubsection{Field of view of higher layers}  

Since the $n>0$ lensing bands all have compact support, it is straightforward to avoid the Gibbs phenomenon in higher-$n$ image layers and visibilities: we just take a FOV that includes the totality of the $n^\text{th}$ photon ring (that is, all the flux in the $n^\text{th}$ image layer).

We briefly describe the effects that windowing the emission profile as discussed above has on higher visibility amplitudes.
The top right panel of Fig.~\ref{fig:FieldOfView} shows that the windowed profile produces roughly the same visibility (solid red curve) as the original one (dashed black curve) up to baselines $u\sim100\,$G$\lambda$, which are necessary to resolve image features of size
\begin{align}
	100\,{\rm G}\lambda=\frac{10^{11}}{{\rm rad}}
	\approx\frac{1}{2\,\mu{\rm as}}.
\end{align}
For M87*, whose photon ring has a diameter $\sim10M\approx40\,\mu$as, this corresponds to a size of $0.5M$, or half the $n=1$ ring width.

Thus, differences in the visibility amplitude $|V_1(u,0^\circ)|$ only become apparent on baselines that start to resolve its width, as can be seen in the small plot of percentage deviation.
On the other hand, truncating the FOV of the $n=1$ image from $400M$ to $50M$ has no discernible effect on its visibility amplitude (solid blue curve) and results in a vanishingly small deviation.

Finally, we can repeat this exercise for the $n=2$ ring.
Again, windowing the profile has a negligible effect of $\lesssim1\%$, as does the truncation.
In fact, the latter deviation in this case ought to vanish entirely and is only nonzero due to numerical error introduced by the discretization of the Fourier transform: extending the FOV of the image to $400M$ by``zero padding'' changes the sampling of the fast Fourier transform, resulting in a minute effect.

We note that these tiny errors are essentially negligible in the full visibilities, which we compare in the bottom right panel of Fig.~\ref{fig:FieldOfView}, where most of the percentage error can be attributed to the direct image, as expected.

\section{Measuring the photon ring shape}
\label{sec:Forecast}

Using black hole imaging to test general relativity in a regime where gravity is strong and yet non-dynamical poses significant challenges: one must disentangle the astrophysical effects of the radiating source from the purely gravitational ones \cite{CardenasAvendano2020,Bauer2021}.
A promising approach is to focus on the $n=2$ photon ring, whose interferometric signature is expected to dominate the time-averaged radio visibility on long baselines.

More precisely, Ref.~\cite{GLM2020} proposed a test of strong-field GR based on a shape measurement of the $n=2$ photon ring around M87*, which has been recently reviewed in Ref.~\cite{Paugnat2022}.

Here, we show in the context of our example from Sec.~\ref{sec:Applications} how \texttt{AART} may be used to simulate this test using a source that includes ``realistic'' astrophysical fluctuations (generated with \texttt{inoisy}, as described in Sec.~\ref{sec:Model}).
We offer a brief overview of the key points in the test, and refer the reader to Refs.~\cite{GLM2020,Paugnat2022} for more details.

In the regime \eqref{eq:UniversalRegime}, a ring of projected diameter $d_\varphi$ has a visibility that takes the universal form \eqref{eq:UniversalVisibility}.
Therefore, its amplitude can be well approximated in this regime by \cite{Gralla2020,GrallaLupsasca2020c}
\begin{align}	
	\label{eq:VisibilityAmplitude}
	\ab{V(u,\varphi)}=\sqrt{\pa{\alpha_\varphi^{\rm L}}^2+\pa{\alpha_\varphi^{\rm R}}^2+2\alpha_\varphi^{\rm L}\alpha_\varphi^{\rm R}\sin\pa{2\pi d_\varphi u}},
\end{align}
where the functions
\begin{align}
	\alpha_\varphi^{\rm L/R}(u)=\frac{e_{\rm upper}(u)\pm e_{\rm lower}(u)}{2}
\end{align}
encode the intensity profile around the ring image, while $e_{\rm upper/lower}(u)$ are the upper/lower envelopes of the  visibility amplitude, respectively.

The width $w$ and diameter $d$ of the $n=1$ ring do not always satisfy the condition $w/d\ll1$ needed for the universal regime \eqref{eq:UniversalRegime} to exist, but those of the $n=2$ ring do.
Hence, we can fit Eq.~\eqref{eq:VisibilityAmplitude} to the visibility amplitudes in the regime \eqref{eq:RingCascade} dominated by the $n=2$ ring after numerically computing the functions $\alpha_\varphi^{\rm L/R}(u)$ (a simple cubic interpolation is enough). 

The inset panel in Fig.~\ref{fig:TimeAverage} shows the visibility amplitude for two baseline angles $\varphi=0^\circ$ and $\varphi=90^\circ$ in the baseline range $u\in[425,445]\,$G$\lambda$ (corresponding to an Earth-Moon baseline with an operation band of $332$--$344$\,GHz) with their best fit to Eq.~\eqref{eq:VisibilityAmplitude} overplotted in black dashed lines. 

According to GR, the shape of the $n=2$ photon ring is the sum of a circle, with radius $R_0$, and an ellipse centered at the origin, with radii $R_1$ and $R_2$ \cite{GLM2020}.
This ``circlipse'' shape has a projected diameter given by \cite{GrallaLupsasca2020c}
\begin{align}
	\label{eq:Circlipse}
	\frac{d_\varphi}{2}=R_0+\sqrt{R_1^2\sin^2\pa{\varphi-\varphi_0}+R_2^2\cos^2\pa{\varphi-\varphi_0}},
\end{align}
where $\varphi_0$ is an offset angle to account for the orientation of the ring within the image plane. 

Using the statistical model described in Sec.~\ref{sec:Model} for the emission profile with the parameters shown in Table~\ref{tbl:Parameters}, we (i) average the visibility amplitudes of $N$ snapshot images; (ii) fit Eq.~\eqref{eq:VisibilityAmplitude} to the visibility amplitude for 36 angles $\varphi=\{0^\circ,5^\circ,\ldots,175^\circ\}$ to obtain the projected diameters $d_\varphi$; and (iii) fit the general GR prediction given in Eq.~\eqref{eq:Circlipse} to obtain $\cu{R_0,R_1,R_2,\varphi_0}$.
The best global fits of Eqs.~\eqref{eq:VisibilityAmplitude} and \eqref{eq:Circlipse} were found using a Markov chain Monte Carlo (MCMC) method.\footnote{Since we are neglecting the phase of the oscillation, Eq.~\eqref{eq:VisibilityAmplitude} allows for several values of $d_\varphi$ separated by $\sim1/u$ to provide a good fit \cite{GLM2020}.
However, the global maximum is not always the actual diameter $d_\varphi$ of the ring as measured directly in the image domain.
As explained in Ref.~\cite{Paugnat2022}, in those cases, one must infer the diameter by fitting at multiple angles.}

\begin{table}
	\begin{tabular}{|c@{\hspace*{10pt}}c@{\hspace*{10pt}}c@{\hspace*{10pt}}c@{\hspace*{10pt}}c@{\hspace*{10pt}}c|}
	\hline 
	$N$ & $R_{0}$ & $R_{1}$ & $R_{2}$ & $\varphi_{0}$ & RMSE \\
	\hline\hline
	$5$ & $37.045$ & $1.289$ & $0.879$ & $1.319^\circ$ & $4.5\times10^{-4}$ \\
	$10$ & $36.250$ & $2.077$ & $1.679$ & $1.308^\circ$ & $3.2\times10^{-4}$ \\
	$20$ & $36.909$ & $1.419$ & $1.027$ & $0.987^\circ$ & $2.4\times10^{-4}$ \\
	$100$ & $36.052$ & $2.283$ & $1.894$ & $0.426^\circ$ & $1.1\times10^{-4}$ \\
	$\infty$ & $36.115$ & $2.219$ & $1.830$ & $0.202^{\circ}$ & $4.2\times10^{-5}$ \\
	\hline 
	\end{tabular}
	\caption{Best-fit values for the projected diameter $d_\varphi$ [Eq.~\eqref{eq:Circlipse}] obtained after averaging $N$ snapshots.
	The resulting fits are shown in Fig.~\ref{fig:ProjectedDiameter}.
	The normalized residuals RMSE are root-mean-square errors divided by the average value of $d_\varphi$.}
	\label{tbl:BestFit}
\end{table}

We report the best-fit values of the parameters in Eq.~\eqref{eq:Circlipse} and their normalized root-mean-square error (RMSE) for each $N$ in Table~\ref{tbl:BestFit}.
The resulting curves, shown in Fig.~\ref{fig:ProjectedDiameter}, follow the GR prediction very closely, even when only five snapshots are (incoherently) averaged.
The lower panels in Fig.~\ref{fig:ProjectedDiameter} show the difference of the best-fit curve for each case with the best-fit curve for the purely radial profile consisting of the envelope of the emission model (that is, when there are no fluctuations).

The results of this example are very encouraging for future missions targeting measurements of the photon ring shape: with only a few snapshots, one may already start to check whether the data follow the GR prediction. A systematic study including instrumental noise is required to simulate a realistic experimental forecast---we leave this for future work.

\section{Future outlook}
\label{Conclusions}

Here, we have presented \texttt{AART}: a new, publicly available, numerical code designed for precision studies of a black hole's photon rings.
\texttt{AART} exploits the integrability of null geodesics in the Kerr geometry to ray trace images analytically, with no loss of numerical precision even for strongly lensed photons that orbit the black hole multiple times.
The code decomposes the image plane into layers---lensing bands---with increasing grid resolution adapted to the lensing behavior of the hole.

The modular structure of \texttt{AART} can accommodate any time-dependent and non-axisymmetric equatorial emission profile, and the code could be extended to non-equatorial, anisotropic sources.
Also, the components described herein can be used separately: \texttt{AART} is designed to individually export the lensing bands, critical curve, apparent horizon, and redshift factors, which can therefore serve as input for other studies. 

As a prime application, we showed how the experimental forecast for the test of general relativity proposed in Ref.~\cite{GLM2020} could be further refined by including source fluctuations.
We used \texttt{inoisy}~\cite{Lee2021} to simulate a stochastic model of equatorial emission, and we produced high-resolution synthetic images together with their corresponding visibility on long baselines.
We then successfully extracted the GR-predicted shape for the projected diameter of the $n=2$ ring from its interferometric signature.
In a follow-up paper, we will use this framework to carry out a parameter estimation survey that also includes instrument response and noise---this will further validate the test proposed in Ref.~\cite{GLM2020} and bears relevance to SALTUS and other space-VLBI missions targeting the photon ring \cite{Gurvits2022,Kurczynski2022}.

Although we only studied the Kerr geometry, our approach may serve as a useful guide to studying other theories.
For instance, the Kerr lensing bands may offer a starting point for numerically finding those of slightly deformed spacetimes.

Finally, recent EHT observations of Sgr~A* found that only $\sim\!3.5\%$ of the EHT GRMHD models passed the light-curve variability constraint~\cite{EHT2022e}.
Given the uncertainty associated with the variability excess, and the possibility of missing physical ingredients in current astrophysical models, efforts to develop astrophysics-agnostic phenomenological approaches such as the one presented in this work are just as valuable as the improvement of accretion disk simulations.

\acknowledgments

We thank Charles Gammie, Delilah Gates, Samuel Gralla, Aaron Held, Daeyoung Lee, Hadrien Paugnat, Eliot Quataert, Leo Stein, Fr\'ed\'eric Vincent, Maciek Wielgus, and George Wong for their useful comments.
This work used resources provided by Princeton Research Computing, a consortium that includes PICSciE (the Princeton Institute for Computational Science and Engineering) as well as the Office of Information Technology's Research Computing division.

A.C.A. and A.L. gratefully acknowledge support from Will and Kacie Snellings.
A.C.A also acknowledges support from the Simons Foundation.
A.L. also thanks Pam Davis Kivelson for her black hole art inspired from this \texttt{AART}.

\appendix

\section{Explicit form of null geodesics in the Kerr exterior}
\label{app:GeodesicIntegrals}

This appendix lists all the formulas needed to compute the observational appearance of Kerr equatorial sources in the sky of a distant observer.
Readers are referred to Refs.~\cite{GrallaLupsasca2020a,GrallaLupsasca2020b} for derivations and further details.
Throughout, $F(\varphi|k)$, $E(\varphi|k)$, and $\Pi(n;\varphi|k)$ denote the incomplete elliptic integrals of the first, second and third kinds, respectively, defined according to the conventions listed in App.~A of Ref.~\cite{Kapec2020}.
$K(k)\equiv F(\pi/2|k)$ denotes the complete integral of the first kind and we also let $E'(\varphi|k)\equiv \partial_kE(\varphi|k)=\br{E(\varphi|k)-F(\varphi|k)}/(2k)$.

\subsection{Angular geodesic integrals}

The angular motion of a Kerr photon can display two qualitatively different behaviors depending on the sign of its energy-rescaled Carter constant $\eta$.
Since we only consider sources in the Kerr equatorial plane, we may ignore vortical motion with $\eta<0$, which can never reach the equator \cite{Kapec2020}.
We thus restrict our attention to ordinary motion with $\eta>0$, in which case the angular geodesic integrals in Sec.~\ref{sec:Lensing} are \cite{GrallaLupsasca2020a}
\begin{align}
	\label{eq:Gtheta}
	G_\theta^{(n)}&=\frac{1}{a\sqrt{-u_-}}\br{2m(n)K\mp_{\rm o}F_{\rm o}}, \\
	\label{eq:Gphi}
	G_\phi^{(n)}&=\frac{1}{a\sqrt{-u_-}}\br{2m(n)\Pi\mp_{\rm o}\Pi_{\rm o}},\\
	\label{eq:Gt}
	G_t^{(n)}&=-\frac{2u_+}{a\sqrt{-u_-}}\br{2m(n)E'\mp_{\rm o} E_{\rm o}'},
\end{align}
where $\pm_{\rm o}=\sign{\beta}$, $m(n)\equiv n+H(\beta)$ counts the number of angular turning points encountered along the trajectory, $H(x)$ denotes the Heaviside function, and we also introduced
\begin{align}
    \label{eq:K}
	K&=K\pa{\frac{u_+}{u_-}}
	=F\pa{\frac{\pi}{2}\left|\frac{u_+}{u_-}\right.},\\
	\label{eq:Fo}
	F_{\rm o}&=F\pa{\arcsin\pa{\frac{\cos{\theta_{\rm o}}}{\sqrt{u_+}}}\left|\frac{u_+}{u_-}\right.},\\
	E_{\rm o}'&=E'\pa{\arcsin\pa{\frac{\cos{\theta_{\rm o}}}{\sqrt{u_+}}}\left|\frac{u_+}{u_-}\right.},\\
	\Pi_{\rm o}&=\Pi\pa{u_+;\arcsin\pa{\frac{\cos{\theta_{\rm o}}}{\sqrt{u_+}}}\left|\frac{u_+}{u_-}\right.}.
\end{align}

\subsection{Radial geodesic integrals}   

The radial integrals can be decomposed into 
\cite{GrallaLupsasca2020b}
\begin{align}
	\label{eq:Iphi}
	I_\phi&=\frac{2Ma}{r_+-r_-}\br{\pa{r_+-\frac{a\lambda}{2M}}I_+-\pa{r_--\frac{a\lambda}{2M}}I_-},\\
	\label{eq:It}
	I_t&=\frac{(2M)^2}{r_+-r_-}\br{r_+\pa{r_+-\frac{a\lambda}{2M}}I_+-r_-\pa{r_--\frac{a\lambda}{2M}}I_-}\notag\\
	&\phantom{=}+(2M)^2I_0+(2M)I_1+I_2,
\end{align}
with the integrals $I_0$, $I_1$, $I_2$ and $I_\pm$ reducible to Legendre form.

Their precise form depends on the nature of the radial roots $\{r_1,r_2,r_3,r_4\}$: there are four different cases for null geodesics in the Kerr exterior, but only two of these---labeled type (2) and type (3)---arise for light rays that connect the equatorial plane to a distant observer; meanwhile, type (4) geodesics can reach distant observers but never the equator, since they are all vortical, whereas type (1) geodesics are all bound to the black hole and cannot reach a distant observer at infinity \cite{GrallaLupsasca2020b}.

\subsubsection{Type (2)}   

In this case, all the roots are real and the motion is restricted to $r\ge r_4>r_3>r_2>r_1$.

The antiderivatives of the radial geodesic integrals take the manifestly real and smooth forms \cite{GrallaLupsasca2020b}
\begin{align}
	\label{eq:Ir2}
	\mathcal{I}_0&=F^{(2)}(r),\\
	\mathcal{I}_1&=r_3F^{(2)}(r)+r_{43}\Pi_1^{(2)}(r),\\
	\mathcal{I}_2&=\frac{\sqrt{\mathcal{R}(r)}}{r-r_3}-\frac{r_1r_4+r_2r_3}{2}F^{(2)}(r)-E^{(2)}(r),\\
	\label{eq:Ipm2}
	\mathcal{I}_\pm&=-\Pi_\pm^{(2)}(r)-\frac{F^{(2)}(r)}{r_{\pm3}},
\end{align}
with (recall that $r_{ij}=r_i-r_j$)
\begin{align}
	F^{(2)}(r)&=\frac{2}{\sqrt{r_{31}r_{42}}}F\pa{\arcsin{x_2(r)}\Big|k_2}
	\ge0,\\
	E^{(2)}(r)&=\sqrt{r_{31}r_{42}}E\pa{\arcsin{x_2(r)}\Big|k_2}
	\ge0,\\
	\Pi_1^{(2)}(r)&=\frac{2}{\sqrt{r_{31}r_{42}}}\Pi\pa{\frac{r_{41}}{r_{31}};\arcsin{x_2(r)}\bigg|k_2}
	\ge0,\\
	\Pi_\pm^{(2)}(r)&=\frac{2 r_{43} }{r_{\pm3}r_{\pm4}\sqrt{r_{31}r_{42}}}\Pi\pa{\frac{r_{\pm3}r_{41}}{r_{\pm4}r_{31}};\arcsin{x_2(r)}\bigg|k_2},
\end{align}
where the auxiliary function $x_2(r)$ is defined as
\begin{align}
	x_2(r)=\sqrt{\frac{r-r_4}{r-r_3}\frac{r_{31}}{r_{41}}}
	\in\br{0,\sqrt{\frac{r_{31}}{r_{41}}}}
	\subset[0,1),
\end{align}
while the elliptic modulus is
\begin{align}
	k_2=\frac{r_{32}r_{41}}{r_{31}r_{42}}
	\in(0,1).
\end{align}

Finally, the source radius is given in terms of the Jacobi elliptic sine function $\sn(\varphi|k)$ by
\begin{align}
	\label{eq:SourceRadius2}
	r_{\rm s}^{(2)}(\tau)&=\frac{r_4r_{31}-r_3r_{41}\sn^2\pa{X_2(\tau)\big|k_2}}{r_{31}-r_{41}\sn^2\pa{X_2(\tau)\big|k_2}},\\
	X_2(\tau)&=\frac{1}{2}\sqrt{r_{31}r_{42}}\tau-F\pa{\arcsin{x_2(r_{\rm o})}\bigg|k_2}.
\end{align}

\subsubsection{Type (3)}   

In this case, only $r_1$ and $r_2$ are real roots while $r_3=\bar{r}_4$ are complex-conjugate roots, and the range of radial motion is $r\ge r_+>r_->r_2>r_1$.

The antiderivatives of the radial geodesic integrals take the manifestly real and smooth forms \cite{GrallaLupsasca2020b}
\begin{align}
	\label{eq:Ir3}
	\mathcal{I}_0&=F^{(3)}(r),\\
	\mathcal{I}_1&=\pa{\frac{Ar_1+Br_2}{A+B}}F^{(3)}(r)+\Pi_1^{(3)}(r),\\
	\mathcal{I}_2&=\pa{\frac{Ar_1+Br_2}{A+B}}^2F^{(3)}(r)+2\pa{\frac{Ar_1+Br_2}{A+B}}\Pi_1^{(3)}(r)\notag\\
	&\phantom{=}+\sqrt{AB}\Pi_2^{(3)}(r),\\
	\label{eq:Ipm3}
	\mathcal{I}_\pm&=-\frac{A+B}{Ar_{\pm1}+Br_{\pm2}}F^{(3)}(r)\notag\\ 
	&\phantom{=}+\frac{2r_{21}\sqrt{AB}}{\pa{Ar_{\pm1}}^2-\pa{Br_{\pm2}}^2}R_1\pa{\alpha_\pm;\arccos{x_3(r)}\Big|k_3},
\end{align}
with (recall that $r_{ij}=r_i-r_j$)
\begin{align}
	F^{(3)}(r)&=\frac{1}{\sqrt{AB}}F\pa{\arccos{x_3(r)}\Big|k_3},\\
	\Pi_\ell^{(3)}(r)&=\pa{\frac{2r_{21}\sqrt{AB}}{B^2-A^2}}^\ell R_\ell\pa{\alpha_0;\arccos{x_3(r)}\Big|k_3},\\
	R_1\pa{\alpha;\varphi\big|j}&=\frac{1}{1-\alpha^2}\br{\Pi\pa{\left.\frac{\alpha^2}{\alpha^2-1};\varphi\right|j}-\alpha f_1},\\
	R_2\pa{\alpha;\varphi|j}&=\frac{1}{\alpha^2-1}\br{F\pa{\varphi\big|j}-\frac{\alpha^2E\pa{\varphi\big|j}}{j+\pa{1-j}\alpha^2}}\notag\\
	&\phantom{=}+\frac{\alpha^3\sin{\varphi}\sqrt{1-j\sin^2{\varphi}}}{\pa{\alpha^2-1}\br{j+\pa{1-j}\alpha^2}\pa{1+\alpha\cos{\varphi}}}\notag\\
	&\phantom{=}+\pa{2j-\frac{\alpha^2}{\alpha^2-1}}\frac{R_1\pa{\alpha;\varphi\big|j}}{j+\pa{1-j}\alpha^2},\\
	f_1&=\frac{p_1}{2}\log\ab{\frac{p_1\sqrt{1-j\sin^2{\varphi}}+\sin{\varphi}}{p_1\sqrt{1-j\sin^2{\varphi}}-\sin{\varphi}}},\\
	p_1&=\sqrt{\frac{\alpha^2-1}{j+\pa{1-j}\alpha^2}},
\end{align}
where the auxiliary function $x_3(r)$ is defined as
\begin{align}
	x_3(r)=\frac{A\pa{r-r_1}-B\pa{r-r_2}}{A\pa{r-r_1}+B\pa{r-r_2}}
	\in\pa{-\frac{1}{\alpha_0},1}\subset(-1,1),
\end{align}
while the elliptic modulus is
\begin{align}
	k_3=\frac{(A+B)^2-r_{21}^2}{4AB}
	\in(0,1),
\end{align}
and the other parameters are
\begin{align}
	\alpha_0&=\frac{B+A}{B-A}
	>1,
	&&\alpha_\pm=\frac{Br_{\pm2}+Ar_{\pm1}}{Br_{\pm2}-Ar_{\pm1}}
	=-\frac{1}{x_3(r_\pm)},\\
	A&=\sqrt{r_{32}r_{42}}
	>0,
	&&B=\sqrt{r_{31}r_{41}}
	>0.
\end{align}

Finally, the source radius is given in terms of the Jacobi elliptic cosine function $\cn(\varphi|k)$ by
\begin{align}
	\label{eq:SourceRadius3}
	r_{\rm s}^{(3)}(\tau)&=\frac{\pa{Ar_1-Br_2}-\pa{Ar_1+Br_2}\cn\pa{X_3(\tau)\big|k_3}}{\pa{A-B}-\pa{A+B}\cn\pa{X_3(\tau)\big|k_3}},\\
	X_3(\tau)&=\sqrt{AB}\tau-F\pa{\arccos{x_3(r_{\rm o})}\bigg|k_{3}}.
\end{align}

\section{Accretion flow four-velocities}
\label{app:FourVelocity}

Let us consider an equatorial four-velocity
\begin{align}
    u=u^t\pd_t+u^r\pd_r+u^\phi\pd_\phi
    =u_t\ed t+u_r\ed r+u_\phi\ed\phi,
\end{align}
with $u^\mu$ and $u_\mu$ the contravariant and covariant components, respectively.
The angular and radial-infall velocities are
\begin{align}
    \Omega=\frac{u^\phi}{u^t},\quad
    \iota=-\frac{u^r}{u^t}.
\end{align}
Meanwhile, the energy and specific angular momentum are
\begin{align}
    \mathcal{E}=-u_t,\quad
    \ell=\frac{u_\phi}{\mathcal{E}}.
\end{align}
For geodesic motion, these quantities are conserved, whereas
\begin{align}
    \nu=\frac{u_r}{u_t}
\end{align}
is not.
These variables parameterize the four-velocity as
\begin{align}
    \label{eq:FourVelocity}
    u=u^t\pa{\pd_t-\iota\pd_r+\Omega\pd_\phi}
    =\mathcal{E}\pa{-\ed t-\nu\ed r+\ell\ed\phi}.
\end{align}
Any two components of $u$ determine it entirely, as the third can be recovered from the normalization condition $u\cdot u=-1$.
We will parameterize the four-velocity using either the pair $(\iota,\Omega)$, the pair $(\nu,\ell)$, or the pair $(\mathcal{E},\ell)$, as follows.
First, define
\begin{align}
    \Pi(r)=r^2g_{\phi\phi}\big|_{\theta=\pi/2}
    =\pa{r^2+a^2}^2-a^2\Delta(r).
\end{align}
Given $(\iota,\Omega)$, unit-normalization fixes
\begin{align}
    \label{eq:NormalizationCondition}
    u^t=\br{1-\pa{r^2+a^2}\Omega^2-\frac{2M}{r}\pa{1-a\Omega}^2-\frac{r^2}{\Delta(r)}\iota^2}^{-1/2},
\end{align}
which is physically admissible provided the quantity in square brackets is positive.
We took the positive square root to ensure $u^t>0$ is future-directed.
Lowering the index of $u^\mu$ yields
\begin{subequations}
\label{eq:LowerIndex}
\begin{gather}
    \mathcal{E}=\chi u^t,\quad
    \nu=\frac{r^2}{\Delta(r)}\frac{\iota}{\chi},\quad
    \ell=\frac{\Pi(r)\Omega-2aMr}{r^2\chi},\\
    \chi=1-\frac{2M}{r}\pa{1-a\Omega}.
\end{gather}
\end{subequations}
Conversely, given $(\nu,\ell)$, unit-normalization fixes the energy to
\begin{align}
    \label{eq:EnergyNormalization}
    \mathcal{E}=\sqrt{\frac{\Delta(r)}{\frac{\Pi(r)}{r^2}-\frac{4aM\ell}{r}-\pa{1-\frac{2M}{r}}\ell^2-\br{\frac{\Delta(r)}{r}\nu}^2}},
\end{align}
which is physically admissible provided the quantity under the square root is positive.
Raising the index of $u_\mu$ then yields
\begin{subequations}
\label{eq:RaiseIndex}
\begin{gather}
    u^t=\frac{\mathcal{E}}{\chi},\quad
    \iota=\frac{\Delta(r)}{r^2}\chi\nu,\quad
    \Omega=\frac{\chi}{r^2}\pa{\ell+aH},\\
    \chi=\frac{1}{1+\frac{2M}{r}(1+H)},\quad
    H=\frac{2Mr-a\ell}{\Delta(r)}.
\end{gather}
\end{subequations}
It is often convenient to express $\Omega$ and $\chi$ in terms of $\ell$ only as
\begin{align}
    \label{eq:SimpleForm}
    \Omega=\frac{a+\pa{1-\frac{2M}{r}}\pa{\ell-a}}{\frac{\Pi(r)}{r^2}-\frac{2aM\ell}{r}},\quad
    \chi=\frac{\Delta(r)}{\frac{\Pi(r)}{r^2}-\frac{2aM\ell}{r}}.
\end{align}
Lastly, given $(\mathcal{E},\ell)$, the full four-velocity can be recovered by solving the normalization condition $u\cdot u=-1$ [or equivalently, Eq.~\eqref{eq:EnergyNormalization}] for
\begin{align}
    \label{eq:RadialNormalization}
    \nu=\frac{r}{\Delta(r)}\sqrt{\frac{\Pi(r)}{r^2}-\frac{4aM\ell}{r}-\pa{1-\frac{2M}{r}}\ell^2-\frac{\Delta(r)}{\mathcal{E}^2}}.
\end{align}

Finally, the observed redshift $g(\iota,\Omega)$ is
\begin{align}
    \label{eq:ObservedRedshiftDefinition}
    g=\frac{E}{-p_\mu u^\mu}
    =\frac{1}{u^t\pa{1\pm_r\frac{\sqrt{\mathcal{R}(r)}}{\Delta(r)}\iota-\lambda\Omega}}.
\end{align}
Alternatively, the redshift can also be expressed in terms of $(\mathcal{E},\nu,\ell)$ as [Eq.~\eqref{eq:RaiseIndex}]
\begin{subequations}
\label{eq:ObservedRedshiftLower}
\begin{align}
    g&=\frac{\Delta(r)}{\mathcal{E}\br{G\pm_r\sqrt{\mathcal{R}(r)}\frac{\Delta(r)}{r^2}\nu}},\\
    G&=\frac{\Pi(r)}{r^2}-\pa{1-\frac{2M}{r}}\ell\lambda-\frac{2aM}{r}\pa{\ell+\lambda}.
\end{align}
\end{subequations}

\subsection{Keplerian circular orbits (geodesic motion)}
\label{subapp:KeplerianFlow}

We let $u=\mathring{u}$ denote the four-velocity of timelike, circular-equatorial geodesics.
These orbits define Keplerian motion in the Kerr spacetime; they are only stable if (Eq.~(2.20) of \cite{Bardeen1972})
\begin{align}
	\label{eq:DefinitionISCO}
    r^2-6Mr+8a\sqrt{Mr}-3a^2\ge0,
\end{align}
or equivalently, if $r\ge r_{\rm ms}$, where $r_{\rm ms}$ denotes the radius of the (marginally stable) innermost stable circular orbit (ISCO),
\begin{subequations}
\label{eq:ISCO}
\begin{align}
	r_\mathrm{ms}&=M\br{3+Z_2-\sqrt{\pa{3-Z_1}\pa{3+Z_1+2Z_2}}},\\
	Z_1&=1+\sqrt[3]{1-a_*^2}\br{\sqrt[3]{1+a_*}+\sqrt[3]{1-a_*}},\\
	Z_2&=\sqrt{3a_*^2+Z_1^2},\quad
	a_*=\frac{a}{M}.
\end{align}
\end{subequations}
Inside the ISCO, we must use an inspiraling geodesic motion, as orbits are unstable and particles must fall into the horizon.

\subsubsection{Outside the ISCO radius}

For $r\ge r_{\rm ms}$, we can set $\mathring{\iota}=0$ and the Kerr geodesic equation determines the Keplerian angular velocity (Eq.~(2.16) of \cite{Bardeen1972}):
\begin{align}
    \mathring{\Omega}=\frac{\sqrt{M}}{r^{3/2}+a\sqrt{M}}.
\end{align}
The contravariant four-velocity is then [Eqs.~\eqref{eq:FourVelocity} and \eqref{eq:EnergyNormalization}]
\begin{subequations}
\begin{align}
    \mathring{u}^t&=\frac{r^{3/2}+a\sqrt{M}}{\sqrt{r^3-3Mr^2+2a\sqrt{M}r^{3/2}}},\\
    \mathring{u}^r&=0,\\
    \mathring{u}^\phi&=\frac{\sqrt{M}}{\sqrt{r^3-3Mr^2+2a\sqrt{M}r^{3/2}}},
\end{align}
\end{subequations}
and the Keplerian conserved quantities are [Eqs.~\eqref{eq:LowerIndex}]
\begin{subequations}
\begin{align}
    \mathring{\mathcal{E}}&=\frac{\sqrt{r}\pa{r-2M}+a\sqrt{M}}{\sqrt{r^3-3Mr^2+2a\sqrt{M}r^{3/2}}},\\
    \label{eq:KeplerianAngularMomentum}
    \mathring{\ell}&=\frac{\sqrt{M}\pa{r^2+a^2-2a\sqrt{Mr}}}{\sqrt{r}\pa{r-2M}+a\sqrt{M}}.
\end{align}
\end{subequations}
Likewise [Eqs.~\eqref{eq:LowerIndex}],
\begin{align}
    \mathring{\nu}=0,\quad
    \mathring{\chi}=1-\frac{2M\sqrt{r}}{r^{3/2}+a\sqrt{M}}.
\end{align}
Hence, the covariant four-velocity is [Eq.~\eqref{eq:FourVelocity}]
\begin{subequations}
\begin{align}
    \mathring{u}_t&=-\frac{\sqrt{r}\pa{r-2M}+a\sqrt{M}}{\sqrt{r^3-3Mr^2+2a\sqrt{M}r^{3/2}}},\\
    \mathring{u}_r&=0,\\
    \mathring{u}_\phi&=\frac{\sqrt{M}\pa{r^2-2a\sqrt{Mr}+a^2}}{\sqrt{r^3-3Mr^2+2a\sqrt{M}r^{3/2}}}.
\end{align}
\end{subequations}

Finally, the observed redshift is [Eq.~\eqref{eq:ObservedRedshiftDefinition}]
\begin{align}
	\label{eq:KeplerianRedshiftOutsideISCO}
    \mathring{g}=\frac{\sqrt{r^3-3Mr^2+2a\sqrt{M}r^{3/2}}}{r^{3/2}+\sqrt{M}\pa{a-\lambda}}.
\end{align}
All of the above quantities are physically admissible provided $r^3-3Mr^2+2a\sqrt{M}r^{3/2}>0$, which is the case for all $r\ge r_{\rm ms}$.

\subsubsection{Inside the ISCO radius}

For $r\in[r_+,r_{\rm ms}]$, we follow Cunningham's prescription \cite{Cunningham1975} and smoothly extend the flow past the ISCO using geodesic inspirals with the conserved quantities of the ISCO:
\begin{subequations}
\label{eq:CunninghamPrescription}
\begin{align}
    \mathcal{E}_{\rm ms}=\mathring{\mathcal{E}}\big|_{r=r_{\rm ms}}
    &=\sqrt{1-\frac{2}{3}\frac{M}{r_{\rm ms}}},\\
    \mathcal{\ell}_{\rm ms}=\mathring{\mathcal{\ell}}\big|_{r=r_{\rm ms}}
    &=\frac{\sqrt{M}\pa{r_{\rm ms}^2-2a\sqrt{Mr_{\rm ms}}+a^2}}{\sqrt{r_{\rm ms}}\pa{r_{\rm ms}-2M}+a\sqrt{M}}.
\end{align}
\end{subequations}
This choice results in a nonzero radial component [Eq.~\eqref{eq:RadialNormalization}]
\begin{align}
    \mathring{\nu}=\frac{r^2}{\Delta(r)\mathring{\mathcal{E}}_{\rm ms}}\sqrt{\frac{2}{3}\frac{M}{r_{\rm ms}}}\pa{\frac{r_{\rm ms}}{r}-1}^{3/2}
    >0,
\end{align}
where we used the identity $r_{\rm ms}^2-6Mr_{\rm ms}+8a\sqrt{Mr_{\rm ms}}-3a^2=0$ [Eq.~\eqref{eq:DefinitionISCO}] to simplify the expressions for both $\mathcal{E}_{\rm ms}$ and $\mathring{\nu}$.

It then follows that [Eqs.~\eqref{eq:RaiseIndex}]
\begin{subequations}
\label{eq:InsideISCO}
\begin{gather}
    \mathring{\iota}=\frac{\mathring{\chi}\pa{\frac{r_{\rm ms}}{r}-1}^{3/2}}{\sqrt{\frac{3}{2}\frac{r_{\rm ms}}{M}-1}},\quad
    \mathring{\Omega}=\frac{\mathring{\chi}}{r^2}\pa{\ell_{\rm ms}+a\mathring{H}},\\
    \mathring{\chi}=\frac{1}{1+\frac{2M}{r}\pa{1+\mathring{H}}},\quad
    \mathring{H}=\frac{2Mr-a\ell_{\rm ms}}{\Delta(r)},
\end{gather}
\end{subequations}
so that the covariant four-velocity is explicitly [Eq.~\eqref{eq:FourVelocity}]
\begin{subequations}
\begin{align}
    \mathring{u}_t&=-\sqrt{1-\frac{2}{3}\frac{M}{r_{\rm ms}}},\\
    \mathring{u}_r&=-\frac{r^2}{\Delta(r)}\sqrt{\frac{2}{3}\frac{M}{r_{\rm s}}}\pa{\frac{r_{\rm ms}}{r}-1}^{3/2},\\
    \mathring{u}_\phi&=\frac{\sqrt{M}\pa{r_{\rm ms}^2-2a\sqrt{Mr_{\rm ms}}+a^2}}{\sqrt{r_{\rm ms}^3-3Mr_{\rm ms}^2+2a\sqrt{M}r_{\rm ms}^{3/2}}}.
\end{align}
\end{subequations}
The contravariant four-velocity is then  [Eqs.~\eqref{eq:FourVelocity} and \eqref{eq:RaiseIndex}]
\begin{subequations}
\begin{align}
    \mathring{u}^t&=\sqrt{1-\frac{2}{3}\frac{M}{r_{\rm ms}}}\frac{\Pi(r)-2aM\ell_{\rm ms}r}{r^2\Delta(r)},\\
    \mathring{u}^r&=-\sqrt{\frac{2}{3}\frac{M}{r_{\rm ms}}}\pa{\frac{r_{\rm ms}}{r}-1}^{3/2},\\
    \mathring{u}^\phi&=\sqrt{1-\frac{2}{3}\frac{M}{r_{\rm ms}}}\frac{a+\pa{1-\frac{2M}{r}}\pa{\ell_{\rm ms}-a}}{\Delta(r)},
\end{align}
\end{subequations}
The contravariant components may also be recast in the form
\begin{subequations}
\begin{gather}
    \mathring{u}^t=\frac{\mathcal{E}_{\rm ms}}{\mathring{\chi}},\quad
    \mathring{u}^\phi=\frac{\mathcal{E}_{\rm ms}}{r^2}\pa{\ell_{\rm ms}+a\mathring{H}},
\end{gather}
\end{subequations}
which matches Eqs.~(A12) of \cite{Cunningham1975} since $(\gamma_e,\lambda_e)=(\mathcal{E}_{\rm ms},\mathcal{\ell}_{\rm ms})$.

Finally, the observed redshift is [Eq.~\eqref{eq:ObservedRedshiftLower}]
\begin{subequations}
\label{eq:KeplerianRedshiftInsideISCO}
\begin{align}
    \mathring{g}&=\frac{\Delta(r)}{\sqrt{1-\frac{2}{3}\frac{M}{r_{\rm ms}}}\br{\mathring{G}\pm_r\sqrt{\mathcal{R}(r)}\frac{\pa{\frac{r_{\rm ms}}{r}-1}^{3/2}}{\sqrt{\frac{3}{2}\frac{r_{\rm ms}}{M}-1}}}},\\
    \mathring{G}&=\frac{\Pi(r)}{r^2}-\pa{1-\frac{2M}{r}}\ell_{\rm ms}\lambda-\frac{2aM}{r}\pa{\ell_{\rm ms}+\lambda}.
\end{align}
\end{subequations}

\subsection{Radial infall (geodesic motion)}
\label{subapp:RadialInflow}

We let $u=\bar{u}$ denote the four-velocity of timelike, radially infalling equatorial geodesics.
One class of such trajectories consists of particles that fall in from spatial infinity with zero initial velocity and vanishing (conserved) angular momentum:
\begin{align}
    \bar{\mathcal{E}}=1,\quad
    \bar{\ell}=0.
\end{align}
As particles fall in, they pick up a radial velocity [Eq.~\eqref{eq:RadialNormalization}]
\begin{align}
    \bar{\nu}=\frac{\sqrt{2Mr\pa{r^2+a^2}}}{\Delta(r)}
    >0,
\end{align}
and due to frame-dragging (the off-diagonal term $g_{t\phi}\neq0$), they also acquire an angular velocity $\bar{\Omega}\neq0$ [Eqs.~\eqref{eq:RaiseIndex}]:
\begin{subequations}
\label{eq:RadialInflowAngularVelocity}
\begin{gather}
    \bar{\iota}=\frac{\Delta(r)}{\Pi(r)}\sqrt{2Mr\pa{r^2+a^2}},\quad
    \bar{\Omega}=\frac{2aMr}{\Pi(r)},\\
    \bar{\chi}=\frac{r^2\Delta(r)}{\Pi(r)},\quad
    \bar{H}=\frac{2Mr}{\Delta(r)}.
\end{gather}
\end{subequations}

Explicitly, the covariant four-velocity is then [Eq.~\eqref{eq:FourVelocity}]
\begin{align}
    \bar{u}_t=-1,\quad
    \bar{u}_r=-\frac{\sqrt{2Mr\pa{r^2+a^2}}}{\Delta(r)},\quad
    \bar{u}_\phi=0,
\end{align}
while the contravariant four-velocity is [Eqs.~\eqref{eq:FourVelocity} and \eqref{eq:RaiseIndex}]
\begin{subequations}
\label{eq:RadialInflow}
\begin{align}
    \bar{u}^t&=\frac{\Pi(r)}{r^2\Delta(r)},\\
    \bar{u}^r&=-\frac{\sqrt{2Mr\pa{r^2+a^2}}}{r^2},\\
    \bar{u}^\phi&=\frac{2aM}{r\Delta(r)}.
\end{align}
\end{subequations}

Finally, the observed redshift is [Eq.~\eqref{eq:ObservedRedshiftDefinition}]
\begin{align}
    \bar{g}=\frac{r^2\Delta(r)}{\Pi(r)-2aMr\lambda\pm_r\sqrt{\mathcal{R}(r)}\sqrt{2Mr\pa{r^2+a^2}}}.
\end{align}

\subsection{Sub-Keplerian orbits (non-geodesic motion)}
\label{subapp:SubKeplerianFlow}

We now define a four-velocity $u=\hat{u}$ for timelike, equatorial sub-Keplerian orbits.
Such trajectories cannot be geodesic; following \cite{Penna2013}, we introduce a ``sub-Keplerianity'' parameter $\xi\in(0,1)$ and instead prescribe a non-geodesic motion via
\begin{align}
    \label{eq:SubKeplerianAngularMomentum}
    \hat{\ell}=\xi\mathring{\ell}
    =\xi\frac{\sqrt{M}\pa{r^2+a^2-2a\sqrt{Mr}}}{\sqrt{r}\pa{r-2M}+a\sqrt{M}},
\end{align}
where $\mathring{\ell}$ denotes the Keplerian specific angular momentum \eqref{eq:KeplerianAngularMomentum}.
As for Keplerian motion (with $\xi=1$), we must treat radii outside and inside of the ISCO radius $r_{\rm ms}$ separately.

\subsubsection{Outside the ISCO radius}

For $r\ge r_{\rm ms}$, we demand that, as for Keplerian orbits,
\begin{align}
    \hat{\nu}=0.
\end{align}
This fixes the sub-Keplerian orbital energy to be [Eq.~\eqref{eq:EnergyNormalization}]
\begin{align}
    \hat{\mathcal{E}}=\sqrt{\frac{\Delta(r)}{\frac{\Pi(r)}{r^2}-\frac{4aM\hat{\ell}}{r}-\pa{1-\frac{2M}{r}}\hat{\ell}^2}}
    >0,
\end{align}
which is manifestly real for all $\xi\in(0,1)$, since the quantity in the square root is strictly greater than when $\xi=1$.

It then follows that [Eqs.~\eqref{eq:RaiseIndex}--\eqref{eq:SimpleForm}]
\begin{align}
    \label{eq:SubKeplerianAngularVelocityInsideISCO}
    \hat{\iota}=0,\quad
    \hat{\Omega}=\frac{a+\pa{1-\frac{2M}{r}}\pa{\hat{\ell}-a}}{\frac{\Pi(r)}{r^2}-\frac{2aM\hat{\ell}}{r}},
\end{align}
so that the covariant four-velocity is explicitly [Eq.~\eqref{eq:FourVelocity}]
\begin{align}
    \hat{u}_t=-\hat{\mathcal{E}},\quad
    \hat{u}_r=0,\quad
    \hat{u}_\phi=\hat{\ell}\hat{\mathcal{E}}.
\end{align}
The contravariant four-velocity is then [Eqs.~\eqref{eq:FourVelocity} and \eqref{eq:RaiseIndex}]
\begin{subequations}
\label{eq:SubKeplerianOutsideISCO}
\begin{align}
    \hat{u}^t&=\frac{\hat{\mathcal{E}}}{\Delta(r)}\br{\frac{\Pi(r)}{r^2}-\frac{2aM\hat{\ell}}{r}},\\
    \hat{u}^r&=0,\\
    \hat{u}^\phi&=\frac{\hat{\mathcal{E}}}{\Delta(r)}\br{a+\pa{1-\frac{2M}{r}}\pa{\hat{\ell}-a}}.
\end{align}
\end{subequations}

Finally, the observed redshift is [Eq.~\eqref{eq:ObservedRedshiftLower}]
\begin{align}
    \hat{g}&=\frac{\Delta(r)}{\hat{\mathcal{E}}\br{\frac{\Pi(r)}{r^2}-\pa{1-\frac{2M}{r}}\hat{\ell}\lambda-\frac{2aM}{r}\pa{\hat{\ell}+\lambda}}}.
\end{align}

\subsubsection{Inside the ISCO radius}

For $r\in[r_+,r_{\rm ms}]$, we do not expect orbits to remain circular.
Instead, we use Cunningham's prescription \eqref{eq:CunninghamPrescription} to smoothly extend the sub-Keplerian condition \eqref{eq:SubKeplerianAngularMomentum} past the ISCO using the conserved quantities of the sub-Keplerian ISCO orbit,
\begin{align}
    \hat{\mathcal{E}}_{\rm ms}=\hat{\mathcal{E}}\big|_{r=r_{\rm ms}},\quad
    \hat{\ell}_{\rm ms}&=\hat{\ell}\big|_{r=r_{\rm ms}}
    =\xi\ell_{\rm ms},
\end{align}
or more explicitly,
\begin{subequations}
\begin{align}
    \hat{\mathcal{E}}_{\rm ms}&=\sqrt{\frac{\Delta(r_{\rm ms})}{\frac{\Pi(r_{\rm ms})}{r_{\rm ms}^2}-\frac{4aM\hat{\ell}_{\rm ms}}{r_{\rm ms}}-\pa{1-\frac{2M}{r_{\rm ms}}}\hat{\ell}_{\rm ms}^2}},\\
    \hat{\ell}_{\rm ms}&=\xi\frac{\sqrt{M}\pa{r_{\rm ms}^2-2a\sqrt{Mr_{\rm ms}}+a^2}}{\sqrt{r_{\rm ms}}\pa{r_{\rm ms}-2M}+a\sqrt{M}}.
\end{align}
\end{subequations}
This choice results in a nonzero radial component [Eq.~\eqref{eq:RadialNormalization}]
\begin{align}
    \hat{\nu}=\frac{r}{\Delta(r)}\sqrt{\frac{\Pi(r)}{r^2}-\frac{4aM\hat{\ell}_{\rm ms}}{r}-\pa{1-\frac{2M}{r}}\hat{\ell}_{\rm ms}^2-\frac{\Delta(r)}{\hat{\mathcal{E}}_{\rm ms}}}.
\end{align}

It then follows that [Eqs.~\eqref{eq:RaiseIndex}]
\begin{subequations}
\label{eq:SubKeplerianAngularVelocityOutsideISCO}
\begin{gather}
    \hat{\iota}=\frac{\br{\Delta(r)}^2\hat{\nu}}{\Pi(r)-2aM\hat{\ell}_{\rm ms}r},\quad
    \hat{\Omega}=\frac{\hat{\chi}}{r^2}\pa{\hat{\ell}_{\rm ms}+a\hat{H}},\\
    \hat{\chi}=\frac{1}{1+\frac{2M}{r}\pa{1+\hat{H}}},\quad
    \hat{H}=\frac{2Mr-a\ell_{\rm ms}}{\Delta(r)},
\end{gather}
\end{subequations}
so that the covariant four-velocity is explicitly [Eq.~\eqref{eq:FourVelocity}]
\begin{align}
    \hat{u}_t=-\hat{\mathcal{E}}_{\rm ms},\quad
    \hat{u}_r=-\hat{\nu}\hat{\mathcal{E}}_{\rm ms},\quad
    \hat{u}_\phi=\hat{\ell}_{\rm ms}\hat{\mathcal{E}}_{\rm ms}.
\end{align}
The contravariant four-velocity is then [Eqs.~\eqref{eq:FourVelocity} and \eqref{eq:RaiseIndex}]
\begin{subequations}
\label{eq:SubKeplerianInsideISCO}
\begin{align}
    \hat{u}^t&=\frac{\hat{\mathcal{E}}_{\rm ms}}{\Delta(r)}\br{\frac{\Pi(r)}{r^2}-\frac{2aM\hat{\ell}_{\rm ms}}{r}},\\
    \hat{u}^r&=-\frac{\Delta(r)}{r^2}\hat{\nu}\hat{\mathcal{E}}_{\rm ms},\\
    \hat{u}^\phi&=\frac{\hat{\mathcal{E}}_{\rm ms}}{\Delta(r)}\br{a+\pa{1-\frac{2M}{r}}\pa{\hat{\ell}_{\rm ms}-a}}.
\end{align}
\end{subequations}

Finally, the observed redshift is [Eq.~\eqref{eq:ObservedRedshiftLower}]
\begin{subequations}
\begin{align}
    \hat{g}&=\frac{\Delta(r)}{\hat{\mathcal{E}}_{\rm ms}\br{\hat{G}\pm_r\sqrt{\mathcal{R}(r)}\frac{\Delta(r)}{r^2}\hat{\nu}}},\\
    \hat{G}&=\frac{\Pi(r)}{r^2}-\pa{1-\frac{2M}{r}}\hat{\ell}_{\rm ms}\lambda-\frac{2aM}{r}\pa{\hat{\ell}_{\rm ms}+\lambda}.
\end{align}
\end{subequations}

\subsection{General flow (non-geodesic motion)}
\label{subapp:GeneralFlow}

We can linearly superpose the preceding four-velocities to obtain a general flow $u=\tilde{u}$ combining both circular motion and radial inflow.
Following \cite{Pu2016,Vincent2022}, we do so by defining
\begin{align}
    \tilde{u}^r&=\hat{u}^r+\pa{1-\beta_r}\pa{\bar{u}^r-\hat{u}^r},\\
    \tilde{\Omega}&=\hat{\Omega}+\pa{1-\beta_\phi}\pa{\bar{\Omega}-\hat{\Omega}},
\end{align}
where $\hat{u}$ is the four-velocity of the sub-Keplerian flow, given by Eq.~\eqref{eq:SubKeplerianOutsideISCO} for $r\ge r_{\rm ms}$ (outside the ISCO) and by  Eq.~\eqref{eq:SubKeplerianInsideISCO} for $r_+<r<r_{\rm ms}$ (inside the ISCO), while $\bar{u}$ is the radial inflow four-velocity \eqref{eq:RadialInflow}.
Likewise, $\hat{\Omega}=\hat{u}^\phi/\hat{u}^t$ is the sub-Keplerian angular velocity, given by Eq.~\eqref{eq:SubKeplerianAngularVelocityOutsideISCO} outside the ISCO and by Eq.~\eqref{eq:SubKeplerianAngularVelocityOutsideISCO} inside the ISCO, while $\bar{\Omega}=\bar{u}^\phi/\bar{u}^t$ is the angular velocity \eqref{eq:RadialInflowAngularVelocity} of radial inflow.

Given $\tilde{u}^r$ and $\tilde{\Omega}=\tilde{u}^\phi/\tilde{u}^t$, we can solve the normalization condition $u\cdot u=-1$ for $\tilde{u}^t$, the only missing component of the four-velocity, resulting in a simple modification of Eq.~\eqref{eq:NormalizationCondition}:
\begin{align}
    \label{eq:ur}
    \tilde{u}^t=\sqrt{\frac{1+\frac{r^2}{\Delta(r)}\pa{\tilde{u}^r}^2}{1-\pa{r^2+a^2}\tilde{\Omega}^2-\frac{2M}{r}\pa{1-a\tilde{\Omega}}^2}},
\end{align}
This fully specifies the general four-velocity $\tilde{u}$.

For generic values of the three parameters $(\xi,\beta_r,\beta_\phi)$, the flow $\tilde{u}$ is not geodesic.
However, when $\beta_r=\beta_\phi=0$, $\tilde{u}$ reduces to the radial inflow $\bar{u}$, which is $\xi$-independent and geodesic.
At the opposite end, when $\beta_r=\beta_\phi=1$, $\tilde{u}$ reduces to the sub-Keplerian flow $\tilde{u}(\xi)$, which in turn reduces when $\xi=1$ to geodesic Keplerian motion $\mathring{u}=\tilde{u}(1)$.

\section{More on the Mat\'ern covariance}
\label{app:MaternCovariance}

This appendix offers a field-theoretic interpretation of the Gaussian random field with Mat\'ern covariance.

\subsection{Classical field}

A free, massive scalar field $\Phi(\mathbf{x})$ with mass $m$ in Euclidean spacetime $\mathbb{R}^d$ is described by the Lagrangian
\begin{align}
	\label{eq:ScalarLagrangian}
	\mathcal{L}=\frac{1}{2}\pd_\mu\Phi\pd^\mu\Phi+\frac{1}{2}m^2\Phi^2,
\end{align}
and therefore obeys the Euler-Lagrange field equation
\begin{align}
	\frac{\pd\mathcal{L}}{\pd\Phi}=\pd_\mu\pa{\frac{\pd\mathcal{L}}{\pd\pa{\pd_\mu\Phi}}},
\end{align}
which in this case is the classical Klein-Gordon wave equation
\begin{align}
	\label{eq:KleinGordon}
	\pa{\nabla^2-m^2}\Phi(\mathbf{x})=0,
\end{align}
where $\nabla^2=\pd_\mu\pd^\mu$ is the scalar Laplacian.
The Green function $\Delta_d(\mathbf{x})$ for this equation is defined by
\begin{align}
	\label{eq:GreenFunction}
	\pa{\nabla^2-m^2}\Delta_d(\mathbf{x})=-\delta^{(d)}(\mathbf{x}).
\end{align}
In momentum space, this reads $\pa{k^2+m^2}\tilde{\Delta}_d(\mathbf{k})=1$, so
\begin{align}
	\tilde{\Delta}_d(\mathbf{k})=\frac{1}{k^2+m^2},
\end{align}
where $k^2=\mathbf{k}\cdot\mathbf{k}$.
The inverse Fourier transform \eqref{eq:FourierConventions} yields
\begin{align}
	\label{eq:Propagator}
    \Delta_d(\mathbf{x})=\int\frac{e^{i\mathbf{k}\cdot\mathbf{x}}}{\mathbf{k}^2+m^2}\frac{\ed^d\mathbf{k}}{(2\pi)^d}
    =\frac{m^{d-2}}{(2\pi)^\frac{d}{2}}(mx)^{1-\frac{d}{2}}K_{1-\frac{d}{2}}(mx),
\end{align}
where $x^2=\mathbf{x}\cdot\mathbf{x}$.
We present a full derivation of this nontrivial identity in Eq.~\eqref{eq:FullPropagator} below.

\subsection{Quantum field}

Integrating the Lagrangian \eqref{eq:ScalarLagrangian} by parts recasts it as
\begin{align}
	\mathcal{L}=-\frac{1}{2}\Phi\pa{\nabla^2-m^2}\Phi.
\end{align}
The path integral over this Lagrangian defines the Euclidean quantum field theory of a free, massive scalar field $\Phi(\mathbf{x})$ with mass $m$ in $\mathbb{R}^d$, which is fully characterized by its two-point function: the Feynman propagator \eqref{eq:Propagator}.

Letting $\nu=1-d/2$ and $\lambda=1/m$, the propagator becomes
\begin{align}
    \Delta_d(\mathbf{x})=\frac{m^{d-2}}{(2\pi)^\frac{d}{2}}\pa{\frac{x}{\lambda}}^\nu K_\nu\pa{\frac{x}{\lambda}},
\end{align}
which is (proportional to) a Mat\'ern covariance.

We conclude that a $d$-dimensional Gaussian random field with Mat\'ern covariance of order $\nu=1-d/2$ and correlation length $\lambda$ is equivalent to a (Euclidean) quantum scalar field of Compton wavelength $\lambda$ (and therefore mass $m=1/\lambda$).

\subsection{Higher-derivative fields}

Consider now the higher-derivative theory
\begin{align}
	\mathcal{L}=-\frac{1}{2}\Phi\pa{\nabla^2-m^2}^n\Phi,
\end{align}
where $n$ is an integer power.
This is a free (Gaussian) theory because it remains quadratic in the field for any $n$.
Classically, the field obeys the higher-derivative wave equation
\begin{align}
	\pa{\nabla^2-m^2}^n\Phi(\mathbf{x})=0,
\end{align}
whose Green function $\Delta_{d,n}(\mathbf{x})$ is defined by
\begin{align}
	\pa{\nabla^2-m^2}^n\Delta_{d,n}(\mathbf{x})=-\delta^{(d)}(\mathbf{x}).
\end{align}
In momentum space, this reads $\pa{k^2+m^2}^n\tilde{\Delta}_{d,n}(\mathbf{k})=1$, so
\begin{align}
	\tilde{\Delta}_{d,n}(\mathbf{k})=\frac{1}{\pa{k^2+m^2}^n},
\end{align}
and the inverse Fourier transform \eqref{eq:FourierConventions} yields
\begin{align}
	\label{eq:GeneralPropagator}
    \Delta_{d,n}(\mathbf{x})&=\int\frac{e^{i\mathbf{k}\cdot\mathbf{x}}}{\pa{\mathbf{k}^2+m^2}^n}\frac{\ed^d\mathbf{k}}{(2\pi)^d}
    \propto\pa{mx}^{n-\frac{d}{2}}K_{n-\frac{d}{2}}(mx).
\end{align}
This nontrivial identity, including the proportionality factor, is derived in Eq.~\eqref{eq:FullPropagator} below.

Letting $\nu=n-d/2$ and $\lambda=1/m$, the propagator becomes
\begin{align}
    \Delta_{d,n}(\mathbf{x})=C\pa{\frac{x}{\lambda}}^\nu K_\nu\pa{\frac{x}{\lambda}},
\end{align}
which is (proportional to) a Mat\'ern covariance.

We conclude that a $d$-dimensional Gaussian random field with Mat\'ern covariance of order $\nu=n-d/2$ and correlation length $\lambda$ is equivalent to a (Euclidean) quantum scalar field of Compton wavelength $\lambda$ (and therefore mass $m=1/\lambda$) with a kinetic term including $2n$ derivatives.

In Lorentzian signature, such a kinetic term would lead to problems with causality, but in Euclidean signature this seems like an acceptable statistical field.

\subsection{Connection with the associated SPDE}

A $d$-dimensional Gaussian random field $\Phi(\mathbf{x})$ with Mat\'ern covariance of order $\nu=n-d/2$ and correlation length $\lambda=1/m$ obeys the stochastic PDE
\begin{align}
	\label{eq:StochasticPDE}
	\pa{m^2-\nabla^2}^\frac{n}{2}\Phi(\mathbf{x})=\mathcal{W}(\mathbf{x}),
\end{align}
where the pseudodifferential operator on the LHS is defined via its spectral properties \cite{Lindgren2011}, and the Gaussian random field $\mathcal{W}(\mathbf{x})$ is the standard white noise process defined in Sec.~\ref{subsec:GRF}.

We now wish to reconcile this statement with more familiar facts from quantum field theory, particularly in the case $n=1$.

When $n=1$, the Gaussian random field $\Phi(\mathbf{x})$ is a Euclidean scalar field.
Yet, Eq.~\eqref{eq:StochasticPDE} implies that it also obeys
\begin{align}
	\label{eq:ScalarStochasticPDE}
	\sqrt{m^2-\nabla^2}\Phi(\mathbf{x})=\mathcal{W}(\mathbf{x}),
\end{align}
which seems to impose yet another condition on the field.
We now show how this statement is consistent with the propagator \eqref{eq:Propagator}, which fully characterizes the behavior of the field.

The key is to multiply two copies of this equation inserted at two different points $\mathbf{x}_1$ and $\mathbf{x}_2$, keeping track of which position the derivatives in the Laplacian act upon:
\begin{align}
	\sqrt{m^2-\nabla_1^2}\Phi(\mathbf{x}_1)\sqrt{m^2-\nabla_2^2}\Phi(\mathbf{x}_2)=\mathcal{W}(\mathbf{x}_1)\mathcal{W}(\mathbf{x}_2).
\end{align}
We now massage the LHS as follows.
First, we are free to move the derivatives to the left, since $\nabla_i^2$ acts only on $\Phi(\mathbf{x}_i)$:
\begin{align}
	\sqrt{m^2-\nabla_1^2}\sqrt{m^2-\nabla_2^2}\Phi(\mathbf{x}_1)\Phi(\mathbf{x}_2)=\mathcal{W}(\mathbf{x}_1)\mathcal{W}(\mathbf{x}_2).
\end{align}
Next, we take expectation values of the fields on both sides:
\begin{align}
	\sqrt{\pa{m^2-\nabla_1^2}\pa{m^2-\nabla_2^2}}\av{\Phi(\mathbf{x}_1)\Phi(\mathbf{x}_2)}=\av{\mathcal{W}(\mathbf{x}_1)\mathcal{W}(\mathbf{x}_2)}.
\end{align}
By definition, the two-point function on the LHS is of course the Euclidean quantum propagator
\begin{align}
	\av{\Phi(\mathbf{x}_1)\Phi(\mathbf{x}_2)}=\Delta_d(\mathbf{x}_1-\mathbf{x}_2).
\end{align}
As for the two-point function on the RHS, it is none other than the autocorrelation function of Gaussian white noise,
\begin{align}
	\av{\mathcal{W}(\mathbf{x}_1)\mathcal{W}(\mathbf{x}_2)}=\delta^{(d)}(\mathbf{x}_1-\mathbf{x}_2),
\end{align}
which is by definition a delta function.
We thus have
\begin{align}
	\sqrt{\pa{m^2-\nabla_1^2}\pa{m^2-\nabla_2^2}}\Delta_d(\mathbf{x}_1-\mathbf{x}_2)=\delta^{(d)}(\mathbf{x}_1-\mathbf{x}_2).
\end{align}
Since both two-point functions depend only on $\mathbf{x}=\mathbf{x}_1-\mathbf{x}_2$, we have $\nabla_1^2=\nabla_2^2=\nabla^2$, and thus we are left with
\begin{align}
	\pa{m^2-\nabla^2}\Delta_d(\mathbf{x})=\delta^{(d)}(\mathbf{x}).
\end{align}
This is exactly Eq.~\eqref{eq:GreenFunction}, proving that it is consistent for the Euclidean quantum field $\Phi(\mathbf{x})$ to also obey the SPDE \eqref{eq:ScalarStochasticPDE}.
This is the quantum-statistical analog of the classical field equation \eqref{eq:KleinGordon}: the presence of the white noise source term $\mathcal{W}(\mathbf{x})$ is what lends the Euclidean field its statistical nature.

Finally, when $n=2$, the Gaussian random field $\Phi(\mathbf{x})$ obeys
\begin{align}
	\pa{m^2-\nabla^2}\Phi(\mathbf{x})=\mathcal{W}(\mathbf{x}),
\end{align}
which is a linear SPDE and therefore straightforward to solve.
This is the reason why it is implemented in \texttt{inoisy} \cite{Lee2021}.

\subsection{Derivation of the scalar propagator}

Here, we derive Eqs.~\eqref{eq:Propagator} and \eqref{eq:GeneralPropagator}, which are equivalent to the position-space formula \eqref{eq:MaternCovariance} for the Mat\'ern covariance.
In spherical coordinates,
\begin{align}
    \Delta_{d,n}(\mathbf{x})=\frac{\Omega_{d-1}}{(2\pi)^d}\int_0^\infty\frac{k^{d-1}}{\pa{k^2+m^2}^n}\int_0^\pi e^{ikx\cos{\theta}}\sin^{d-2}{\theta}\ed\theta\ed k,
\end{align}
where $\Omega_n$ denotes the solid angle on the sphere $S^n$, which is
\begin{align}
    \Omega_n=\frac{2\pi^\frac{n}{2}}{\Gamma\pa{\frac{n}{2}}}.
\end{align}
We now need two identities from Gradshteyn and Rhyzik \cite{Gradshteyn2007}: by 3.915-5 (p492), we have for $\re\nu>-1/2$,
\begin{align}
    \int_0^\pi e^{i\beta\cos{\theta}}\sin^{2\nu}\theta\ed\theta=\sqrt{\pi}\pa{\frac{2}{\beta}}^\nu\Gamma\pa{\nu+\frac{1}{2}}J_\nu(\beta),
\end{align}
and by 6.565-4 (p678), we have for $-1<\re\nu<\re(2\mu+2/3)$, $r>0$, and $m>0$,
\begin{align}
    \int_0^\infty\frac{k^{\nu+1}J_\nu(kx)}{\pa{k^2+m^2}^{\mu+1}}\ed k=\frac{m^{\nu-\mu}x^\mu}{2^\mu\Gamma(\mu+1)}K_{\nu-\mu}(mx).
\end{align}
Using the first formula with $\beta=kx$ and $\nu=d/2-1$, we find
\begin{align}
    \Delta_{d,n}(\mathbf{x})=X\int_0^\infty\frac{k^{d-1}}{\pa{k^2+m^2}^n}\pa{\frac{2}{kx}}^{\frac{d}{2}-1}J_{\frac{d}{2}-1}(kx)\ed k,
\end{align}
where the prefactor is
\begin{align}
    X=\frac{\Omega_{d-1}}{(2\pi)^d}\sqrt{\pi}\Gamma\pa{\frac{d-1}{2}}
    =\frac{2\pi^\frac{d-1}{2}}{(2\pi)^d}\sqrt{\pi}
    =\frac{2^{1-\frac{d}{2}}}{(2\pi)^\frac{d}{2}}.
\end{align}
We can thus rewrite
\begin{align}
    \Delta_{d,n}(\mathbf{x})=\frac{x^{1-\frac{d}{2}}}{(2\pi)^\frac{d}{2}}\int_0^\infty\frac{k^\frac{d}{2}}{\pa{k^2+m^2}^n}J_{\frac{d}{2}-1}(kx)\ed k.
\end{align}
The second identity with $\nu=d/2-1$ and $\mu=n-1$ then yields
\begin{align}
    \Delta_{d,n}(\mathbf{x})=\frac{x^{1-\frac{d}{2}}}{(2\pi)^\frac{d}{2}}\frac{m^{\frac{d}{2}-n}x^{n-1}}{2^{n-1}\Gamma(n)}K_{\frac{d}{2}-n}(mx),
\end{align}
so we finally obtain
\begin{align}
    \Delta_{d,n}(\mathbf{x})=\frac{2^{1-n}}{(2\pi)^\frac{d}{2}(n-1)!}\pa{\frac{m}{x}}^{\frac{d}{2}-n}K_{\frac{d}{2}-n}(mx).
\end{align}
Since $K_{-\nu}(x)=K_\nu(x)$, this exactly reproduces Eq.~\eqref{eq:GeneralPropagator}:
\begin{align}
	\label{eq:FullPropagator}
    \Delta_{d,n}(\mathbf{x})=\frac{2^{1-n}m^{d-2n}}{(2\pi)^\frac{d}{2}(n-1)!}\pa{mx}^{n-\frac{d}{2}}K_{n-\frac{d}{2}}(mx),
\end{align}
and setting $n=1$ recovers Eq.~\eqref{eq:Propagator}.

\bibliographystyle{apsrev4-1}
\bibliography{AART}

\end{document}